\documentclass[aps,prd,10pt,twocolumn,superscriptaddress,showpacs,longbibliography]{revtex4-1}

\pdfoutput=1
\usepackage{hyperref}
\usepackage{graphicx}
\usepackage{amsfonts,amsmath,amssymb,bm,bbm}
\usepackage{color}

\hypersetup{
    pdfnewwindow=true,    
    colorlinks=true,      
    linkcolor=blue,       
    citecolor=blue,       
    filecolor=blue,       
    urlcolor=blue         
}

\newcommand{\sag}{SA$\gamma$}

\begin{document}

\title{Unexpected Dip in the Solar Gamma-Ray Spectrum}

\author{Qing-Wen Tang}
\email{qwtang@ncu.edu.cn}
\thanks{\scriptsize \!\!  \href{http://orcid.org/0000-0001-7471-8451}{orcid.org/0000-0001-7471-8451} }
\affiliation{Center for Cosmology and AstroParticle Physics (CCAPP), Ohio State University, Columbus, Ohio 43210, USA}
\affiliation{Department of Physics, Nanchang University, Nanchang 330031, China}

\author{Kenny C. Y. Ng}
\email{ chun-yu.ng@weizmann.ac.il}
\thanks{\scriptsize \!\! \href{http://orcid.org/0000-0001-8016-2170}{orcid.org/0000-0001-8016-2170}}
\affiliation{Department of Particle Physics and Astrophysics, Weizmann Institute of Science, Rehovot 76100, Israel}

\author{Tim Linden}
\email{linden.70@osu.edu}
\thanks{\scriptsize \!\!  \href{http://orcid.org/0000-0001-9888-0971}{orcid.org/0000-0001-9888-0971}}
\affiliation{Center for Cosmology and AstroParticle Physics (CCAPP), Ohio State University, Columbus, Ohio 43210, USA}

\author{Bei~Zhou}
\email{zhou.1877@osu.edu}
\thanks{\scriptsize \!\!  \href{http://orcid.org/0000-0003-1600-8835}{orcid.org/0000-0003-1600-8835}}
\affiliation{Center for Cosmology and AstroParticle Physics (CCAPP), Ohio State University, Columbus, Ohio 43210, USA}
\affiliation{Department of Physics, Ohio State University, Columbus, Ohio 43210, USA}

\author{John F. Beacom}
\email{beacom.7@osu.edu}
\thanks{\scriptsize \!\!  \href{http://orcid.org/0000-0002-0005-2631}{orcid.org/0000-0002-0005-2631}}
\affiliation{Center for Cosmology and AstroParticle Physics (CCAPP), Ohio State University, Columbus, Ohio 43210, USA}
\affiliation{Department of Physics, Ohio State University, Columbus, Ohio 43210, USA}
\affiliation{Department of Astronomy, Ohio State University, Columbus, Ohio 43210, USA}

\author{Annika H. G. Peter}
\email{apeter@physics.osu.edu}
\thanks{ \scriptsize \!\!  \href{http://orcid.org/0000-0002-8040-6785}{orcid.org/0000-0002-8040-6785}}
\affiliation{Center for Cosmology and AstroParticle Physics (CCAPP), Ohio State University, Columbus, Ohio 43210, USA}
\affiliation{Department of Physics, Ohio State University, Columbus, Ohio 43210, USA}
\affiliation{Department of Astronomy, Ohio State University, Columbus, Ohio 43210, USA}

\date{9th May 2018}

\begin{abstract}
\noindent The solar disk is a bright source of multi-GeV gamma rays, due to the interactions of hadronic cosmic rays with the solar atmosphere.  However, the underlying production mechanism is not understood, except that its efficiency must be greatly enhanced by magnetic fields that redirect some cosmic rays from ingoing to outgoing before they interact.  To elucidate the nature of this emission, we perform a new analysis of solar atmospheric gamma rays with 9 years of Fermi-LAT data, which spans nearly the full 11-year solar cycle.  We detect significant gamma-ray emission from the solar disk from 1\,GeV up to $\gtrsim200$\,GeV.  The overall gamma-ray spectrum is much harder~($\sim E_{\gamma}^{-2.2}$) than the cosmic-ray spectrum~($\sim E_{\rm CR}^{-2.7}$).  We find a clear anticorrelation between the solar cycle phase and the gamma-ray flux between 1--10~GeV.  Surprisingly, we observe a spectral dip between $\sim$30--50\,GeV in an otherwise power-law spectrum.  This was not predicted, is not understood, and may provide crucial clues to the gamma-ray emission mechanism.  The flux above 100\,GeV, which is brightest during the solar minimum, poses exciting opportunities for HAWC, LHAASO, IceCube, and KM3NeT.

\end{abstract}

\maketitle


\section{Introduction}
\label{sec:introduction}

The solar disk is a bright, hard-spectrum gamma-ray source due to its constant bombardment by hadronic cosmic rays, which interact inelastically with gas in the solar atmosphere (including deep under the optical photosphere) and produce gamma rays. If magnetic fields are ignored, the expected gamma-ray flux is small, as only cosmic rays skimming the solar surface and directed towards Earth (the solar limb) produce unabsorbed gamma rays~\cite{Zhou:2016ljf}.  However, the pioneering work of Seckel, Stanev, and Gaisser~\cite{Seckel:1991ffa}~(hereafter SSG1991) showed that solar atmospheric magnetic fields can redirect some cosmic rays from ingoing to outgoing before they interact.  
This greatly enhances the gamma-ray flux, as the entire Earth-facing solar disk can emit gamma rays.  These gamma rays from the solar disk, which we denote as the solar atmospheric gamma rays~(\sag), are a new probe of cosmic rays in the solar system~\cite{Potgieter:2013pdj}, solar atmospheric magnetic fields~\cite{2014A&ARv..22...78W}, and potentially new physics~\cite{2017PDU....17...13B, Leane:2017vag, Arina:2017sng}.  

In addition to \sag, two other processes produce gamma rays from the solar position. Inverse-Compton~(IC) scattering between cosmic-ray electrons and sunlight~\cite{Orlando:2006zs, Moskalenko:2006ta, Orlando:2013pza} produces a diffuse gamma-ray halo that is most luminous just outside of the solar disk.  Also, particles accelerated during energetic solar events, such as solar flares and coronal mass ejections~\cite{Kafexhiu:2018wmh}, can produce $\sim$\,GeV gamma rays.  These, however, are transients events with much lower maximum energy~($<$4\,GeV~\citep{Fermi-LAT:2013cla, Pesce-Rollins:2015hpa, Ackermann:2017uer}) compared to the \sag\ flux.

The first evidence of \sag\ were reported in Ref.~\cite{Orlando:2008uk} using EGRET data, following limits from earlier studies~\cite{ROG:ROG56,JGRA:JGRA13592}.  Shortly after the launch of the Fermi Gamma-Ray Space Telescope, the Sun was detected with the LAT (Large Area Telescope)~(\cite{Abdo:2011xn}, hereafter Fermi2011) from 0.1\,GeV to 10\,GeV, with much better background separation.  Importantly, this analysis found that the \sag\ flux is significantly higher than even the magnetically enhanced flux predicted by SSG1991.  Later, a new analysis~(\cite{Ng:2015gya}, hereafter NBPR2016), using six years of Fermi data, confirmed this result, extended the \sag\ detection up to 30\,GeV, and found hints of a signal up to 100\,GeV, with a spectrum that is significantly harder than predicted by SSG1991.  In addition, NBPR2016 also found that the \sag\ flux anticorrelates with the solar activity with a large amplitude, a result that was hinted at by Refs.~\cite{Orlando:2008uk, Abdo:2011xn}, but was not theoretically predicted.

In Refs.~\cite{Ng:2015gya, Zhou:2016ljf}, we showed that powerful new tests of \sag\ will be possible with TeV gamma-ray experiments.  The Sun is a unique target for large ground-based gamma-ray experiments, such as ARGO-YBJ~\cite{Aielli:2006cj}, Tibet AS-$\gamma$~\cite{Hibino:1988er}, HAWC~\cite{Abeysekara:2013tza, Abeysekara:2017mjj}, and LHAASO~(\cite{Zhen:2014zpa, He:2016del}, under construction).  The TeV regime is important for \sag, as it likely contains the transition between the low-energy regime, where solar magnetic fields enhance the \sag\ production, and the high-energy regime, where such effects terminate.  Even a limit on the solar flux at TeV energies will provide critical information on the magnetic environment that is responsible for \sag\ production~\cite{Ng:2015gya, Zhou:2016ljf}.

The production of \sag\ necessitates the production of solar atmospheric neutrinos~\cite{Moskalenko:1991hm, Seckel:1991ffa, Moskalenko:1993ke, Ingelman:1996mj, Hettlage:1999zr, Fogli:2006jk, Arguelles:2017eao, Ng:2017aur, Edsjo:2017kjk, Masip:2017gvw}.  The Sun can be detected by large neutrino telescopes~\cite{Ng:2017aur}, such as IceCube, where a search is ongoing~\cite{icecubesolar}, and KM3NeT~\cite{Adrian-Martinez:2016fdl}, in the future.  A detection of these neutrinos will provide important and complementary information concerning the physics of solar cosmic-ray interactions.  Additionally, these neutrinos are backgrounds for solar dark matter searches, and could constitute an impenetrable sensitivity floor~\cite{Arguelles:2017eao, Ng:2017aur, Edsjo:2017kjk}.  It is therefore important to have a robust prediction for the solar atmospheric neutrino flux, which can only be achieved with an understanding of the gamma-ray flux. 

The continuous operation of Fermi provides a steady stream of data probing the evolution of solar gamma rays over the entire solar cycle.  In this work, we revisit solar observations with an improved data set~(\texttt{Pass 7} to \texttt{Pass 8}) and an improved analysis technique, aiming to address several unanswered questions from NBPR2016. In particular, we aim to extend the detection of \sag\ above 30~GeV, in search of a spectral softening that may reveal the termination of magnetic field enhancements from the SSG1991 mechanism. Second, we utilize the additional 3~years of data provided by Fermi to definitively test the anticorrelation between the \sag\ flux and solar activity.  
Addressing both issues will lay the foundation for building a better model of cosmic-ray interactions in the Sun.  

The paper is structured as follows.  We describe the analysis method in Sec.~\ref{sec:analysis} and the physics results in Sec.~\ref{sec:results}.  We investigate an apparent spectral dip in Sec.~\ref{sec:dip}, and study the \sag\ flux during solar minimum in Sec.~\ref{sec:vhe}.  Lastly, we discuss the implications of this work in Sec.~\ref{sec:implications}, and then conclude in Sec.~\ref{sec:conclusions}.  Here we focus on the \sag\ spectrum above 1\,GeV and its time variation.  In a companion work~\cite{Linden:2018}, we pay special attention to the highest part of the energy spectrum~(above 100\,GeV) and, for the first time, study resolved images of the solar disk in gamma rays.


\section{Fermi data Analysis}\label{sec:analysis}
\subsection{Data selection}

Our analysis procedure is similar to that of NBPR2016, except that we use an improved photon-position calculation. Specifically, we utilize 468 weeks of \texttt{Pass 8} data collected between \texttt{2008-8-7} and \texttt{2017-7-27}. The data are processed with the Fermi Science Tools version \texttt{10r0p5}, and we utilize events belonging to the \texttt{Source} event class in our default analysis. In several later analyses and cross-checks we restrict our analysis to \texttt{UltraCleanVeto} class events, to further reduce cosmic-ray contamination. 

To track the solar position and perform the event selection, we produce 4.2-hr temporal bins (40/week), during which the Sun moves $\sim0.2$\,degrees. We utilize \texttt{gtselect} to extract photons within a 9-degree region of interest (ROI) around the Sun, and then calculate the exact instantaneous distance of each photon from the solar position using the {
\tt sunpos.py} program included in the Fermi tools. This final step is a significant improvement with respect to NBPR2016, which utilized average positions in each temporal bin. In particular, calculating the exact solar distance of each photon is necessary to take advantage of the $\sim$0.1$^\circ$ angular resolution provided by Fermi at high energies. 
 
We process the data using standard cuts in \texttt{gtmktime}, setting keyword DATA$\_$QUAL==1 and LAT$\_$CONFIG==1 to ensure that the data quality are suitable for scientific analysis.  The maximum zenith angle is set to be 90$^\circ$ to avoid the bright Earth limb. To reduce the background due to bright point sources and the diffuse emission from the Galactic plane, we select data when the Sun is at latitude $|b|> 30^\circ$.  Given the size of the ROI, this effectively removes photons coming from $|b|< 21^\circ$.   We conservatively removed data from any week (selected using the weekly photon data) that contains a significant solar flare (detected by Fermi at $>10\sigma$)~\cite{flare_list}.  Compared to NBPR2016, only one more flare, in week labeled \texttt{326}, is removed.
 
We calculate the Fermi exposure with \texttt{gtltcube} and \texttt{gtexpcube2} using the P8R2$\_$SOURCE$\_$V6 instrumental response function with pixel size 0.1$^\circ \times$0.1$^\circ$.  The exposure is calculated in identical 4.2-hr temporal bins, which are summed to determine the total solar exposure. The fine temporal binning ensures that the Sun moves only $\sim$\,0.2$^\circ$ within each exposure bin. This is negligible compared to the angular dependence of the Fermi-LAT effective area. We additionally apply the same solar flare and Galactic plane cuts described above to the exposure calculation. 

\subsection{Likelihood analysis}\label{sec:likelihood}

To determine the \sag\ signal from the data extracted above, we follow a similar procedure as NBPR2016. In our default analysis, we focus on gamma rays above $1$\,GeV, dividing the data into 4 equal logarithmic bins per decade.  The 1\,GeV lower bound is motivated by the Fermi-LAT point spread function; at lower energies, it is difficult to separate the \sag\ from background emission (see Fermi2011 for an analysis down to 0.1~GeV).

There are two background components in the ROI that need to be taken into account for the \sag\ analysis. The first is the IC halo from cosmic-ray electrons scattering with solar photons~\cite{Orlando:2006zs, Moskalenko:2006ta, Orlando:2013pza}. The second is the collective gamma-ray emission from point sources and diffuse emission near the ecliptic plane, which all merge into a single diffuse component due to the motion of the Sun. We thus denote this background simply as the ``diffuse background.'' 

To separate the signal and the two background components, we exploit their different angular distributions in the ROI. The \sag\ flux, by definition, comes from the atmosphere of the Sun, and has an apparent angular radius $\theta_{\odot}\simeq 0.27^{\circ}$, where $\theta_{\odot}$ is the angular distance from the center of the Sun.  

For the IC halo, we assume the intensity falls linearly with $\theta_{\odot}$.  This follows from the assumption that the cosmic-ray electron density in the solar system is homogeneous, which was found to be reasonable~\cite{Abdo:2011xn} and is conservative for our purpose.  The IC intensity thus peaks at small angles, except for $\theta_{\odot} \lesssim 0.27^{\circ}$, where we set the IC component to 0 due to the strong suppression of IC emission stemming from the anisotropy of the solar photons and the shadow of the Sun~\cite{Orlando:2006zs, Moskalenko:2006ta, Orlando:2013pza}. 

\begin{table}
\begin{ruledtabular}
\begin{tabular}{lcccccc}
\hline
Energies 	& {Total cts.} 		& {\sag\ cts.}	& $\sqrt{\rm TS}$	 		& 	{\sag\ Flux} 		\\	
{[GeV]}		&  				& 					& 	 				&   {$[10^{-6} {\rm\ MeV\ cm}^{-2}\ {\rm s}^{-1}$]} 		\\	
\hline
1.0--1.8 			&	2903			&	1425			&20.1 				 &	$17.5^{+0.9}_{-0.9}$	\\
1.8--3.2 			&	1574			&	922			&18.6 				 &	$19.2^{+1.1}_{-1.1}$	\\
3.2--5.6 			&	776			&	527			&16.1 				&	 $20.0^{+1.3}_{-1.3}$	\\
5.6--10			&	345			&	214			&9.7					&	 $14.6^{+1.6}_{-1.6}$	\\
\hline
10--18 			&	162			&	116			&9.8					&	 $14.0^{+1.7}_{-1.6}$		\\
18--32 			&	84			&	71			&9.3 					&	 $15.2^{+2.1}_{-2.0}$		\\
32--56 			&	23			&	17			&4.1 					&	 \phantom{x}$6.4^{+2.0}_{-1.8}$			\\
56--100 			&	23			&	20			&6.1					&	 $13.4^{+3.4}_{-3.0}$			\\
\hline
100--178   		&	3			&	2			&1.2  				 &	$2.2^{+2.5}_{-1.9}$		\\
178--316    		&	1			&	1			&0.5    				&	$2.1^{+2.3}_{-2.1}$		\\

316--562   		&	0			&	--			&--	  			&	$<$8.7						\\
562--1000   		&	0			&	--			&--    			&	\phantom{x}$<$16.5				\\

\end{tabular}
\end{ruledtabular}
\caption{The total photon count and the corresponding best-fit photon count~(rounded) produced by the \sag\ in the angular bin considered. The bin size is $1.5^{\circ}$ below 10\,GeV and $1^{\circ}$ above.  The value  $\sqrt{TS}$ represents the significance of the \sag\ detection, as described in the text. The last column shows the best-fit \sag\ flux, and 90\% upper limits in the case of a non detection.  The 32--56\,GeV bin is lower than the adjacent bins at about 
4$\sigma$ significance; see text for a detailed discussion about this feature.  
}
\label{tab:result}
\end{table}
 
\begin{figure}[!t]
\includegraphics[width=\columnwidth]{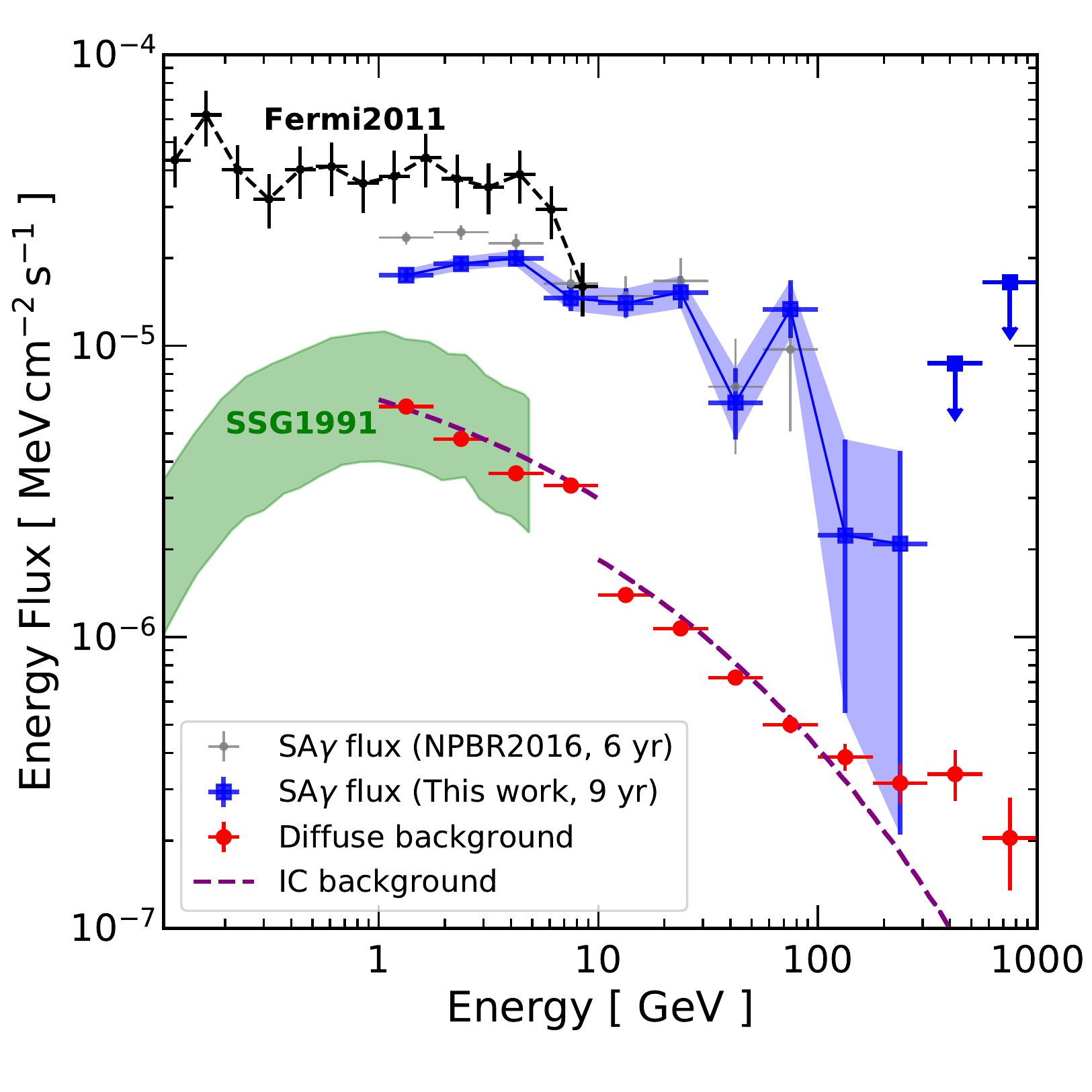}
\caption{The 9-year averaged \sag\ spectrum.  The Sun is detected significantly~($>5\sigma$) for all energy bins between {1--100\,GeV}, except the 32--56\,GeV bin, which is still above $4\sigma$. Above 100\,GeV, the apparent cutoff stems from the fact that high-energy photons are only observed during solar minimum~(see Sec.~\ref{sec:vhe} for details).  For comparison, we also show the 1.5-year spectrum from Fermi2011, the 6-year spectrum from NBPR2016, and the SSG1991 prediction.  The differences between the intensities of these measurements are explained by the time-variability of the \sag\ emission, as described in Section~\ref{sec:time}. The estimated background contribution in the corresponding angular bin~(which changes at 10\,GeV) from the IC halo~(purple dashed line) and the diffuse backgrounds~(red points) are also shown.   }
\label{fig:diskflux}
\end{figure}

Due to the motion of the Sun, the diffuse emission is strongly smeared out, and assumed to be constant over the ROI. This is supported by estimates of the background using the ``fake Sun'' method~(see also Appendix~\ref{appendix:analysis_systematics}). We perform the same analyses and cuts for 3 ROIs that are displaced from the true time by $+90$, $+180$, and $+270$ days, which is roughly $90^{\circ}$, $180^{\circ}$, and $-90^{\circ}$ away from the true Sun position.  As a result, we can accurately estimate the diffuse background intensity and avoid contamination from the IC halo.  The diffuse background flux obtained from these three fakes Suns is used directly in our \sag\ analysis.

We utilize a binned profile likelihood analysis to extract the \sag\ signal from both background contributions~(see NBPR2016 for a technical description). We divide the ROI into annuli of equal angular width. The widths of the annuli are chosen to be $1.5^{\circ}$ and $1.0^{\circ}$ for photons between $1$--$10$\,GeV and $> 10$\,GeV, respectively.  These widths are chosen such that the first angular bin is larger than the Fermi angular resolution in the energy range,  and thus the \sag\ signal is fully contained in the first angular bin to better than 95\%. In each energy bin, we fit the \sag\ component and the background components using their distinct angular distributions.  Because we are only interested in the \sag\ signal, we treat the IC halo and diffuse background normalizations as nuisance parameters.  For the diffuse background, we add a Gaussian term with a 20\% systematic error to incorporate the information we obtained from the fake-Sun studies.  The normalization of the IC halo is completely free, thus our \sag\ result also includes the uncertainty in the IC halo flux.  The resultant \sag\ flux, its uncertainties and significance are then obtained through the standard profile likelihood procedure~\cite{Rolke:2004mj, Cowan:2010js}.  We note that the best fitting diffuse flux is consistent with the expectation from studies of the diffuse Fermi-background~\citep{Ackermann:2014usa}.

\begin{figure*}
	\begin{center}
		\includegraphics[width=0.99\columnwidth]{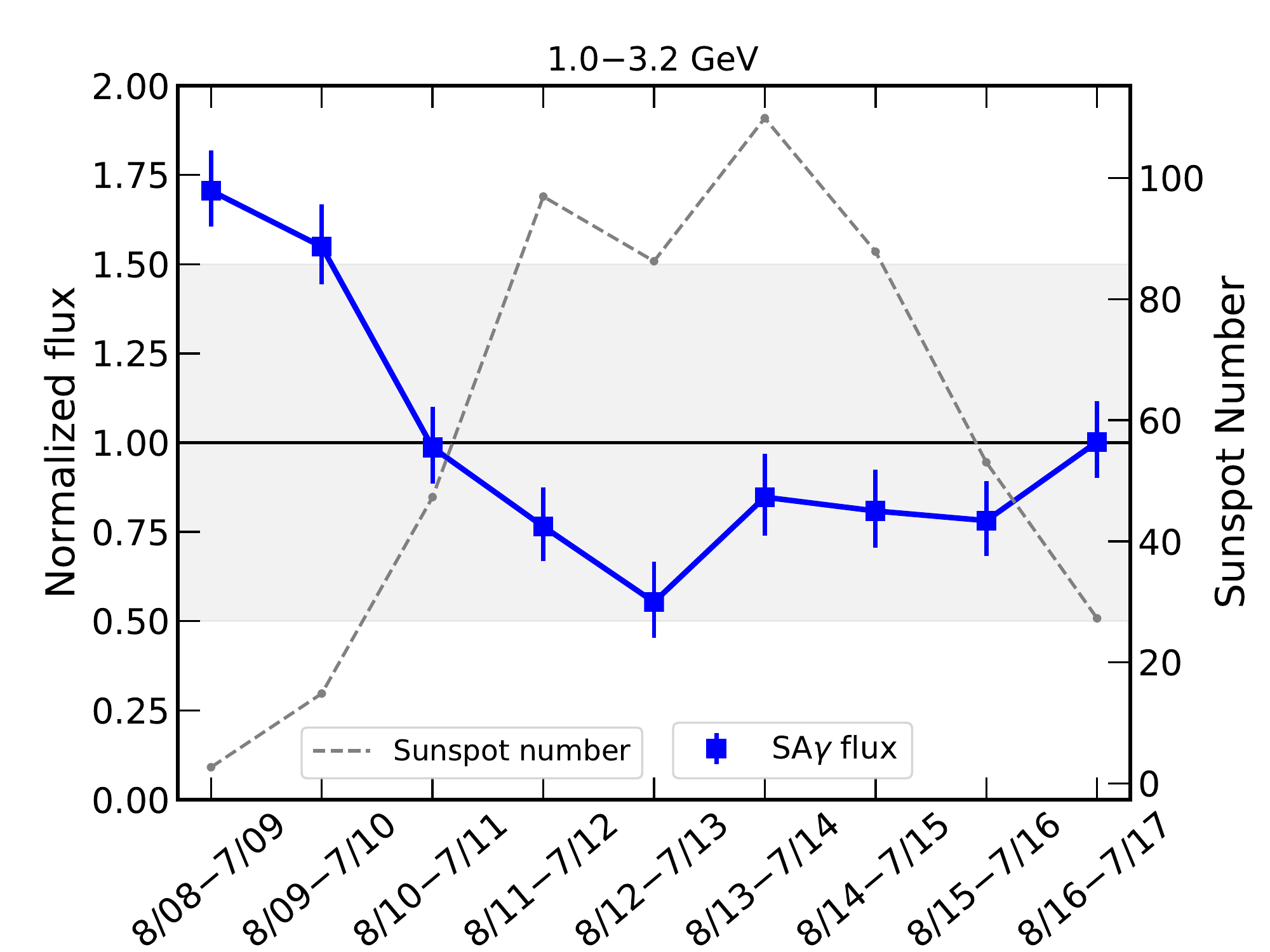}\vspace{-0.05cm}
		\includegraphics[width=0.99\columnwidth]{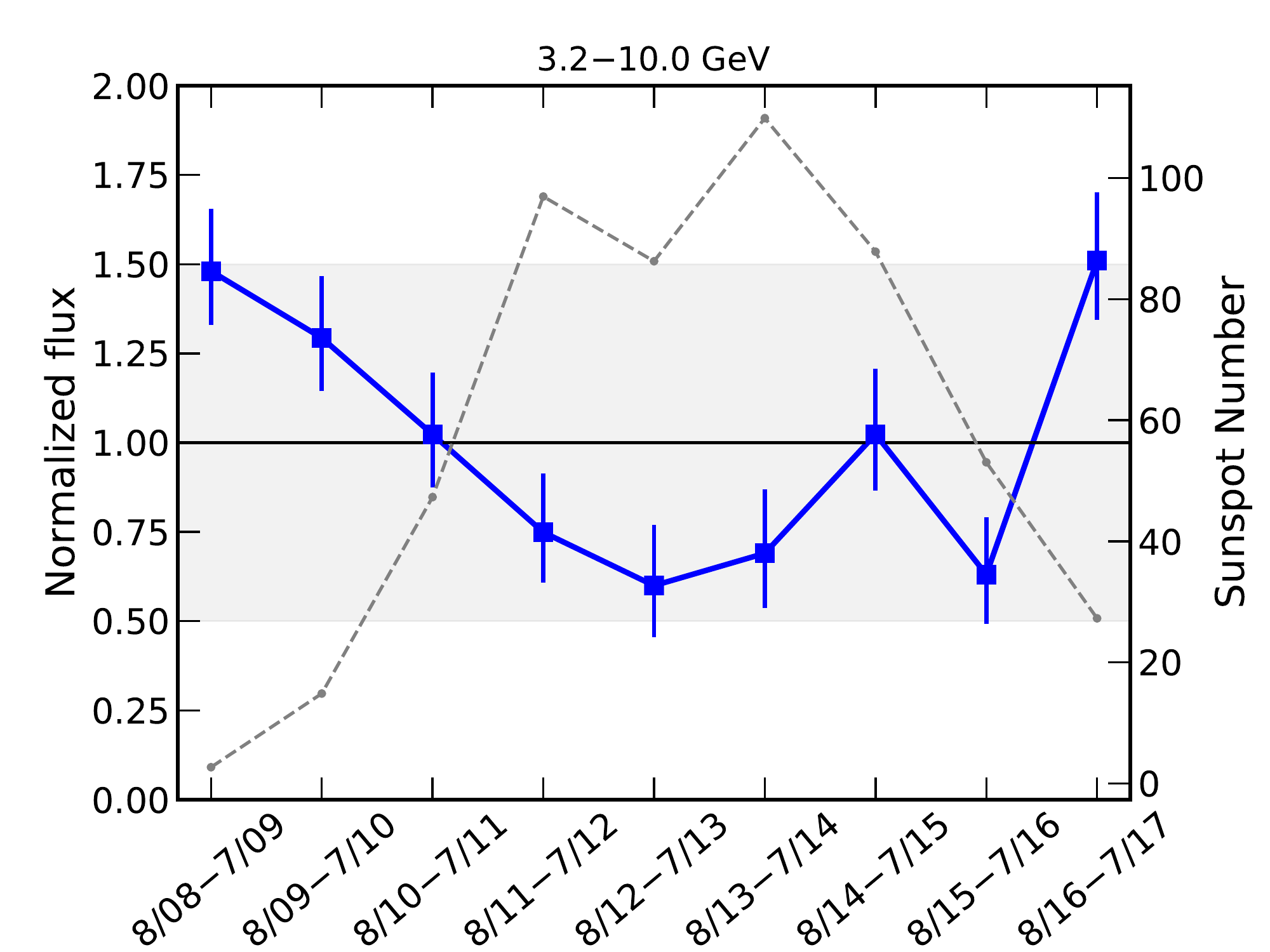}
		\includegraphics[width=0.99\columnwidth]{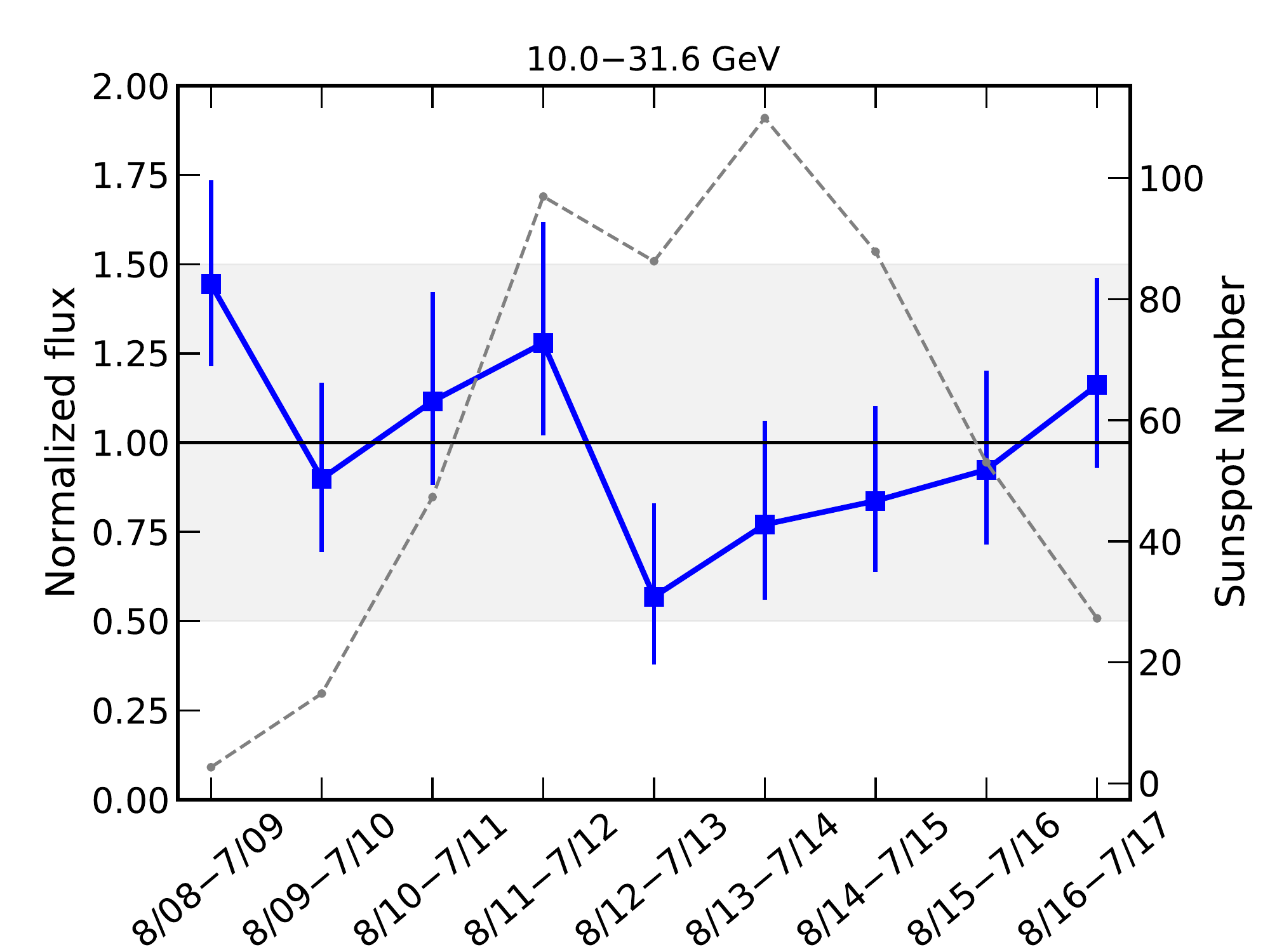}\vspace{-0.05cm}
		\includegraphics[width=0.99\columnwidth]{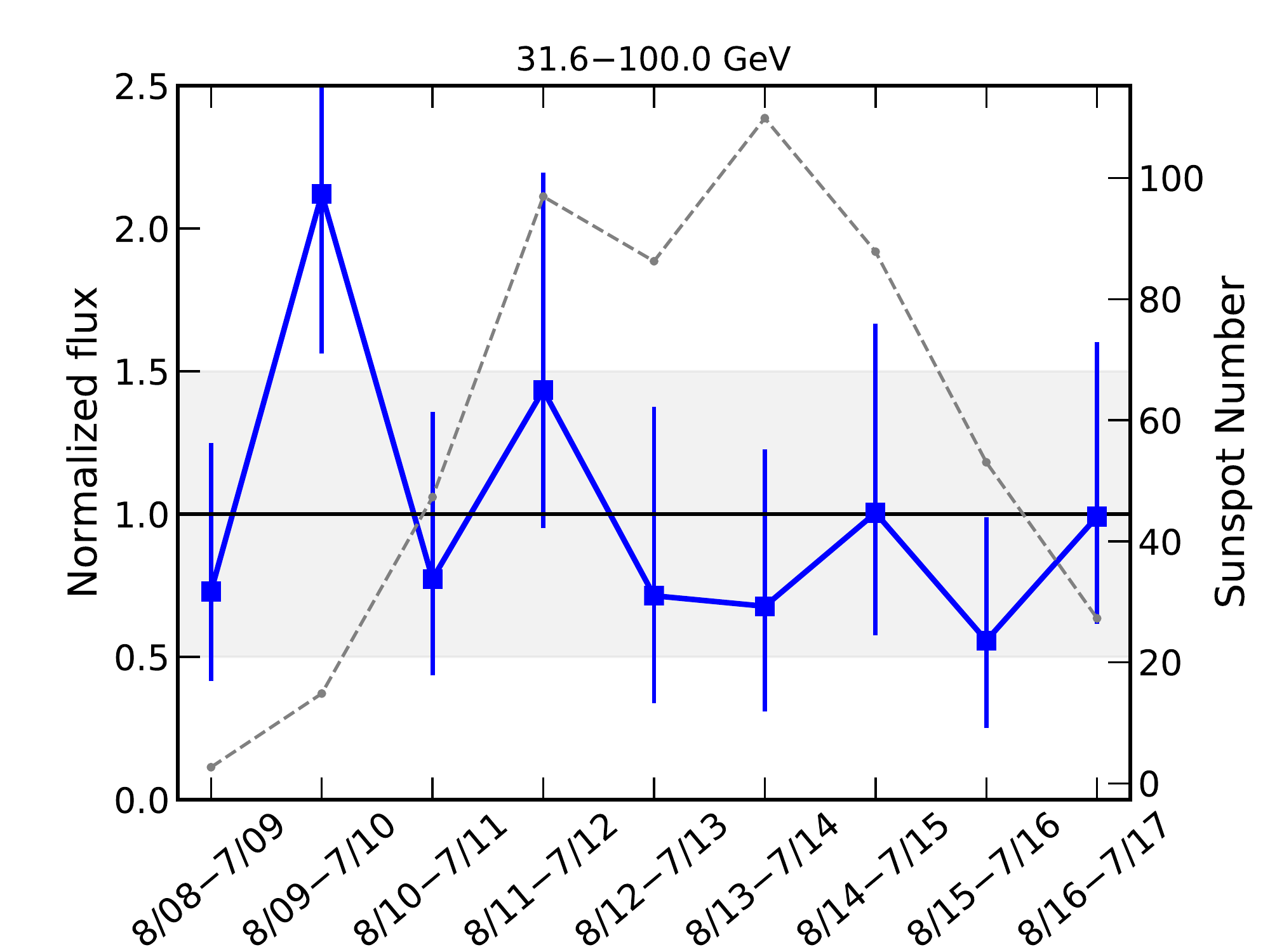}
		\caption{Time variation of the \sag\ flux, calculated as the annual flux divided by the 9-year averaged flux.  Each panel covers half of a decade in energy; versions with four energy bins per decade are shown in Appendix~\ref{appendix:time_variation}.  The \sag\ flux between 1--3.2\,GeV and 3.2--10\,GeV show a consistent anticorrelation with solar activity, with a comparable amplitude of $\pm 50\%$.  For the first time, the rise of the \sag\ flux correlating with the recent decrease in solar activity is observed.	The sunspot number is shown as a tracer of the solar activity~(minimum at $\sim2008$ and maximum at $\sim2014$). }\label{fig:timevar}
	\end{center}
\end{figure*}

\section{\sag\ Analysis Results}\label{sec:results}

\subsection{Time-Averaged Spectrum}\label{sec:spectrum}

Table~\ref{tab:result} summarizes the results of our \sag\ analysis.  The 9-year averaged \sag\ flux is detected above $>$5$\sigma$ significance in all bins between 1--100\,GeV, except for the 32--56\,GeV bin, which is 4.1$\sigma$.  This substantially improves on NBPR2016, where the \sag\ flux was only detected at $\simeq 2\sigma$ in the 32--56\,GeV and 56--100\,GeV bins.  

Above 100\,GeV, only a handful of events are observed in our data, and the \sag\ detection is only marginally significant. However, in a companion paper~\citep{Linden:2018}, we have shown that the emission above 100~GeV is highly significant during solar minimum, while no emission is detected over the remaining solar cycle.  Thus, averaging over these two different time-periods artificially produces a softer flux in the full dataset. We compare these analyses in detail in Section~\ref{sec:vhe}. Above 316\,GeV, we observe zero events in our default data set, and thus derive the 90\% upper limit using the total exposure within $1^{\circ}$ from the Sun, conservatively assuming no background events.

In Figure~\ref{fig:diskflux}, we compare the \sag\ flux with previous theoretical and observational results. We confirm the key findings of Fermi2011 and NPBR2016, which indicated that the \sag\ flux significantly exceeds the theoretical predictions of SSG1991. Given that their prediction is already a significant enhancement over the limb-only flux~\cite{Zhou:2016ljf}, it is surprising that the observed data are even brighter, and have a harder spectrum than that of the parent cosmic rays.  We extend these results to higher energies, finding that the \sag\ spectrum has a significantly harder flux than SSG1991 models up to an energy of at least 100~GeV. We additionally compare the \sag\ flux with the estimated contributions from IC and diffuse backgrounds.  The theoretically predicted IC flux is calculated using a modified version of the {\tt StellarICs} code~\cite[see Ref.~\cite{Zhou:2016ljf} for a detailed description of inputs and modifications]{Orlando:2013pza, Orlando:2013nga}, while the predicted diffuse background is based on an identical analysis centered on the positions of three ``fake Suns" that trail the position of the real sun by 90, 180 and 270 days. Our analysis clearly indicates that the \sag\ flux is significantly brighter and has a harder spectrum than either the IC or diffuse components. Finally, we find evidence for a spectral dip in the 32--56~GeV bin, which we discuss in detail in Section~\ref{sec:dip}.
   
\subsection{Time Variability}\label{sec:time}

In Figure~\ref{fig:diskflux}, we note that the extracted \sag\ flux is significantly dimmer than the analysis of Fermi2011 and slightly dimmer than NPBR2016. This is not due to any tension between the analyses, but rather due to the intrinsic time-variability of the \sag\ flux. This time variation was first robustly detected in NBPR2016, which found that the \sag\ flux anti-correlates with solar activity. With 9 years of data, covering most of the solar cycle, we can now study the time variation in more detail. 

We repeat the likelihood analysis described in Sec.~\ref{sec:analysis}, but divide the gamma-ray emission into individual years, which we define to be 52 weeks, starting from the beginning of Fermi-LAT operations. Because the Sun traverses the ecliptic once in each period, every analysis window includes the same average diffuse background. To focus on the time-variation in each analysis period, we combine the \sag\ flux into two logarithmic bins per decade spanning from 1--100~GeV (see Appendix~\ref{appendix:time_variation} for an analysis with finer energy bins).

In Figure~\ref{fig:timevar}, we show the results of this analysis. The flux in each time period is shown as a fraction of the time-averaged flux at each energy. We overlay the averaged sunspot number~\footnote{http://www.sws.bom.gov.au/Solar/1/6} as a tracer for the solar activity.  (See Appendix~\ref{appendix:time_variation} for time variations of the \emph{total flux}, defined in the next section.)

Between 1--10\,GeV, we find that the \sag\ flux clearly anti-correlates with the sunspot number, peaking during solar minimum and reaching a minimum flux near solar maximum. To formally test the statistical significance of this time-variation in each energy bin, we calculate the $\chi^2$ fit obtained under the assumption that the flux is time-independent, finding $\chi^2$ values of 91, 35, 8 and 6 with 9 degrees of freedom in each respective energy bin. These results thus provide clear evidence of time-variability below 10~GeV. While there is no statistically significant detection of time-variability at higher energies, we note that this is primarily due to insufficient statistics, rather than any significant shift in the degree of time-variability. At low energies, we find that the amplitude of this time variability is approximately 50\% of the time-averaged flux, and find no evidence that the time variability decreases as a function of the gamma-ray energy. This final observation is surprising, given that higher-energy cosmic rays are less likely to be affected by solar magnetic fields.  We investigate this further in the discussion.

Last but not least, as we have recently shown in Ref.~\cite{Linden:2018}, the degree of time-dependence is very different for the highest energy gamma rays (above 100~GeV), where 100\% of the gamma-ray flux is observed during the first 1.5~yr of solar minimum activity. This degree of temporal variation is significant at nearly the 5$\sigma$ level.

\section{Investigating the 30--50~GeV Dip}\label{sec:dip}

The most striking feature of Figure~\ref{fig:diskflux} is the peculiar dip in the \sag\ flux in the 32--56~GeV energy bin, which falls more than a factor of two below adjacent energy bins (see Table~\ref{tab:result}). The feature appears to be statistically significant~($\sim$4$\sigma$ compared to adjacent bins), and may correspond to a dip in the underlying gamma-ray spectrum between 30--50\,GeV.  In Section~\ref{sec:implications}, we discuss possible interpretations of this feature. Here, we attempt to examine its robustness and quantify its statistical significance.

In order to more sensitively and straightforwardly ascertain the statistical significance of the spectral dip, we refine our standard analysis method in three ways. First, we more finely bin the solar gamma-ray flux into eight logarithmic energy bins per decade, in order to more accurately determine the spectrum of the dip. Second, we simplify our analysis routine by examining the total gamma-ray flux within 0.5$^\circ$ from the Sun, rather than utilizing the \sag\ analysis techniques that fit the gamma-ray components based on their angular profiles. This eliminates potential systematics in the likelihood analysis stemming from the assumed background emission morphologies.  
At high energies the effect of this choice is marginal, because the IC and diffuse backgrounds are highly subdominant to the true \sag\ flux above 10~GeV (see, e.g., Figure~\ref{fig:diskflux}).

Third, we enhance our analysis by including a new independent dataset from photons recorded when the Sun lies in the latitude range 5$^{\circ}< |b| <30^{\circ}$~(applying equivalent solar flare cuts). This latitude cut avoids only the brightest regions of the Galactic plane. While regions near the Galactic plane would suffer from significant diffuse contaminations at low energies, the contamination at energies above 10~GeV is minimal due to the soft spectrum of diffuse emission and the smaller angular acceptance window above 10~GeV.  We note that the 5$^{\circ}< |b| <30^{\circ}$ sample has roughly 60\% of the exposure of the $|b|>30^{\circ}$ sample, and thus provides a significant boost in statistics to the combined sample.

\begin{figure}[!t]
	\includegraphics[width=\columnwidth]{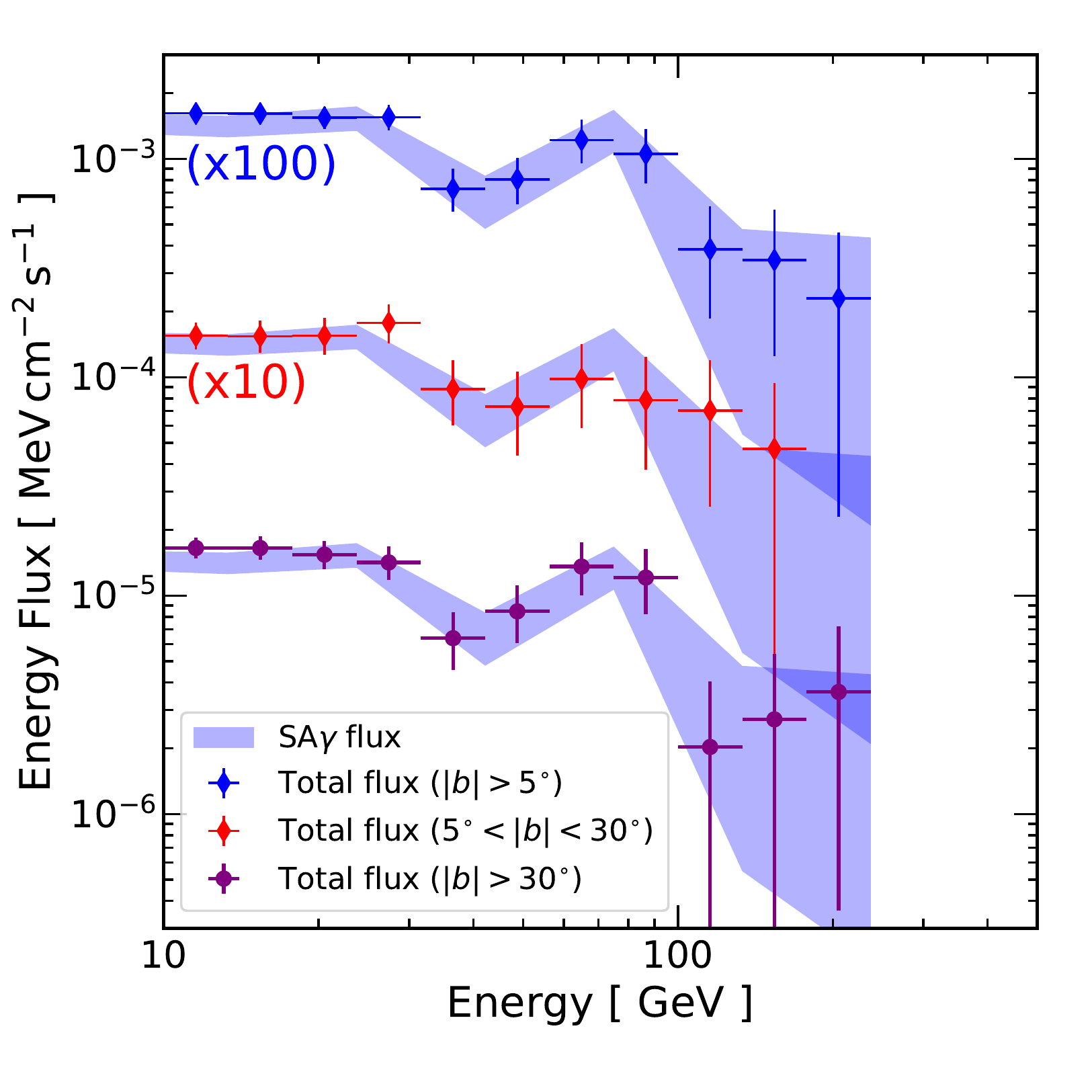}
	\caption{ The total flux spectrum from within $0.5^{\circ}$ of the solar position, using three latitude selections: high-latitude~\mbox{($|b|>30^{\circ}$)}, low-latitude~\mbox{($5^{\circ}<|b|<30^{\circ}$)}, and the combined data~\mbox{($|b|>5^{\circ}$)}.  Here the $|b|>30^{\circ}$ data is independent to the $5^{\circ}<|b|<30^{\circ}$ one; the 30--50\,GeV dip can be seen in both.  Our default \sag\ flux, derived from the $|b|>30^{\circ}$ data, is also shown for comparison. 
	}
	\label{fig:totalflux}
\end{figure}

\begin{figure}[t]
	\includegraphics[width=\columnwidth]{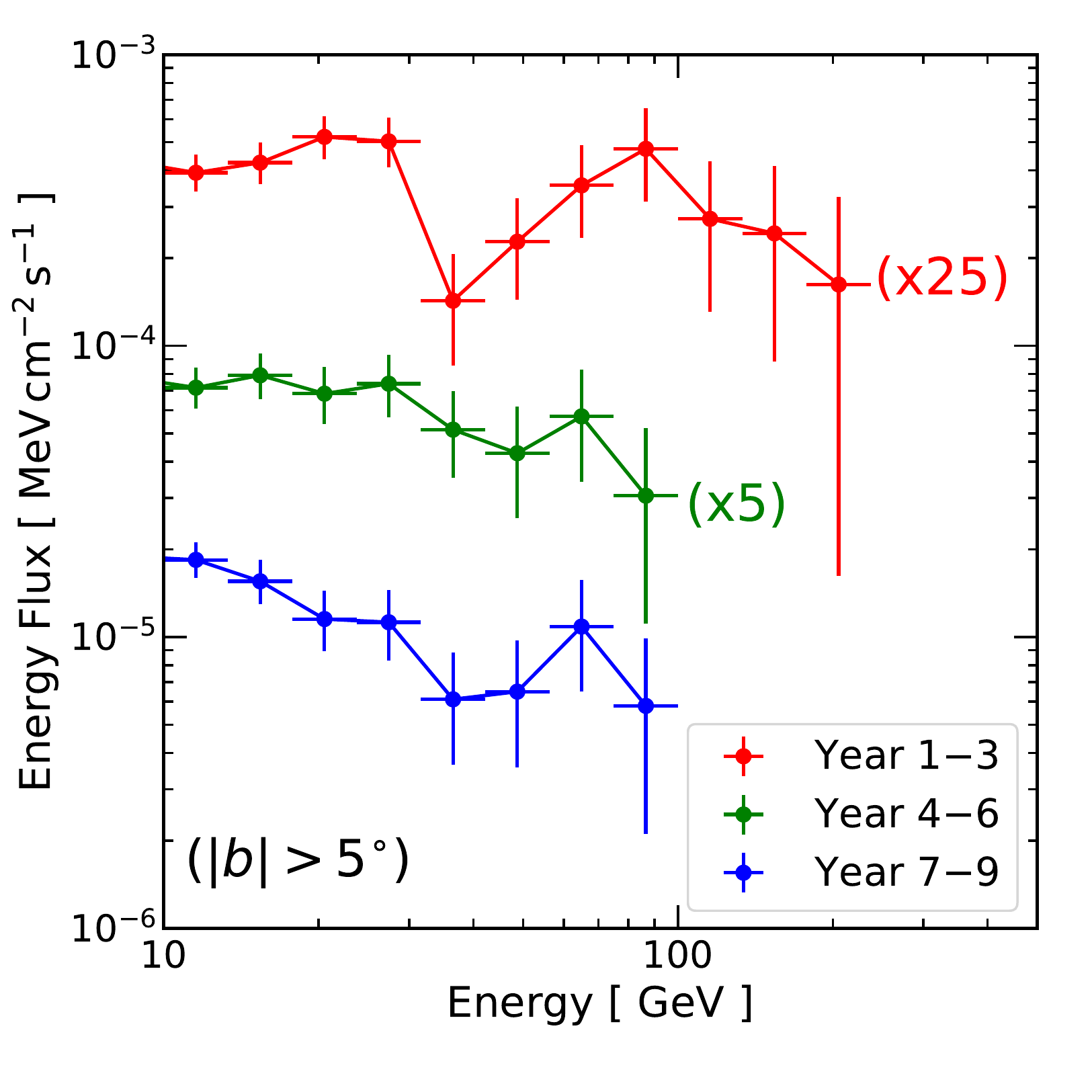}
	\caption{ 
		The total flux spectra within 0.5$^\circ$ of the solar position for the $|b|>5^{\circ}$ data selection, each covering 3 years of exposure. The dip between 30--50\,GeV can be seen for all three spectra.  While the dip may have appeared shallower for year 4--6, the location of the dip does not change much with time. 
	}
	\label{fig:threeyear}
\end{figure}

In Figure~\ref{fig:totalflux}, we show the gamma-ray spectrum contained within 0.5$^\circ$ of the solar position in eight logarithmic energy bins per decade. Spectra are shown for photons recorded for both our default latitude cut, and the low-latitude analysis, along with the combined spectrum from both event selections. We find evidence for the spectral dip in both datasets and note that the gamma-ray dip now spans two distinct energy bins, spanning the same energy range of 32--56~GeV as above. This energy range is not altered by utilizing more finely binned energy spectra. We note that the energy range of the spectral feature is significantly larger than the $\sim$7\% energy resolution of the Fermi-LAT at 40~GeV, and is thus consistent with a real spectral feature. We quantitatively test the effect of the Fermi-LAT energy resolution in Appendix~\ref{appendix:dispersion}.

We apply several different techniques to calculate the statistical significance of this dip, and describe each in Appendix~\ref{appendix:dip_significance}. Here, we note that by fitting a broken power law to the remaining dataset (i.e., removing the two ``dip" datapoints), we can calculate the $\chi^2$ fit of the dip to this model.  We find that the dip is statistically significant at 4.6$\sigma$ in the $|b|>30^{\circ}$ dataset, 2.1$\sigma$ in the $5^{\circ}<|b|<30^{\circ}$ data, and 5.7$\sigma$ in the combined dataset.

In Figure~\ref{fig:threeyear}, we examine the time evolution of the dip feature by dividing the combined ($|b|>5^{\circ}$) data into three periods spanning three years (156 weeks) of data. This corresponds to the periods \texttt{8/08-7/11}, \texttt{8/11-7/14}, and \texttt{8/14-7/17}.  We note that the dip is observed in each temporal bin, and the location of the dip does not appear to vary in energy. We find that the dip is most significant during years 1--3~($4.8\sigma$), but is weak in years 4--6~($1.4\sigma$) and years 7--9~($2.1\sigma$). This may provide an indication that the dip is stronger during periods of reduced solar activity, a phenomenon that can be tested by upcoming solar-minimum data.

Finally, we note that there are several systematic issues intrinsic to solar analyses. In particular, the position of the Sun in Fermi-LAT instrumental coordinates is unique, as the x-axis of the satellite is oriented towards the Sun in order to maximize the solar panel efficiency~\cite{2012ApJS..203....4A}.  It is possible that this could imbue a systematic feature in the response of the instrument, and cause such a dip feature.  In the Appendix~\ref{appendix:check_dip}, we further check the robustness this feature against several potential instrumental effects, including the azimuthal dependence of the LAT effective area and energy reconstruction, additional checks regarding the fake-Sun background estimation, and finally the event selection.  
We have not found any systematic and instrumental effects that can produce the spectral dip.  
(We do find a spectral feature, previously identified by the Fermi-LAT collaboration~\cite{fermifl8y, fermicaveat} that affects solar events  $\lesssim$\,10\,GeV, but does not influence the observed spectral dip feature.)
We thus move on and discuss possible interpretations in Sec.~\ref{sec:implications}.

\section{The solar minimum \sag\ flux}\label{sec:vhe}
\begin{figure*}[t]
	\includegraphics[width=2\columnwidth]{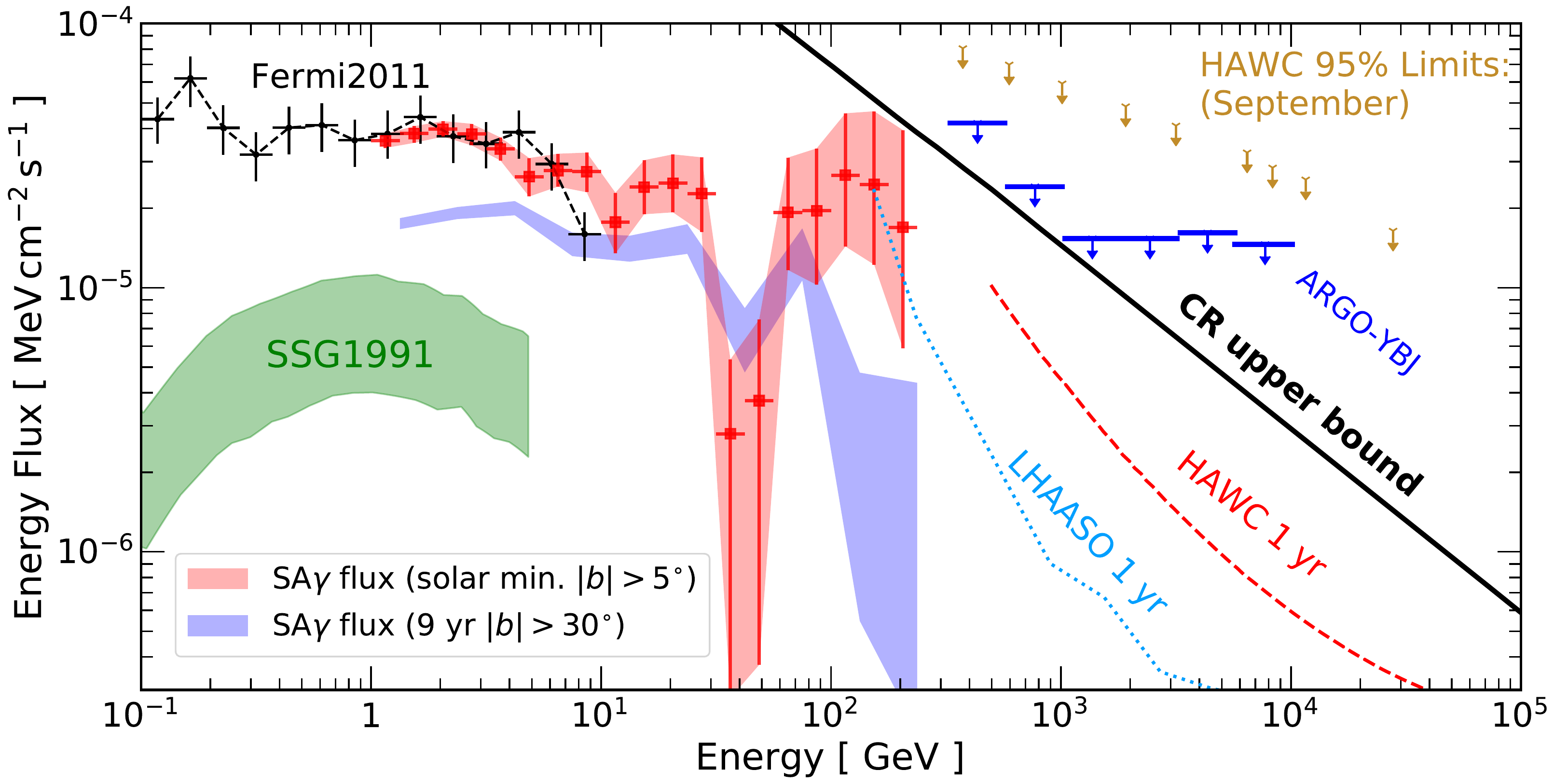}
	\caption{ The 9-year averaged \sag\ flux and the solar minimum \sag\ flux~(76 weeks of data from \texttt{2008-8-7} to \texttt{2010-1-21}), which is significantly harder above 100\,GeV. We also show the point source sensitivities of HAWC~\cite{Abeysekara:2017mjj} and LHAASO~\cite{Zhen:2014zpa, He:2016del}, as well as the preliminary limits from HAWC~(\cite{Nisa:2017xpf}, one month of data) and ARGO-YBJ~\cite{argo:2016}.  	The theoretical maximum gamma-ray flux that the Sun can produce with cosmic rays~(CR upper bound, see text for detail) is shown by the black solid line.  }
	\label{fig:vhe}
\end{figure*}

In this section, we focus on the \sag\ spectrum during solar minimum.  As first pointed out in Ref.~\cite{Linden:2018}, the solar minimum \sag\ spectrum is much harder and extends to higher energies than the 9-year averaged spectrum. This has important implications for the results of this study. For example, the soft gamma-ray spectrum $>100$~GeV in our default 9-year analysis is actually due to the combined contributions of hard-spectrum gamma-ray emission during solar minimum, and very soft-spectrum gamma-ray emission over the remaining solar cycle.

Motivated by this result, we apply our \sag\ analysis to the solar-minimum period, defined here as the 76 weeks of data between 
 \texttt{2008-8-7} and \texttt{2010-1-21}. In this section, we again make several modifications to our default event selection in order to better represent the solar-minimum data. First, we again choose to utilize a more-lenient solar latitude cut of $|b|>5^\circ$ to maximize the statistics during this smaller time window. Second, we allow the diffuse background to have a free normalization, rather than utilizing fake-Sun constraints that were optimized for the $|b|>30^\circ$ analysis. Lastly, we apply the UltraCleanVeto event selection cuts to minimize cosmic-ray contamination, a cut which reduces the exposure by 20--30\%. This cut is employed to remove a systematic spectral feature that we (and the Fermi-LAT collaboration~\cite{fermifl8y, fermicaveat}) identify in finely binned solar data at an energy of $\sim$10~GeV. Unlike the 30--50~GeV feature discussed here, the 10~GeV feature appears to be due to cosmic-ray contamination, and disappears entirely in the UltraCleanVeto class events. We discuss this feature in detail in Appendix~\ref{appendix:phidependence}.

In Figure~\ref{fig:vhe}, we show the resulting solar minimum \sag\ spectrum, noting three important results. First, we find close agreement with the Fermi2011 result between 1--10~GeV, verifying that differences between the \sag\ flux in these analyses is due to the intrinsic time variability of the solar flux. Second, we find close agreement with the results of Ref.~\cite{Linden:2018} above 10~GeV, which is independent of this analysis and utilizes a slightly different data selection.  Third, we find that the solar minimum analysis provides a more significant detection of \sag\ emission above 100~GeV (3.7$\sigma$) than is found in our default analysis. The higher statistical significance comes from the utilization of gamma-rays during periods when the Sun resided at 5$^\circ<|b|<$30$^\circ$, and not including the non-Solar minimum period in the analysis, where no photons are observed while the expected background increases with time.  If we also narrow the angular bin size to be $0.5^{\circ}$, the detection significance reaches 5.2$\sigma$, in agreement with the findings in Ref.~\cite{Linden:2018}. 

Finally, we note that the dip between 30--50~GeV is extremely significant (8.4$\sigma$) in this analysis. At first, this result may be confusing, as it exceeds the 5.7$\sigma$ significance of the dip over the full 9-year dataset. This disparity is partly due to the harder gamma-ray flux above the spectral dip during the solar minimum period, which provides support for a harder gamma-ray spectrum (without a significant cutoff) on both sides of the spectral dip, and prevents the broken power law from absorbing a portion of the spectral dip.  The high significance is also partly due to the dip being less significant in periods outside the solar minimum~(see Figure~\ref{fig:threeyear}), which lowers the dip amplitude in the full 9-year dataset. 

\subsection{Implications for the \sag\ production efficiency  }

The hard \sag\  spectrum observed at solar minimum has important implications for the underlying \sag\ production mechanism. Even without a detailed model, it is informative to compare the observed \sag\ flux with the maximum flux allowed from cosmic-ray interactions in the solar photosphere. This extremal prediction can be robustly calculated by maximizing several relevant parameters in the gamma-ray production mechanism. These include the fraction of the solar surface that includes magnetic fields capable of reversing the direction of incoming cosmic-rays, the fraction of cosmic-rays that effectively have their directions reversed before undergoing a hadronic interaction, and the fraction of cosmic-rays which subsequently undergo a hadronic interaction before escaping the solar surface. We describe this calculation in detail in Appendix E of Ref.~\cite{Linden:2018}. By setting these three terms to unity, we can calculate an upper bound on the gamma-ray flux, which can be approximately parameterized as $4\times 10^{-4} \left( E_{\gamma}/{\rm 10\,GeV} \right)^{-2.72}\,{\rm MeV\,cm^{-2}\,s^{-1}} $ above 1\,GeV. 

In Figure~\ref{fig:vhe}, we compare the solar minimum \sag\ spectrum to this maximal gamma-ray flux, finding that the observed \sag\ flux has a significantly harder spectrum ($\sim E_{\gamma}^{-2.2}$) relative to the cosmic-ray spectrum~($\sim E_{\rm CR}^{-2.7}$), allowing it to approach the upper bound at high energies. This suggests two seemingly contradictory results. On one hand, the \sag\ production efficiency must be near unity at high energies to produce the observed flux. On the other hand the \sag\ production efficiency must be increasing rapidly at high energies in order to maintain the hard \sag\ spectrum. Moreover, while it is not surprising that the \sag\ production efficiency is energy dependent, it is somewhat surprising that the efficiency increases with energy, given that it is more difficult to confine and reflect higher energy particles in the solar magnetic field. It is also important to note that the SSG1991 model, if continued to high energy, follows the cosmic-ray spectrum.  Thus the gamma-ray data implies qualitatively different physical effects compared to SSG1991.  

The proximity of the \sag\ flux at solar minimum to the maximal cosmic-ray flux also presents significant issues.  Above 100\,GeV, the \sag\ flux is close to 30\% of the cosmic-ray upper bound. It is not clear how to achieve such an efficiency in a realistic setup. In Ref.~\cite{Linden:2018}, we discuss several possible methods of further enhancing the \sag\ flux, including magnetic focusing from the large scale magnetic field, long-term cosmic-ray confinement in the coronal magnetic field, anisotropies in the \sag\ emission mechanism, or some unknown intrinsic solar gamma-ray production mechanism. 

Finally, because the highest-energy gamma-ray photon from a typical hadronic interaction has an energy of $\sim$\,0.1$E_p$, the observation of a $>$\,100\,GeV \sag\ flux suggests that the enhanced gamma-ray production mechanism applies to even $>$\,TeV cosmic rays, at least during solar minimum. This observation is partially supported by the cosmic-ray shadow measurements by Tibet AS$\gamma$ measurement~\cite{Amenomori:2013own}, which showed that solar coronal magnetic fields can affect cosmic rays $\gtrsim$\,10\,TeV.   These will have important implications for searches of solar atmospheric neutrinos and dark matter signals from the Sun; we discuss further below.

\subsection{Detection Prospects at TeV Energies } 

The high-energy \sag\ flux during solar minimum has immediate implications for TeV solar gamma-ray searches. In this regime, only ground-based air shower experiments and water Cherenkov telescopes, such as ARGO-YBJ~\cite{Aielli:2006cj}, Tibet AS-gamma~\cite{Hibino:1988er}, HAWC~\cite{Abeysekara:2013tza, Abeysekara:2017mjj}, and LHAASO~\cite{Zhen:2014zpa, He:2016del} are capable of solar observations. Figure~\ref{fig:vhe} also compares the \sag\ flux with the 1-year differential sensitivity of HAWC and LHAASO.  If the hard spectrum observed during solar minimum continues to the TeV range, then it will fall within the reach of HAWC's 1-year sensitivity.  Given that HAWC is already operating, the current solar minimum~(Cycle 25, beginning presently) provides the perfect opportunity for HAWC to test the TeV \sag\ flux.  We also show the preliminary upper limits of \sag\ from ARGO-YBJ~\cite{argo:2016} and HAWC~\cite{Nisa:2017xpf}.  The latter is obtained with only 1 month of data, and the sensitivity is expected to improve significantly. 

In the near future, LHAASO is expected to provide even better sensitivity and a lower energy threshold.  Excitingly, the projected sensitivity is already reaching the solar minimum spectrum at around 200\,GeV.  LHAASO may even be able to probe the considerably softer \sag\ flux outside of the solar minimum.  

\section{Discussion and Implications}\label{sec:implications}

\subsection{Interpretations of the 30--50\,GeV dip}

In this work we discover a spectral dip between approximately 30--50\,GeV, with a local significance exceeding $5\sigma$. This is surprising given that the cosmic-ray energy spectrum is incredibly smooth. Moreover, though SSG1991 predicts a spectral cutoff between 30--2000\,GeV, a hard spectral component above the cutoff is not expected.  While we cannot rule out any unknown systematic effects that may produce this feature, it has survived several sensitive tests detailed in the Appendix.  In Appendix~\ref{appendix:dispersion}, we find that the spectral dip is consistent with approximately a 50\% reduction in the gamma-ray flux over an energy range of approximately 20~GeV, which is then smeared by the Fermi-LAT energy resolution. These numbers guide our search for a potential source of the spectral dip feature.

In this subsection, we consider and quickly investigate several potential interpretations of the dip feature, with additional details presented in the appendix.  

\subsubsection{Hypothesis I: Two Cosmic-Ray Components }

One model capable of producing a spectral dip includes two separate cosmic-ray proton components, which dominate \sag\ emission below and above the spectral dip. Given that the interstellar cosmic-ray spectrum is smooth, and the efficiency of \sag\ production is nearly maximized at high energies, this split would be most reasonably produced by magnetic-field structures that inhibit cosmic-ray interactions within a specific, and narrow, energy range. For example, it is possible that one class of magnetic field structures are responsible for gamma-ray production below $\sim$\,30~GeV (e.g. flux tubes, as suggested by SSG1991), which become ineffective at high energies and cause a cutoff. The rise in the gamma-ray flux above 50~GeV would then be caused by a separate class of magnetic structures that are \emph{only} accessible to high-energy cosmic-rays.

However, we find that it is difficult to produce both the depth and the width of the observed spectral dip with a two-component proton spectrum.  This is due to the fact that the gamma-ray kernel for a given proton interaction is broad, which we detail in Appendix.~\ref{appendix:proton_gap}. The kernel may be sharpened, if only a narrow kinematic region is considered~(e.g., a tiny angular cone from the parent proton direction).  However, it is not clear how such a scenario can be realized in the Sun, where the cosmic-ray direction is expected to be roughly isotropic, and the gamma-ray efficiency must be nearly maximal, in order to explain the high  \sag\ flux.

\subsubsection{Hypothesis II: Gamma-Ray Absorption}
Another possible mechanism capable of producing the spectral dip is the preferential absorption of gamma-rays in the 30--50~GeV energy range.
Gamma rays can be absorbed by gas through the interaction $\gamma + p \rightarrow e^+ + e^- + p$, with an absorption length of $\rm \simeq 80\, g\, cm^{-2}$, which is comparable to that of p-p interactions~\cite{Patrignani:2016xqp}.
However, above the threshold ($\simeq$\,1~MeV), the cross section is smooth, and thus unlikely to cause a dip.
While $\gamma + p \rightarrow {\rm resonances}$ (e.g., $\Delta^+$) could produce a dip feature, the cross sections are too small~\cite{Patrignani:2016xqp}.  Similar processes occur for electron targets with higher energy thresholds, but the cross-sections are even smaller.
Absorption on photon targets, i.e. $\gamma + \gamma\rightarrow e^+ + e^-$ interaction, may produce a dip feature. The reaction threshold is $\simeq 250( 1\,{\rm eV}/ E_{\gamma}$)~GeV, taking the typical target photon energy at the photosphere~($\sim$\,1\,eV).  The threshold can be lower and thus match the energy of the dip if the interaction occur below the photosphere, where the Sun is hotter.  However, in this region the matter density would be orders of magnitude higher than the photon density, and thus matter effects should provide the dominant source of gamma-ray absorption.  Therefore, we find that it is unlikely for gamma-ray absorption to produce the observed dip feature.

\subsubsection{Hypothesis III: Solar Neutrons }
\label{sec:neutronhypothesis}
One intriguing possibility is that the dip is produced by solar neutrons, rather than solar gamma-rays. Unlike the case of more distant sources, solar neutrons can travel unimpeded to Earth before decaying. Assuming that the \sag\ flux is produced by hadronic processes, the solar neutron flux recorded by the Fermi-LAT should be similar to the gamma-ray flux (see e.g., SSG1991). Solar neutrons have, in fact, been observed by Earth-bound observatories and correlated with gamma-ray emission during solar flares~\cite{2016SoPh..291.1241M, Muraki:2017mhm}.

While the Fermi-LAT event reconstruction algorithms are finely tuned to eliminate the large contaminating flux from charged cosmic-rays, the efficiency in rejecting high-energy neutrons is not publicly available. Additionally, it may be difficult to definitively ascertain the efficiency of this cut, due to the very low flux of solar neutrons from non-solar or terrestrial sources. Furthermore, while analysis cuts of gamma-rays and hadrons have been carefully tuned to avoid inducing spectral features in the resulting gamma-ray spectrum, sharp features may potentially be induced from neutron sources. In Appendix~\ref{appendix:neutrondip}, we test several potential event-level quantities that may potentially distinguish neutron and $\gamma$-ray events, finding no obvious evidence for a neutron component. However, without utilizing non-public instrument-level data, we do not have the ability to rigorously test this hypothesis. 

Assuming for a moment that solar neutrons are present in the Fermi-LAT data, we note that these events would be reconstructed utilizing analysis algorithms tuned to accurately determine the energy of gamma-ray showers. Thus, the energy reconstruction of neutron events may be highly biased. In this scenario, the observed ``dip" in the solar gamma-ray spectrum may result from a population of 30--50~GeV neutrons, which are mis-identified by the Fermi-LAT analysis algorithm as $\sim$\,50--80~GeV gamma-rays. This would explain several features of the gamma-ray dip, including (1) its sharp spectral fall-off and rise, and (2) the large gamma-ray flux at energies immediately above the spectral dip. Additionally, if the neutron rejection efficiency were found to fall smoothly as a function of energy, it may explain the hard spectrum of observed \sag\ emission, although the neutron rejection efficiency would need to be fine-tuned in order to replicate power-law observations over more than two decades in energy.

We note, however, that this explanation does not explain the time-variation of \sag\ emission, or the significant spectral differences observed from \sag\ emission above 100~GeV during solar minimum. As pointed out by Ref.~\citep{Linden:2018}, photons observed during solar minimum have unique spectral and morphological characteristics compared to photons observed during the remaining solar cycle. These morphological shifts, in particular, cannot be modeled by misidentified neutrons, and instead indicate a significant shift in the solar gamma-ray generation mechanism.

\subsubsection{Hypothesis IV: \\Statistical fluke or Time-dependent instrumental effects}

We note that the dip is most significant during the solar minimum and becomes much weaker in later time periods.  While this may be a hint that the dip is related to solar minimum activity, we cannot fully rule out the possibly of an unlikely statistical fluke, given that only one solar minimum was observed by Fermi. 

More importantly, we note that the first solar minimum corresponds to data taken just after the launch of the Fermi-LAT. Though contrived, we cannot  rule out the possibility of a time-dependent instrumental effect that causes a dip only in the instrumental phase space of the Sun.  The time-dependent effect could be caused by changes of Fermi operation or event reconstruction during its lifetime.  For example, in September 2009, the rocking angle of Fermi was increased from $35^{\circ}$ to $50^{\circ}$~\cite{2012ApJS..203....4A}. However, it is not clear show such changes would produce a spectral dip feature. Fortunately, these two possibilities can be tested by Fermi observations of the Sun during the on-going solar minimum. 

\subsubsection{Hypothesis V: Intrinsic Solar Emission}

Lastly, we must also note that we are unaware of any intrinsic solar mechanism that can produce such energetic gamma-ray emission.  But in light of a two-component \sag\ scenario, we cannot rule out any intrinsic solar component interpretation. 

\subsection{Time variations of \sag\ flux }

In this work, we have observed moderate time variations~($\pm 50\%$) in the \sag\ flux between 1--10\,GeV.  Interestingly, the amplitude of this variation does not seem to vary significantly with energy. Naively, one would expect that the amplitude of the time variation decreases with energy, as seen in the solar modulation of cosmic rays observed at Earth~\cite{Potgieter:2013pdj}.  However, simple models, where the amplitude of the time variation is proportional to either  $E_\gamma$ or $E_\gamma^{-1}$, are ruled out by our \sag\ analysis.  The relatively energy-independent time variation amplitude could be a hint that additional solar modulation from within the Earth's orbit plays a relatively minor role in affecting the cosmic-ray spectrum at the solar surface. Alternatively, the observed relationship could be produced via two competing effects. For example, if the efficiency of solar atmospheric magnetic fields in producing outgoing gamma rays from incoming cosmic rays increases with energy, this could offset the effect of solar modulation in decreasing the cosmic ray flux.  Finally, the observed time variation could be dominated by other time-varying solar properties, such as a time-dependent scale height for cosmic-ray reversal that affects cosmic rays equally at all energies.

We note two potential observables that could illuminate these possibilities. First, a detailed analysis of the IC halo may be able to probe the effect of cosmic-ray electron propagation inside the Earth's orbit, isolating any effects due to the solar atmosphere. Second, while the solar minimum \sag\ flux is bright up to 200\,GeV, no $>100$\,GeV events were observed outside the solar minimum. This extreme time variation may provide important clues for the underlying gamma-ray production mechanism.

\subsection{Implication for solar atmospheric neutrinos}

Understanding the production mechanism of the \sag\ flux is extremely important for solar atmospheric neutrino searches by IceCube and KM3NeT. 
Normally, the calculations of solar atmospheric neutrinos assume that the Sun has zero magnetic fields~\cite{Ingelman:1996mj, Arguelles:2017eao, Edsjo:2017kjk}, implying that Earth-bound neutrinos are produced from the far side of the Sun. Importantly, neutrino absorption in the Sun starts to be important above 100\,GeV.   
Even in this case, the harder solar atmospheric neutrino spectrum (compared to the Earth's atmospheric neutrino background) makes the Sun a potentially detectable source for large neutrino telescopes above 1\,TeV~\cite{Ng:2017aur}. 
It was, however, not clear how magnetic fields could affect the solar atmospheric neutrino flux. 

The observed $>100$\,GeV \sag\ flux during the solar minimum implies that solar magnetic fields must be able to significantly  affect cosmic rays $\gtrsim$\,TeV to boost the gamma-ray production.  If there is a large component of reflected TeV cosmic rays that interacts in the atmosphere, this may enhance the solar atmospheric neutrino flux by producing neutrinos in lower density regions that have a longer charged pion interaction length, and where the neutrinos suffer negligible absorption.

Importantly, the observed solar minimum \sag\ flux provides a model-independent estimate of the \emph{minimal} solar atmospheric neutrino flux at this energy range. This estimate assumes no gamma-ray absorption. If gamma rays are efficiently absorbed by the solar atmosphere, it would increase the predicted neutrino flux.

We take the solar minimum \sag\ flux above $50$\,GeV to be $\Phi_{\gamma} \simeq 2\times 10^{-8}\,{\rm GeV\,cm^{-2}\,s^{-1} }(E_{\gamma}/{\rm 100\,GeV})^{-\Gamma}$, with $\Gamma\simeq 2$. The corresponding muon neutrino flux is about a factor of $2^{1-\Gamma}$ smaller, and thus $\Phi_{\nu} \simeq 0.5 \Phi_{\gamma}(E_{\gamma} \rightarrow E_{\nu})$.  This takes into account that the number of photons is roughly the same as each flavor of neutrino~(including antineutrinos) from hadronic interaction, but photons have twice the averaged energy~\cite{Kistler:2006hp}.  Approximating the neutrino effective area of IceCube as $20\,{\rm cm^{2}}(E_{\nu}/100\,{\rm GeV})^{2}$~\cite{Aartsen:2016zhm}, and considering $\Phi_{\nu}$ from 100\,GeV to 1\,TeV, the number of events expected by IceCube is about 1 event per 1.5 year.  It is thus extremely interesting to see if IceCube can detect these neutrinos in the upcoming solar minimum. 

If the high-energy \sag\ flux is transient in nature (e.g., correlating with the formation and destruction of specific magnetic field structures), then a coordinated search using IceCube, Fermi and HAWC will be extremely beneficial. In particular, employing variability information can improve the sensitivity of IceCube by allowing it to relax some data selection criteria to potentially lower the energy threshold, while maintaining a high signal-to-background ratio within a smaller time window. It is fortunate that IceCube and HAWC are located in different hemispheres, as IceCube is significantly more sensitive to solar neutrinos when the Sun is below the horizon. 

\vspace{-0.3cm}
\subsection{Implication for solar dark matter searches}

The detection of $>100$\,GeV \sag\ flux could significantly affect solar dark matter searches with neutrino telescopes.  The fact that magnetic fields strongly affect cosmic rays above 1 TeV adds considerable uncertainty to current solar atmospheric neutrino flux calculations.  If the neutrino flux were similarly enhanced, compared to the calculations without magnetic fields, this could raise the dark matter sensitivity floor~\cite{Arguelles:2017eao, Ng:2017aur, Edsjo:2017kjk}, and reduce the dark matter parameter space that can be probed by neutrino telescopes~\cite{Bagnaschi:2017tru, Costa:2017gup}. 

Most dark matter models do not produce a significant neutrino flux above 1\,TeV, due to neutrino absorption within the solar core.  However, if dark matter accumulated in the solar core annihilates first into some metastable mediator, that mediator could subsequently decay outside the solar core and produce observable neutrinos, gamma rays, and other messengers at TeV energies~(for some examples, see Refs.~\cite{Bell:2011sn, Feng:2016ijc, Leane:2017vag, Arina:2017sng}).  Correlated high-energy neutrino and gamma ray observations were proposed to search for these metastable mediator dark matter model, as the cosmic-ray induced gamma-ray flux was expected to be negligible at $\sim$\,TeV energies~\cite{Leane:2017vag}.   If solar magnetic fields can enhance both gamma-ray and neutrino production above TeV, it would be difficult to differentiate this emission from metastable mediator dark matter signals, at least in the TeV range. 

Lastly, while it is tempting to associate the \sag\ component above the 30--50\,GeV dip with dark matter or another new physics origin, it is not clear how exotic physics can accommodate the extreme time variation and morphological shift~\cite{Linden:2018} observed in \sag events $>100$\,GeV. On the other hand, the extreme time variation in the high-energy \sag\ flux potentially provides a method for removing the astrophysical contribution to reveal any underlying dark matter component.

\vspace{-0.5cm}
\section{Conclusions and Outlook}
\label{sec:conclusions}
\vspace{-0.3cm}
In this work, we analyzed 9 years of Fermi data to detect the \sag\ flux and study its spectrum and time variation. We detect significant \sag\ emission from 1\,GeV up to $\gtrsim$\,200\,GeV, extending our previous results in NPBR2016 that detected \sag\ emission only up to 30\,GeV.   

Using data covering most of the solar cycle, we find that the \sag\ flux anticorrelates with solar activity between 1--10~GeV. The \sag\ flux peaks during the 2008 solar minimum, and decreases as the Sun becomes more active.  Additionally, we observe, for the first time, the increase of the \sag\ flux corresponding to the decreasing solar activity since 2013.  We find that the amplitude of this time variation does not change significantly with energy between 1--10\,GeV, which may shed light on the underlying gamma-ray production mechanism. 

Interestingly, we discover a significant spectral dip between $\sim$\,30--50\,GeV in multiple datasets. We find no evidence that this dip is a systematic or instrumental artifact, and find no obvious theoretical interpretation.  The dip appears to be more significant when the Sun is less active, thus Fermi observations during the upcoming solar minimum may shed light on the nature of this feature.

Motivated by our companion paper~\cite{Linden:2018}, we compute the \sag\ flux during the Cycle 24 solar minimum.  The solar minimum \sag\ flux exhibits a hard and bright spectrum up to at least 200\,GeV. This hard \sag\ component could be tested by Fermi and HAWC during the upcoming Cycle 25 solar minimum, which is beginning presently.  In the future, LHAASO can probe the \sag\ flux at $\sim$\,TeV energies with even better sensitivity. 

The hard spectrum of the solar minimum \sag\ flux suggests that solar magnetic fields affect cosmic rays above 1\,TeV in the solar atmosphere, which suggests that magnetic fields may significantly affect the expected solar atmospheric neutrino flux.  In addition, the observed \sag\ flux during solar minimum implies a minimal detectable neutrino flux, which may be constrained by IceCube.  A coordinated search of IceCube, HAWC, and Fermi data may improve the sensitivity of these observations and test whether this component is transient in nature.  

The results of this work lay the foundation for building a more accurate model for cosmic-ray induced high-energy emission from the Sun.  This is crucial for a robust prediction of high-energy solar neutrinos, which is, in turn, important for solar dark matter searches.  Ultimately, high-energy emission from the Sun may provide a novel probe of the solar atmosphere, cosmic rays in the solar system, and new physics.
	
\vspace{-0.5cm}
\section*{Acknowledgments}
\vspace{-0.5cm}
We thank members of the Fermi-LAT collaboration, and in particular Regina Caputo, for discussions concerning potential instrumental systematics near the solar position.
We also thank Markus Ackermann, Bill Atwood, Xiaojun Bi, Keith Bechtol, Zhen Cao, Ofer Cohen, Federico Fraschetti, Hongbo Hu, and Carsten Rott for helpful discussions.
TQW is supported by the National Natural Science Foundation of China under grants No. 11547029.
KCYN is supported by Croucher Fellowship and Benoziyo Fellowship.
TL, BZ, and AHGP are supported in part by NASA Grant No. 80NSSC17K0754. 
BZ is also supported by a University Fellowship from The Ohio State University.
JFB is supported by (and BZ is partially supported by) NSF grant PHY-1714479.


\setcounter{figure}{0}
\renewcommand{\thefigure}{A\,\arabic{figure}}

\appendix
\section*{Supplemental Material}

Here we provide more details concerning the analysis techniques utilized throughout the main text, and perform systematic checks to evaluate the possibility that instrumental systematics affect our determination of the \sag\ flux and spectrum. 
In Appendix~\ref{appendix:analysis_systematics}, we investigate potential systematics stemming from the non-standard data analysis techniques required to evaluate the flux and spectrum of the Sun, which is rapidly moving compared to diffuse backgrounds.  
In Appendix~\ref{appendix:check_dip}, we investigate potential systematic issues in the Fermi-LAT effective area or energy reconstruction stemming from the unique position of the Sun in instrumental phase space.
In Appendix~\ref{appendix:dip_significance}, we describe the methodology used to quantify the significance of the 30--50\,GeV dip in the \sag\ spectrum.   In Appendix~\ref{appendix:dispersion}, we check the effect of instrumental energy dispersion regarding the gamma-ray absorption hypothesis of the dip.  In Appendix~\ref{appendix:neutrondip}, we check the quality of the photons regarding the solar neutron hypothesis. And in Appendix~\ref{appendix:proton_gap}, we investigate the potential two-component cosmic-ray interpretation of the dip. 
Lastly, in Appendix~\ref{appendix:time_variation}, we show the time variation of the \sag\ flux in finer energy bins.

\section{Systematic Effects in the Data Analysis Methodology}
\label{appendix:analysis_systematics}
\subsection{Background Estimation Using ``Fake Suns"}
\label{appendix:fakesuns}

\begin{figure}[tbp]
\centering
\includegraphics[width=0.48\textwidth]{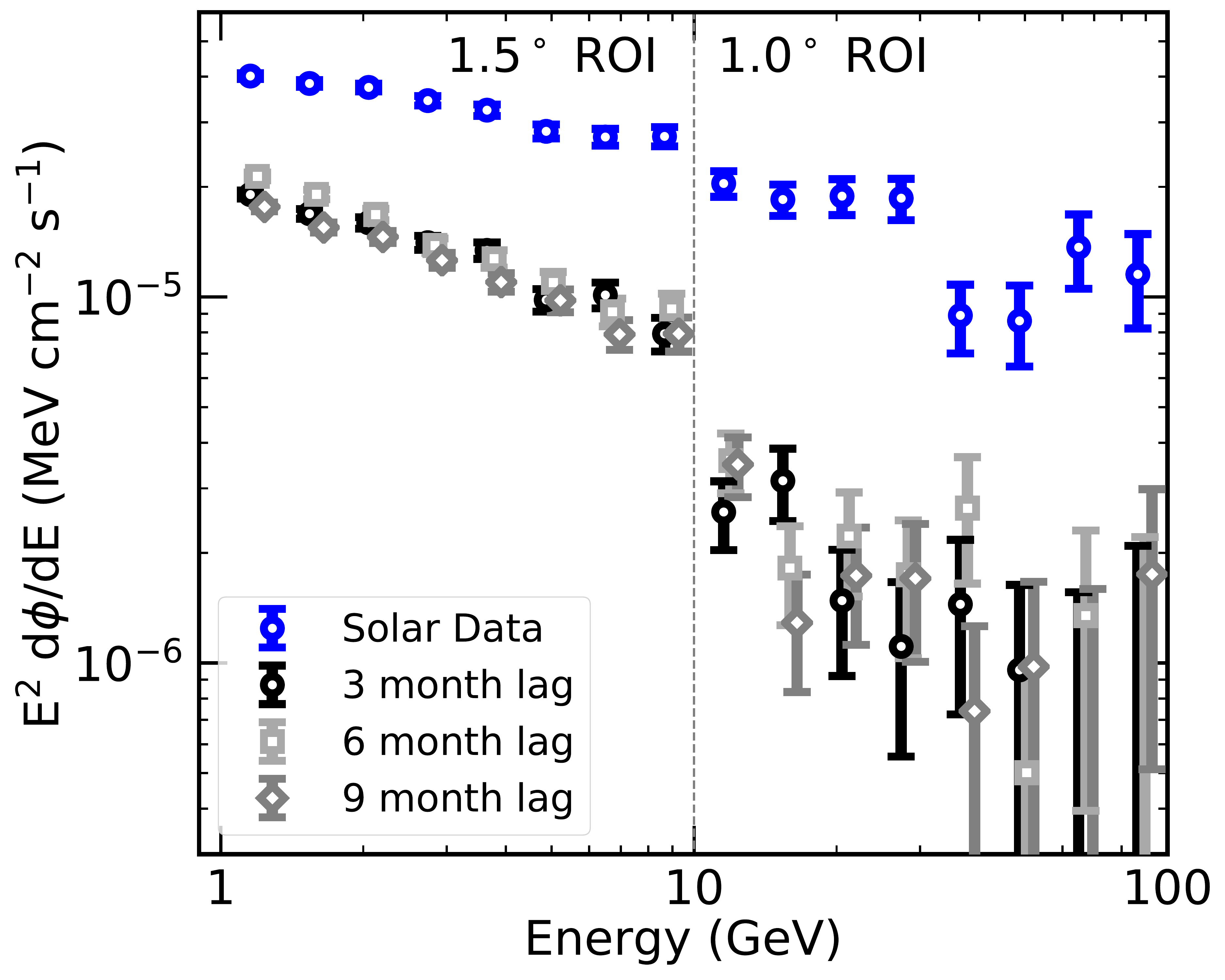}
\caption{The total gamma-ray flux from within 1.5$^{\circ}$ and 1.0$^{\circ}$ of the Sun~(blue) with $|b|>5^{\circ}$ selection, compared to the spectrum of ``fake Suns" characterized by a position that trails the Sun by 3-months (black), 6 months (dark gray) and 9 months (light gray). The mechanism utilized to calculate the flux from fake-Sun positions is identical to that of the real-Sun, except that periods with solar flares are not removed from the data. The flux observed from fake-Sun positions characterizes the diffuse contribution to the \sag\ flux. The sudden dip in the ``fake-sun" gamma-ray flux at 10\,GeV stems from the more stringent angular cut of 1$^\circ$ imposed on the data above this energy. The lack of any observed dip in the 30--50\,GeV energy range argues against any systematic effect relating to our data-extraction algorithm in producing an anomalous gamma-ray spectral feature, though we note the large error-bars make it difficult to rigorously rule out this possibility.}
\label{fig:fakesuns}
\end{figure}

In this section, we examine the gamma-ray flux and spectrum from three ``fake Sun" models, which trail the position of the Sun by 3 months, 6 months, and 9 months, respectively. This test serves two purposes. First, it provides an accurate estimate of the diffuse background at the solar position, because each fake Sun moves (over the course of a year) through the same array of coordinates as the real Sun. Because the vast majority of gamma-ray emission occurs in steady state, this provides an exposure-weighted calculation of the background flux. Second, this test allows us to evaluate our data extraction methodology, to ensure that our analysis of a quickly moving solar region does not produce any spectral features in the position of power-law backgrounds.

In Figure~\ref{fig:fakesuns}, we compare the total flux from within 1.5$^{\circ}$ and 1.0$^{\circ}$ of the Sun with the calculated gamma-ray flux from each fake-Sun position, both with the $|b|>5^{\circ}$ selection. We immediately note two conclusions: (1) the diffuse gamma-ray emission accounts for approximately half of the observed gamma-ray flux near the solar position at low-energies, but contributes only $\sim$10\% of the total emission at high energies. These results are consistent with our \sag\ analysis (shown in Table~\ref{tab:result}) (2) there is no spectral feature in the diffuse gamma-ray flux of the fake-Sun positions in the 30--50\,GeV energy range, which argues against any contribution from either diffuse emission or our analysis routines to the observed solar spectral dip.

\subsection{Background Estimation Using Stacked ``Fake Suns"}
\label{appendix:stackedfakesuns}

\begin{figure}[tbp]
\centering
\includegraphics[width=0.48\textwidth]{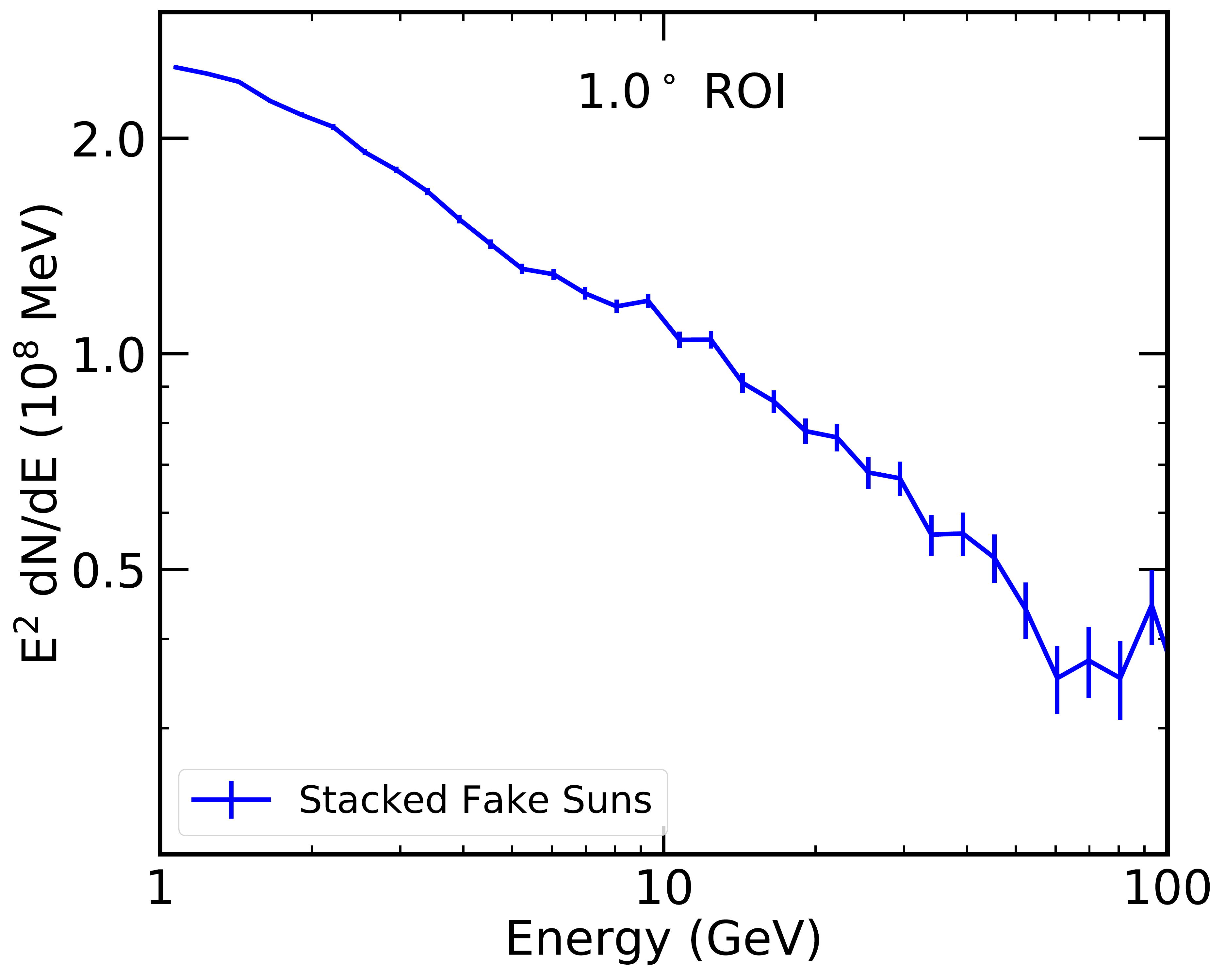}
\caption{The gamma-ray energy spectrum calculated from stacking the emission at 177 ``fake Sun" positions that trail the real sun by between 6-360 days (even numbered days only). This provides a large statistical sample of the diffuse gamma-ray spectrum obtained at the solar position. We find no evidence for a spectral dip between 30--50\,GeV, validating both our analysis method, and the lack of any significant contribution from diffuse gamma-ray sources. Unlike Figure~\ref{fig:fakesuns}, we utilize a consistent 1$^\circ$ angular cut for all photons above 1\,GeV, in order to remove the artificial dip at 10\,GeV, and to focus our interest on the 30--50\,GeV feature.}
\label{fig:stackedfakesuns}
\end{figure}

At high energies, the statistical information that can be obtained from the fake-Sun positions analyzed in Section~\ref{appendix:fakesuns} is minimal, due to the very small diffuse gamma-ray flux at high energies. In particular, the small statistical sample limits our ability to conclusively determine whether or not a dip in the gamma-ray spectrum exists between $\sim$\,30--50\,GeV in our fake-Sun sample. 

In this section, we increase the statistical sample by evaluating the total emission from 177 fake sky positions, including ``fake-Suns" that trail the position of the real sun by between 6--360 days in two-day increments. We choose the two degree increment because the Sun moves approximately 1$^\circ$ per day, and remove positions within 5$^\circ$ of the Sun due to the non-negligible solar ICS component close to the real Sun. Because this analysis is focused on the origin of the 30--50\,GeV dip, we use a consistent radial cut of 1.0$^\circ$ at all energies in this analysis. 

\begin{figure*}[tbp]
	\centering
	\includegraphics[width=1.0\textwidth]{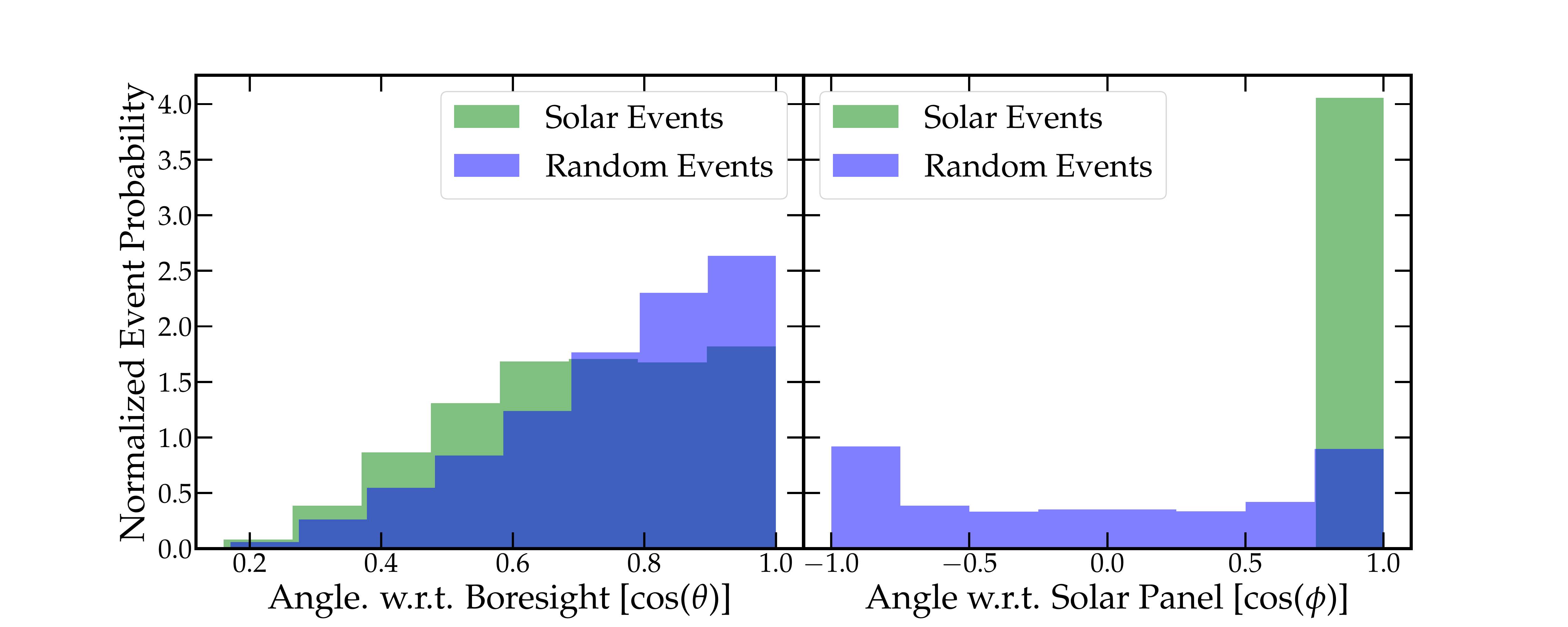}
	\caption{The instrumental phase-space distribution of solar photons that pass our energy ($>$1~GeV) and angular ($<$1.5$^\circ$ below 10~GeV and $<$1$^\circ$ above 10~GeV) event selection cuts (green), compared to the distribution of an equal number of randomly chosen events with energy $>$1~GeV and an angular separation of $>$5$^\circ$ from the solar position (blue). The instrumental phase space is divided into the angle $\theta$, which describes the incident angle of the photon with respect to the LAT boresight, and the angle $\phi$, which describes the incident angle with respect to the \emph{Fermi} solar panels. While the distribution of events in $\theta$-space is relatively normal (and is carefully corrected for in calculating the Fermi-LAT effective area), the distribution of events in $\phi$-space is unique for solar photons. We note that the $\phi$-dependence of solar events is somewhat more extreme than indicated by our binning choices, as more than 90\% of the events are concentrated within 3$^\circ$ of $\phi$~=~0, (i.e.~cos($\theta$)~$>$~0.998).}
	\label{fig:instrumentalspace}
\end{figure*}

Unfortunately, due to the large number of solar positions we analyze, it is not computationally feasible to calculate the Fermi-LAT exposure at each fake-sun position. Thus, we plot the spectrum in units E$^2$~dN/dE, utilizing the result from Section~\ref{appendix:effectivearea} which indicates that their is no peculiar energy dependence in the Fermi-LAT effective area near the position of the gamma-ray dip.  For the spectrum of a few fake Suns in physical units, they can be inferred from Fig.~\ref{fig:diskflux} for $|b|>{30}^{\circ}$ and Fig.~\ref{fig:fakesuns} for $|b|>5^{\circ}$.

In Figure~\ref{fig:stackedfakesuns} we show the resulting spectrum from our analysis, finding no evidence for a unique spectral feature between $\sim$ 30-50~GeV. The combination of this test, alongside the fluxes of individual fake suns in Appendix~\ref{appendix:fakesuns}, indicate that our analysis technique does not introduce any spectral dips in the 30-50~GeV energy range. 

\section{Systematic Effects in the Instrumental Response Near the Solar Position} 
\label{appendix:check_dip}

\subsection{The $\phi$-dependence of the Effective Area}
\label{appendix:effectivearea}

\begin{figure}[tbp]
\centering
\includegraphics[width=0.48\textwidth]{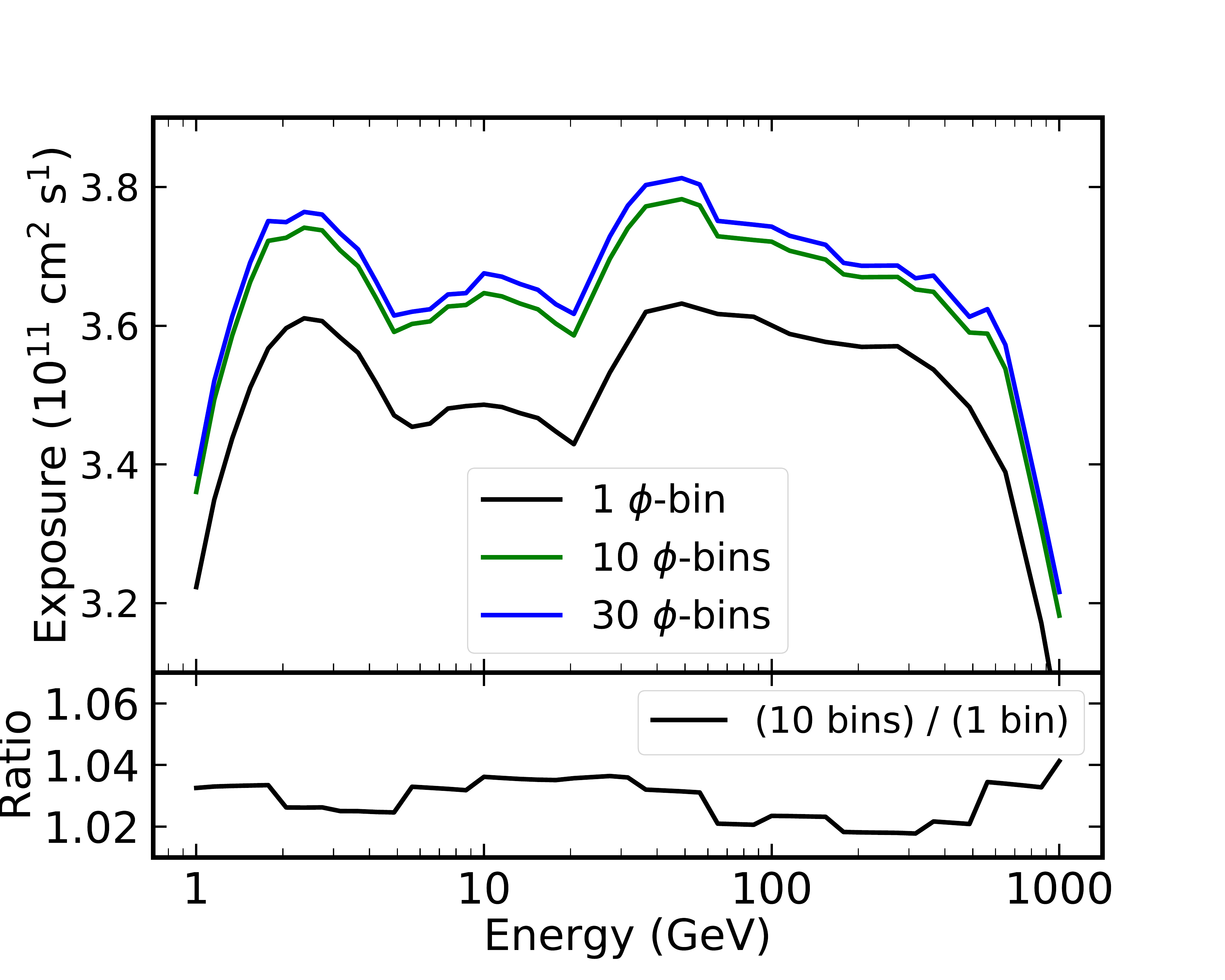}
\caption{The integrated Fermi-LAT exposure over our analysis time period at the Solar position, utilizing all periods when the Sun exceeds a latitude of b$>$5$^\circ$. The exoposure is calcualted for our default assumption, which averages over the $\phi$-distribution of solar photons ({\tt phibins}=1, black), as well as for two improved exposure calculations which finely bin the $\phi$-dependence of the Fermi-LAT effective area ({\tt phibins}=10, green), ({\tt phibins}=30, blue). We find that the finer $\phi$-bining increases the overall instrumental exposure by approximately 4\%, but does not induce any significant energy-dependent features above the $\sim$1\% level. This precludes \emph{known} systematic effects in the $\phi$-dependence of the Fermi-LAT exposure from explaining the amplitude of the solar gamma-ray dip.}
\label{fig:phieffectivearea}
\end{figure}

We utilize the Fermi-LAT tools {\tt gtltcube} and {\tt gtexpcube2} to calculate the energy-dependent effective area and exposure of solar observations. In our default analysis, we follow the standard procedure, which finely bins the effective area in the angular coordinate denoting the distance of the observed photon from the Fermi-LAT boresight ($\theta$), but does not finely bin the effective area in the second angular coordinate, which describes the recorded angle of photons with respect to the Fermi solar panels ($\phi$). For standard target analyses, these choices are correct - because the effective area varies strongly with $\theta$ and only marginally with $\phi$. 

However, because the Fermi-LAT solar panels are typically aligned with the Sun, it occupies a unique position in $\phi$-space. Thus, it is likely that any small $\phi$-dependent systematic will appear first in solar observations. In Figure~\ref{fig:instrumentalspace}, we plot the distribution of solar photons that pass our angular and energy reconstruction cuts in instrumental phase-space, compared to a randomly selected sample of photons located more than 5$^\circ$ from the solar position. We find that the Sun has a slightly-skewed distribution in $\theta$-space, but note that we carefully take this into account by weighting the effective area in bins of cos($\theta$)~=~0.025. However, the $\phi$-distribution of solar photons differs drastically from the average, and this is not taken account in our default analysis. In fact, the distribution is somewhat more peaked than portrayed in Figure~\ref{fig:instrumentalspace}, as nearly all solar events are located within 3$^\circ$ of $\phi$~=~0 (which gives cos($\phi$)~$>$~0.998).

While the Fermi-LAT collaboration notes that the overall change in the effective area at 10~GeV as a function of $\phi$ is only $\sim$1\%~\cite{fermi_performance}, they do not discuss the energy-dependence of this effect. Here, we utilize the hidden parameter {\tt phibins} in {\tt gtltcube} to directly calculate the $\phi$-weighted energy-dependent exposure of solar photons. This accounts for any \emph{known} systematic issue affecting the Fermi-LAT effective area or energy reconstruction near the solar position that was not taken into account in our standard analysis. In Figure~\ref{fig:phieffectivearea}, we plot the Fermi-LAT effective area as a function of energy for models where we set {\tt phibins=1} (our default assumption) as well as {\tt phibins=10} and {\tt phibins=30}. While even the choice of 30 $\phi$-bins produces $\sim$12$^\circ$ bins that are somewhat larger than the distribution of solar photons in $\phi$-space, we found that this was the largest computationally feasible measurement. Moreover, because events near $\phi$~=~0 pass directly through the flat side of the instrument, we expect the $\phi$-dependence to be mild at values very close to $\phi$~=~0 itself. 

We note two effects. First, the calculated effective area increases by about 4\% when the analysis is increased to 30 bins. This result is similar to the Fermi-LAT collaboration analysis, and implies that the calculated solar gamma-ray flux is decreased by about 4\% relative to the values calculated in the main portion of the text. Second, the modification to the Fermi-LAT effective area is relatively energy-independent, inducing spectral changes only at the level of $\sim$1\%. The result of this analysis indicates that no \emph{known} systematic issue in the Fermi-LAT event reconstruction at values near $\phi$~=~0 contributes to the dip in the solar gamma-ray spectrum.

\subsection{Spectral Features in Events Near $\phi$=0}
\label{appendix:phidependence}

\begin{figure*}[tbp]
\centering
\includegraphics[width=1.0\textwidth]{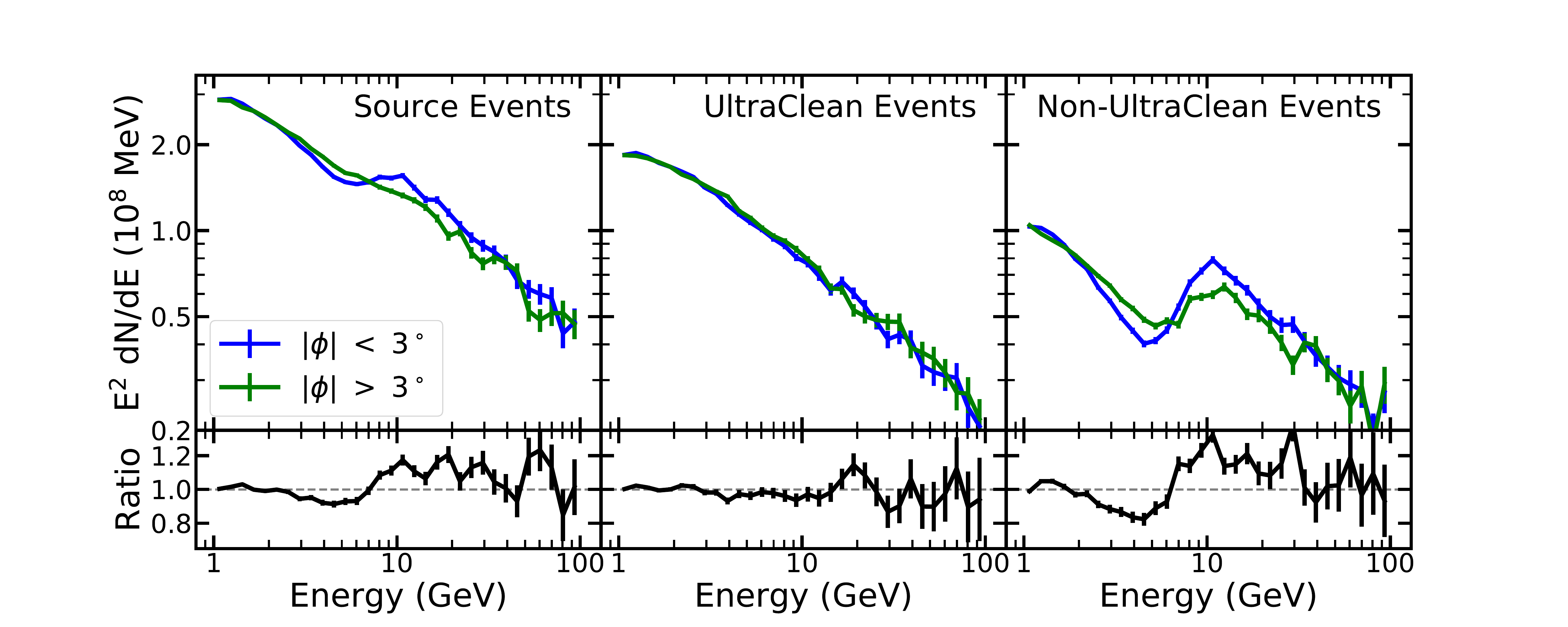}
\caption{The spectrum of gamma-rays found at least 5$^\circ$ from the contemporaneous solar position, and recorded within 3$^\circ$ of $\phi$~=~0 in instrumental coordinates ($\phi$~$<$~3$^\circ$ or $\phi$~$>$~357$^\circ$), compared to the spectrum of random ``partner" events recorded at similar R.A. and dec positions to each event, but outside of this $\phi$-range in instrumental coordinates. Results are shown for our default Source event class (left), the more conservative UltraCleanVeto event class (center), which is designed to remove mis-identified charged cosmic-rays, and the ensemble of Source class events that do not pass the UltraCleanVeto event selection (right). We find a statistically significant feature in the energy-reconstruction of Source class events near $\phi$~=~0 that likely indicates a $\phi$-dependent systematic error in the LAT-reconstruction algorithm at the level of $\sim$10-20\%. This effect disappears in UltraCleanVeto events, signaling that it is likely due to cosmic-ray misidentification. This effect potentially explains the sharp peak in solar gamma-ray photons near directly below 10~GeV, as shown in Figure~\ref{fig:totalflux}. However, we note that this feature does not appear to produce any significant spectral feature between 30--50~GeV, where the spectra of both $\phi$~$\sim$~0 and partner events appear similar. Additionally, the amplitude of this effect is significantly smaller than the observed 30--50~GeV spectral dip.}
\label{fig:phidependence}
\end{figure*}

\begin{figure*}[tbp]
	\centering
	\includegraphics[width=1.0\textwidth]{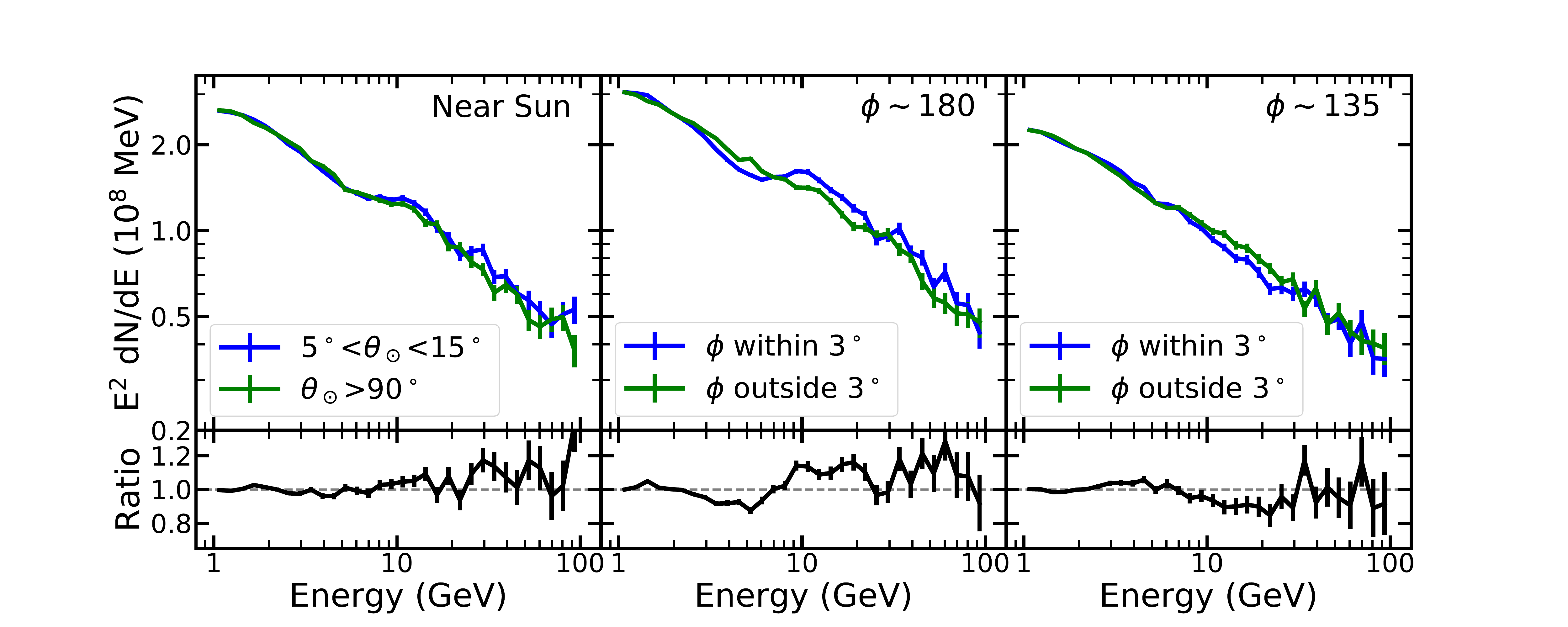}
	\caption{Same as Figure~\ref{fig:phidependence}, but for three separate divisions of the Fermi-LAT Source class events. First, we compare events observed between 5$^\circ$--15$^\circ$ from the contemporaneous solar position with events recorded more than 90$^\circ$ from the solar position (left). Second, we compare events observed within 3$^\circ$ of $\phi$~=~180$^\circ$, which corresponds to events observed opposite to most solar gamma-rays (center). Third, we compare events observed within 3$^\circ$ of $\phi$~=~135$^\circ$, which corresponds to a corner in the Fermi-LAT instrument rather than the center of the instruments face. We find that the sun itself does not appear to significantly affect the rate of cosmic-ray contamination, while $\phi$-positions corresponding to the center of the Fermi-LAT instrument do appear to produce such an effect.}
	\label{fig:phidependenceextra}
\end{figure*}

In the previous section, we ruled out the possibility that ``known" features in the $\phi$-dependence of the Fermi-LAT effective area could account for the observed spectral-dip in the Solar gamma-ray flux. However, it is also possible that ``unknown" features --- those not represented in models of the Fermi-LAT effective area or energy dispersion --- could account for the spectral dip observed in the solar gamma-ray flux. Because most sources have relatively uniform exposures in $\phi$, it is possible that a significant effect has gone unnoticed in previous analyses of the Fermi-LAT data -- only to appear first in high-energy analyses at the solar position.

We develop the following test of this hypothesis. We first extract every gamma-ray photon that passes our default event class and temporal cuts with the following characteristics: (1) a recorded energy exceeding 1~GeV, (2) a distance of at least 5$^\circ$ from the contemporaneous solar position, (3) an observed $\phi$-angle within 3$^\circ$ of $phi$~=~0. These cuts produce a dataset of 223,707 events with similar instrumental reconstructions as solar photons, but which do not have a solar origin. For each event, we select a single random ``partner" event with the following characteristics: (1) an energy exceeding 1~GeV, (2) an observed $\phi$-angle exceeding 3$^\circ$ from $\phi$~=~0.0, and (3) a small angular separation from the original event in r.a. and dec. 

The final selection criteria ensures that the average spectrum of all source and partner events is similar, as they are produced in equivalent regions of the Fermi-LAT sky. We pick angular separations following a tiered approach, we first randomly select any partner event within 0.1$^\circ$ of a given $\phi$~$\sim$~0.0 event, noting that this works for $\sim$90\% of our population and the separation is much smaller than the Fermi-LAT angular resolution, indicating that the origin of these two events is nearly identical. If no such photon is found, we widen our angular acceptance to 0.2$^\circ$ and then 0.5$^\circ$ to ensure that every photon has a partner. In combination with our observation (in the previous section) that the $\phi$-dependence of the instrumental effective area is energy-independent to within 1\%, this implies that the counts spectrum of gamma rays from both our source events and our random ``partner" events should be identical. 

First, we verify that this hierarchical event-selection procedure does not induce any spectral features into the  $\phi$~$\sim$~0.0 and partner events. We test this by repeating the analysis routine 10 times, with randomly chosen seed events. We find no spectral differences between the parent and partner events at the level of $\sim$1\%, indicating that our hierarchical event selection technique does not introduce any systematic biases into the energy-distribution of events.

In Figure~\ref{fig:phidependence} (left) we show the results of this analysis. Overall, we find that the spectra of the $|\phi|$~$<$~3$^\circ$ and  $|\phi|$~$>$~3$^\circ$ event selections are similar. However, there is a highly statistically significant feature extending from approximately 3~GeV to nearly 25~GeV in energy. In this range, we see an initially smaller population of $|\phi|$~$<$~3$^\circ$ gamma rays shift to a larger population of $|\phi|$~$<$~3$^\circ$ gamma rays at an energy of $\sim$8~GeV. We note that, because there are an equal number of events in each selection, an excess of photons at one energy must correspond to a deficit elsewhere. The magnitude of this effect is nearly 20\% at 10~GeV.  In the center panel, we show that this effect disappears when we select only the subset of gamma rays that pass the UltraCleanVeto event selection, while in the right panel, we show that this effect is more extreme for gamma rays that pass the Source, but not the UltraCleanVeto event classes. This indicates that the spectral feature is likely due to differences in the cosmic-ray rejection (or cosmic-ray energy reconstruction) between events observed near $\phi$~=~0 and events observed elsewhere in instrumental coordinates.

We note that this spectral feature could contribute to the observed solar gamma-ray spectrum at 5-15~GeV, shown in Figure~\ref{fig:diskflux}. In particular, there are hints of a rise in the observed solar gamma-ray spectrum between 5-8~GeV. However, we do not see any evidence in Figure~\ref{fig:phidependence} for a statistically significant spectral change between 30--50~GeV for photons observed near $\phi$~=~0. Because the photon count is equivalent in both datasets, we note that spectral changes should appear as changes in the gamma-ray ratio between selection cuts, rather than in the overall normalization of this selection ratio.

One potential explanation for the peculiar spectrum of gamma rays observed near $\phi$~=~0 is the influence of the Sun itself. This would be a particularly concerning possibility, as it would have an even more acute effect on our solar analysis compared to $\phi$~=~0 events, many of which are not located near the Sun. While the mechanism for such an effect is not known, the influence of solar radiation may potentially affect the performance of the Fermi-LAT. In Figure~\ref{fig:phidependenceextra} (left), we test this possibility by comparing the spectrum of photons observed between 5--15$^\circ$ from the contemporaneous solar position with photons observed more than 90$^\circ$ from the solar position, finding no statistically significant evidence for such an effect. We note that this observation tightly constrains the systematic shown in Figure~\ref{fig:phidependence}, because many photons located between 5--15$^\circ$ from the Sun have rather small $\phi$ values. Thus the systematic effect is strongly peaked only toward events with $\phi$~$\lesssim$~3$^\circ$.

Additionally, in the center panel of Figure~\ref{fig:phidependenceextra}, we examine the distribution of events recorded near $\phi$~=~180$^\circ$, finding a similar spectral feature at $\sim$10~GeV. Because $\phi$~=~180$^\circ$ events tend to point away from the contemporaneous solar position, but also are observed in an orientation normal to the solar panel plane, this implies that the systematic issue may appear in certain locations of instrumental $\phi$-space. Finally (right), we test photons recorded near $\phi$~=~135$^\circ$, which occupies a corner of the Fermi-LAT. We do not find the same systematic issue, but do observe a small dip in the gamma-ray spectrum between 10-25~GeV with an amplitude of $\sim$5\%.

The conclusion of this analysis is uncertain. There does appear to be a statistically significant effect, with an amplitude of nearly 20\%, affecting the reconstruction of Source class (but not UltraCleanVeto class) events near $\phi$~=~0. The fact that this event disappears at angles between 5$^\circ$--15$^\circ$ from the Sun, suggests that this effect disappears very rapidly for larger values of $\phi$. While this effect may influence our Solar spectrum near an energy of 10~GeV (where a similar feature is observed), there does not appear to be any lingering effect between 30--50~GeV. Moreover, the amplitude of this effect is insufficient to explain the much larger spectral dip observed in our data. While more analysis is necessary to understand this effect -- at present we find no indication that it explains the observed data. In the next section, we verify this result by re-running our analysis technique on the UltraCleanVeto event class, which shows no evidence for any peculiar $\phi$-dependent effects.

\subsection{Analysis of UltraCleanVeto Events}
\label{appendix:ultracleanveto}
In the last subsection, we showed that a systematic spectral feature appears to be induced for events near 10~GeV in the Source event class, but does not appear to be present in events which pass the more stringent UltraCleanVeto event reconstruction. Here, we re-run our analysis on events that pass the more stringent UltraCleanVeto cuts. This decreases the size of our statistical sample, but likely removes any remaining effects due to cosmic-ray misidentification. 

\begin{figure}[tbp]
	\centering
	\includegraphics[width=0.48\textwidth]{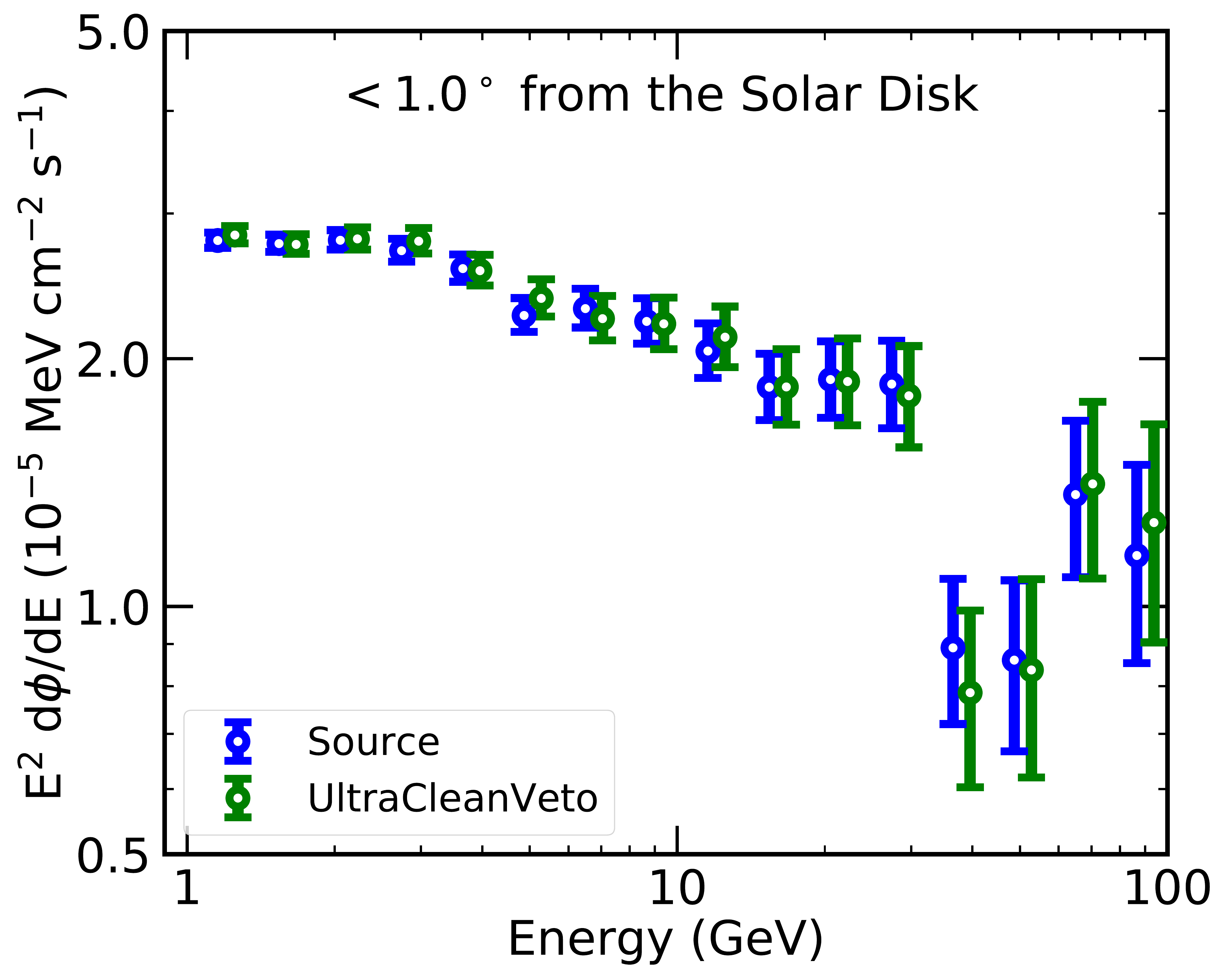}
	\caption{The gamma-ray spectrum obtained from analyzing only photons which belong to the more stringent UltraCleanVeto event class, which signifies a high-level of confidence that these events are not contaminated by charged cosmic rays. We stack all events observed while the Sun was located at a latitude $|$b$|$~$>$~5$^\circ$, in order to maintain comparable statistics to the Source event class utilized in the main text. We see similar evidence for the continuation of the spectral dip from $\sim$30-50~GeV, indicating that the misidentification of charged cosmic rays is unlikely to be responsible for the dip.  The same analysis, but for Source event class, is shown for comparison. }
	\label{fig:ultracleanveto}
\end{figure}

In Figure~\ref{fig:ultracleanveto} we show the resulting gamma-ray spectrum, utilizing the full angular ROI ($|b|>$5$^\circ$) with solar flare cuts implemented.  In order to minimize cosmic-ray background, we repeat the same analysis routine, restricting our analysis to photons which pass the UltraCleanVeto analysis cuts. This event class places a significantly more stringent rejection cut on possible charged cosmic-ray contamination, compared to the Source Event class utilized in the main text -- but does so at the cost of a 20--30\% rejection of gamma-ray events. This latter effect can be accounted for in the re-calculation of the instrumental effective area.  Our analysis shows that the spectral dip remains a resilient feature in this event class, providing strong evidence that charged cosmic-rays events are unlikely to be responsible for the gamma-ray spectral dip.

\begin{figure}[tbp]
	\centering
	\includegraphics[width=0.48\textwidth]{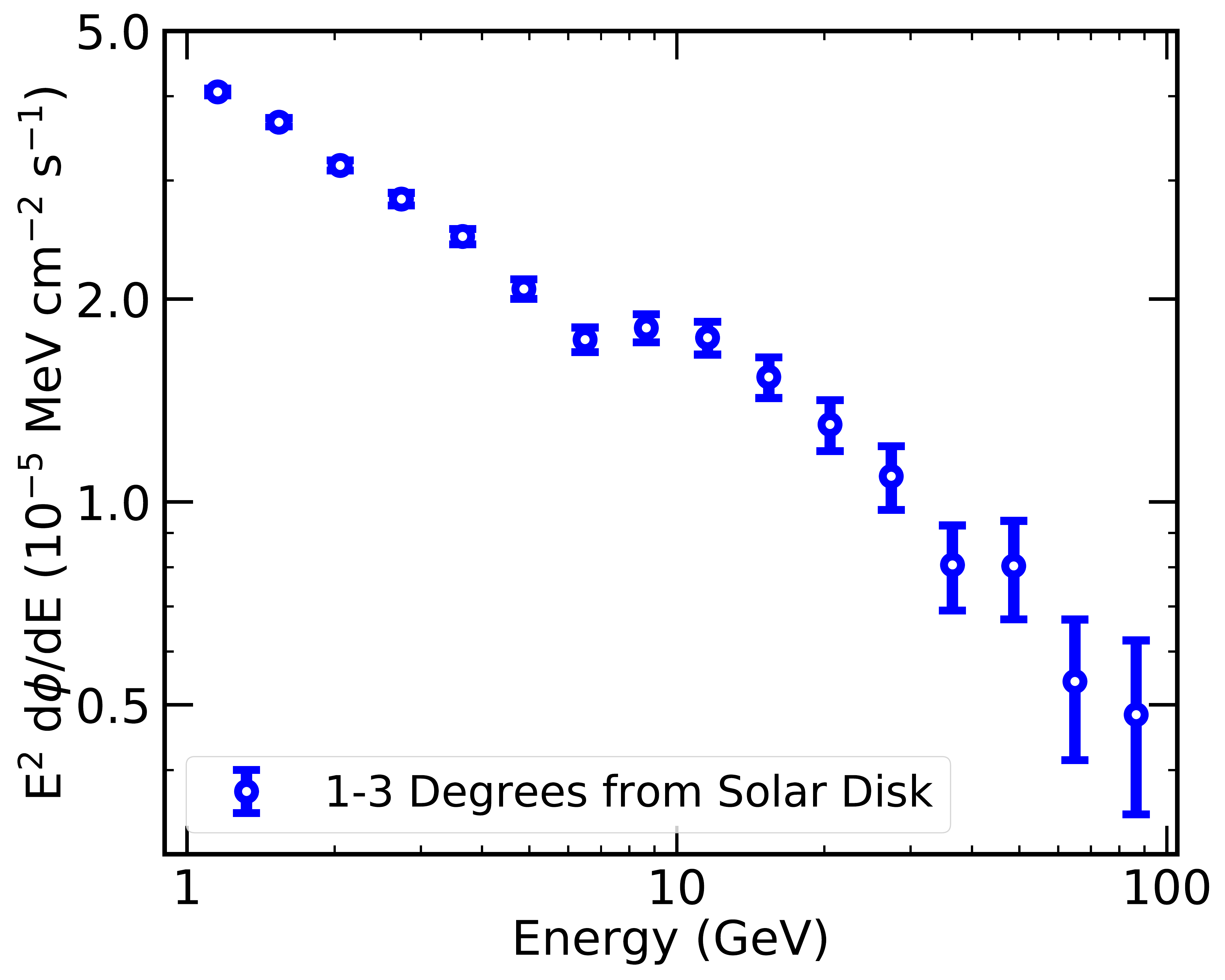}
	\caption{The gamma-ray spectrum obtained utilizing source Class photons observed at angular distances between 1$^\circ$--3$^\circ$ from the contemporaneous solar position. This emission is dominated by a combination of solar IC events, and background diffuse emission sources, and is not expected to produce observable spectral features at either 10~GeV or between 30--50~GeV. We clearly observe the 10~GeV spectral feature observed in Figures~\ref{fig:phidependence}, providing further evidence that this effect is systematic, while we observe no similar systematic feature between 30--50~GeV.}
	\label{fig:between13}
\end{figure}

\subsection{Analysis of Events Immediately Outside of the Solar Disk}
\label{appendix:between13}

\begin{figure}[tbp]
	\centering
	\includegraphics[width=0.48\textwidth]{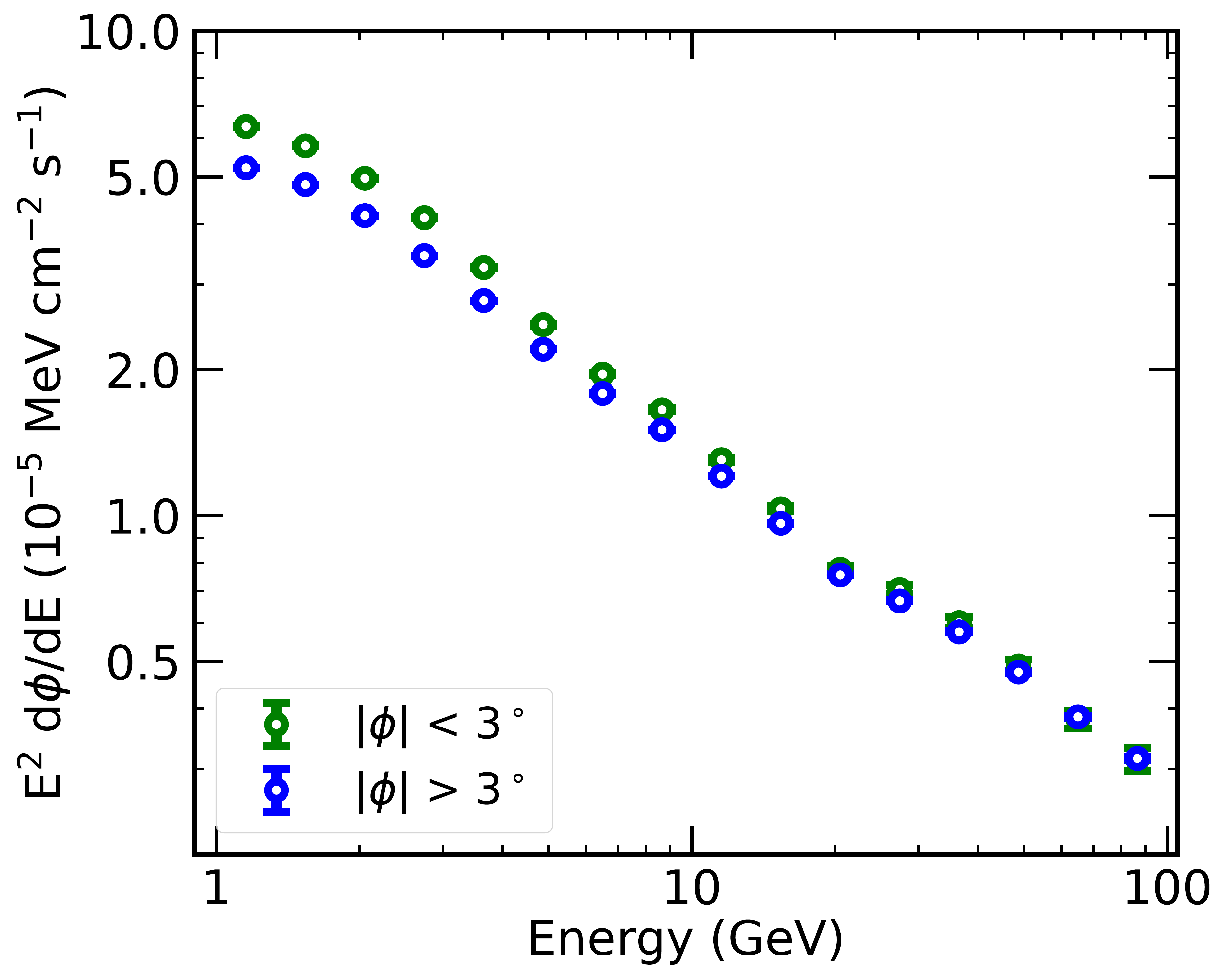}
	\caption{The gamma-ray spectrum obtained from events recorded with a zenith angle exceeding 110$^\circ$, which have a high likelihood of being produced by the bright Earth-limb. The events are divided into $|\phi|<3^{\circ}$~(green) and $|\phi|>3^{\circ}$~(blue, arbitrarily normalized for comparison). The spectrum of events near $\phi$~=~0 is found to be slightly softer than the spectrum of more distant events, possibly due to differences in the instrumental pointing towards the Earth-plane. However, no evidence for a spectral feature is found between 30--50~GeV in either dataset.}
	\label{fig:earthlimb}
\end{figure}

\begin{figure*}[tbp]
	\centering
	\includegraphics[width=0.90\textwidth]{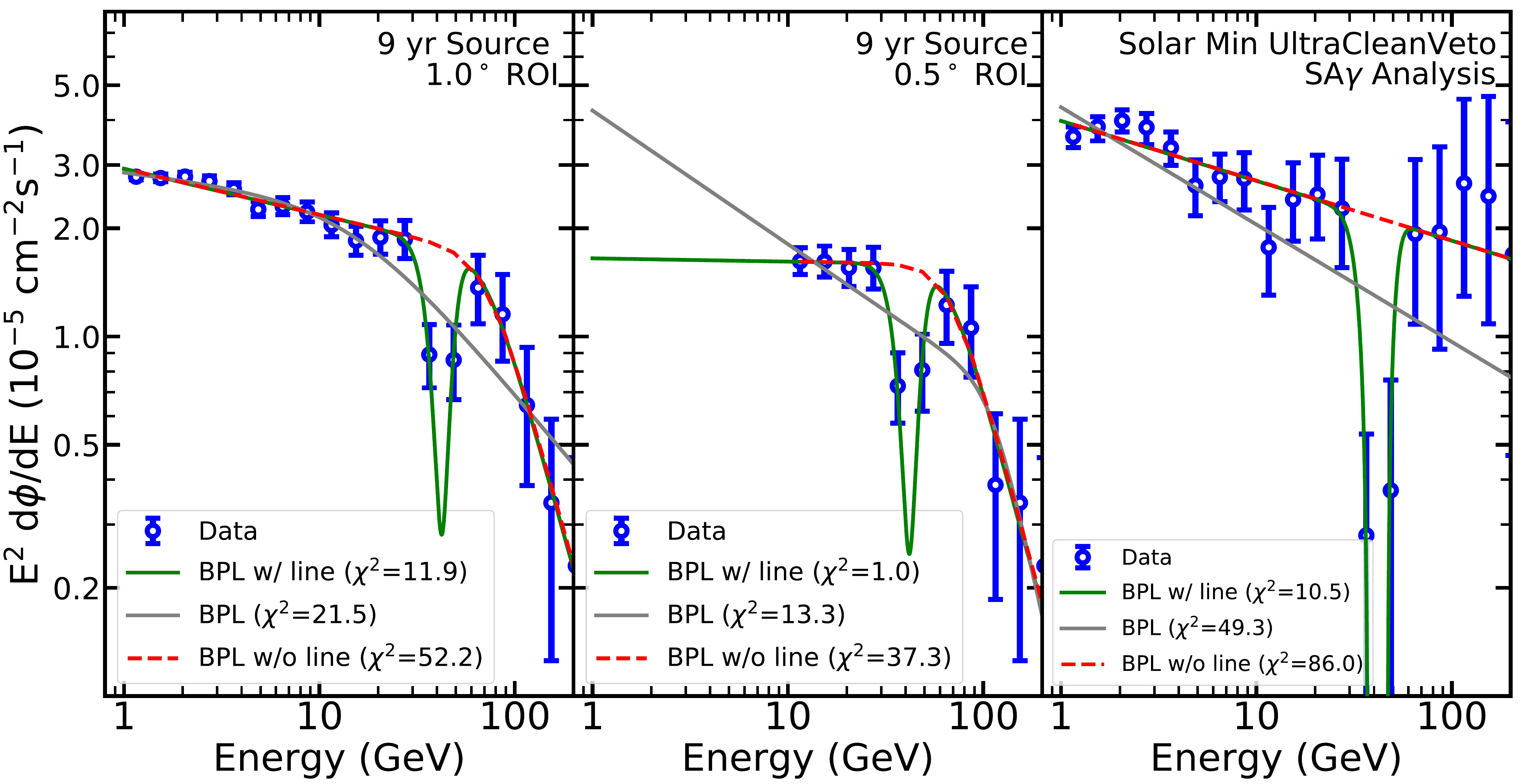}
	\caption{ {\it Left:} The 9-year averaged total flux~($\theta < 0.5^{\circ}$, $|b|>5^{\circ}$), and fits using a broken power-law with a negative line~(BPL-line).  The dashed line~(BPL w/o line) is the is the broken power law component in BPL-line.  The solid grey line just fitting the data with a broken power law~(BPL).  {\it Middle:} Same as left, but for $\theta < 1.0^{\circ}$ and considering only above 10\,GeV.   {\it Right:} Same as the left, but for the solar minimum \sag\ flux. }
	\label{fig:line_significance}
\end{figure*}

Thus far, we have found a significant systematic artifact in the reconstruction of Fermi-LAT events near $\phi$~=~0, which affects photons near 10~GeV, but does not affect events near the 30--50~GeV dip. Here, we produce another test, by calculating the spectrum of gamma-rays observed immediately outside of the solar disk. In particular, we analyze the gamma-ray spectrum in an annulus between 1$^\circ$--3$^\circ$ from the center of the Sun. This lies outside the Fermi-LAT PSF for all but the lowest energy gamma-rays, and is thus relatively free of contamination from the solar disk. Instead, these gamma-rays are expected to be produced by a combination of solar IC emission and diffuse background sources. Thus, this emission would not be expected to show any matching spectral features at either 10~GeV or between 30--50~GeV. 

\begin{figure*}[tbp]
	\centering
	\includegraphics[width=1.0\textwidth]{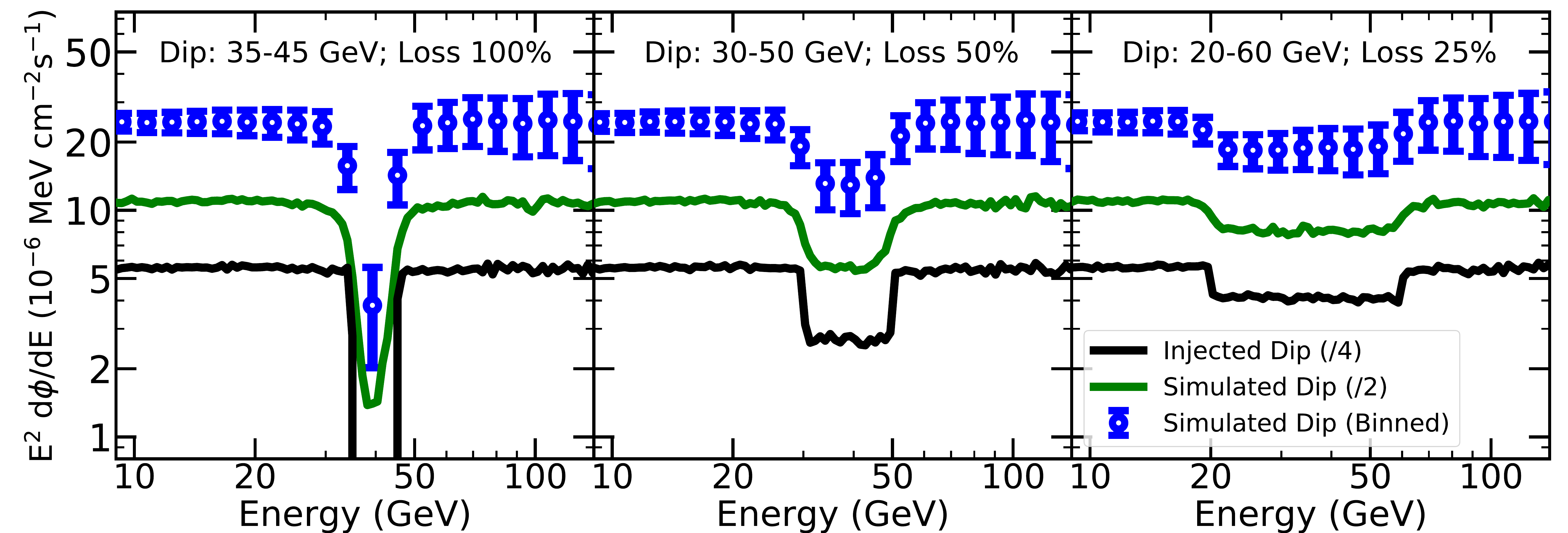}
	\caption{The injected (black) and reconstructed (green) energy spectrum for an dN/dE~$\propto$E$^{-2}$ gamma-ray emission spectrum with ``holes" centered at 40~GeV with depths and widths of 100\%, 10~GeV (left), 50\% and 20~GeV (center) and 25\% and 40~GeV (right), respectively. The binned-version of the reconstructed energy is shown in blue, with error bars corresponding to the 1$\sigma$ fluctuations expected in each bin for a single realization of the data. We find that the observed spectral dip in the Fermi data is not more narrow than the instrumental energy-dispersion, and is best fit by a feature where approximately 50\% of photons are removed over an energy range of approximately 20~GeV (center panel).}
	\label{fig:energydispersion}
\end{figure*}

In Figure~\ref{fig:between13} we show the resulting gamma-ray spectrum between 1$^\circ$--3$^\circ$ from the solar center, using Source class photons over the full 9~yr exposure. We find clear evidence for the $\sim$10~GeV spectral feature found in Figure~\ref{fig:phidependence}, providing additional evidence for its systematic origin. On the other hand, we find no evidence of a 30--50~GeV spectral dip in this spectrum. The statistical sample of these events is large enough that the observed emission is statistically inconsistent with the spectral dip observed across the solar disk. This implies that if the 30--50~GeV spectral dip is a systematic feature, the effect is confined to within 1$^\circ$ of the solar position, which we believe to be unlikely. Thus, this argues against a systematic origin for the 30--50~GeV feature.

\subsection{Analysis of the Earth Limb}
\label{appendix:earthlimb}

The brightest $\gamma$-ray source observable by the Fermi-LAT is the Earth limb, which produces gamma-ray emission through the interaction of cosmic-rays with the solar atmosphere. Because this emission source is known to be hadronic and feature-less, it serves as an optimal test location for instrumental artifacts. In this section, we compare the spectrum of Earth-limb events (those recorded with zenith angles exceeding 110$^\circ$) from $|\phi|<3^{\circ}$ with events at larger azimuthal angles $|\phi|>3^{\circ}$. 

In Figure~\ref{fig:earthlimb}, we show the results of this analysis, which provides no evidence for a spectral feature between 30--50~GeV in either dataset. We do note a slight (but highly statistically significant) spectral softening in the $|\phi|<3^{\circ}$ data below $\sim$10~GeV, compared to background events.  This spectral feature may result from the different exposure of $\phi$~=~0 events towards the Earths surface, stemming from the preferred orientation of the solar panels towards the Sun. While we are unable to conclusively test this hypothesis, we stress that this feature is extremely smooth, most prominent at relatively low energies, and does not have an amplitude capable of significantly affecting the measurements presented in the main text.

\subsection{Discussion}
In this section, we have discussed both systematic issues relating to the instrumental effective area of Fermi-LAT events recorded near the solar position, as well as systematic issues relating to the data-analysis procedure utilized in the text. While we have found a peculiar systematic issue involving events recorded near $\phi$~=~0$^\circ$ and near an energy of 10~GeV, we have found no plausible systematic issue which might produce a dip in the solar gamma-ray spectrum in the 30-50~GeV energy range under investigation.  In addition, we do not detect any 30--50\,GeV dip in photons recorded very close to the contemporaneous solar position.  We stress that the results here do not conclusively rule out the possibility that a systematic issue produces the peculiar 30--50~GeV spectral dip, as we are unable to adequately account for the extremely unique role of the Sun in both the coordinate system and the operation of the Fermi-LAT. Further studies are necessary on an event-reconstruction level in order to verify the presence and characteristics of the 30-50~GeV spectral feature.

\section{Statistical significance of the dip} 
\label{appendix:dip_significance}

In this section, we describe our calculation for the significance of the spectral dip between 30--50\,GeV. 

We first consider the case of the total flux~(Figure~\ref{fig:totalflux} and Figure~\ref{fig:threeyear}), using the 9-year averaged flux with the $\theta<0.5^{\circ}$ and $|b|>5^{\circ}$ selection as an example. For this data set, we only consider events $E_{\gamma}>10$\,GeV, as point-spread-function effect causes the spectrum to depart from the power-law behavior at low energies.  This choice is conservative.  To estimate the effect of adding the low energy information, we also consider the $\theta<1.0^{\circ}$ selection with photons $E_{\gamma}>1$\,GeV, where the power-law behavior is restored, with is also a reasonably approximation to the \sag\ flux. 

In each case, we first fit the data with a broken power-law plus a negative Gaussian line to approximate the dip feature~(BPL w/ line).  Because the energy, amplitude and width of the line is allowed to vary, the negative Gaussian completely absorbs the two points belonging to the spectral dip, yielding  $\chi^{2}$ values of nearly zero for the two dip points. Next, we compute the $\chi^{2}$ using the obtained power-law parameters, but without the negative line~(BPL w/o line).  The significance of the line can be obtained by the difference in the $\chi^{2}$ value.  

\begin{figure*}[tbp]
	\centering
	\includegraphics[width=1.0\textwidth]{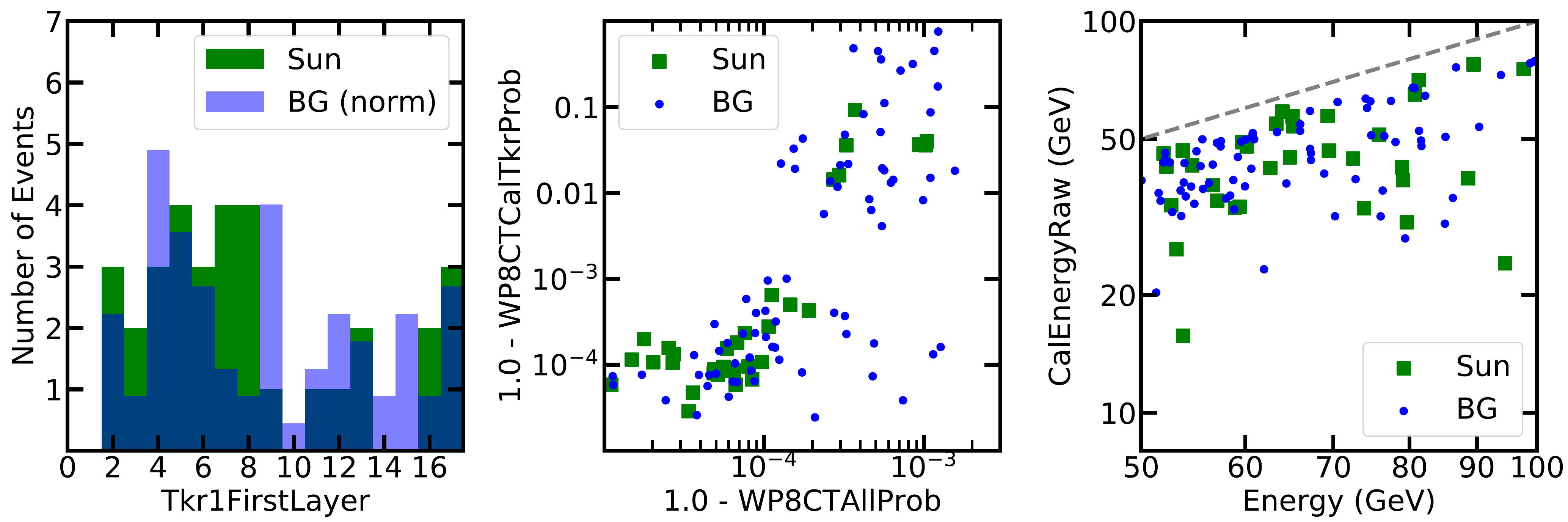}
	\includegraphics[width=1.0\textwidth]{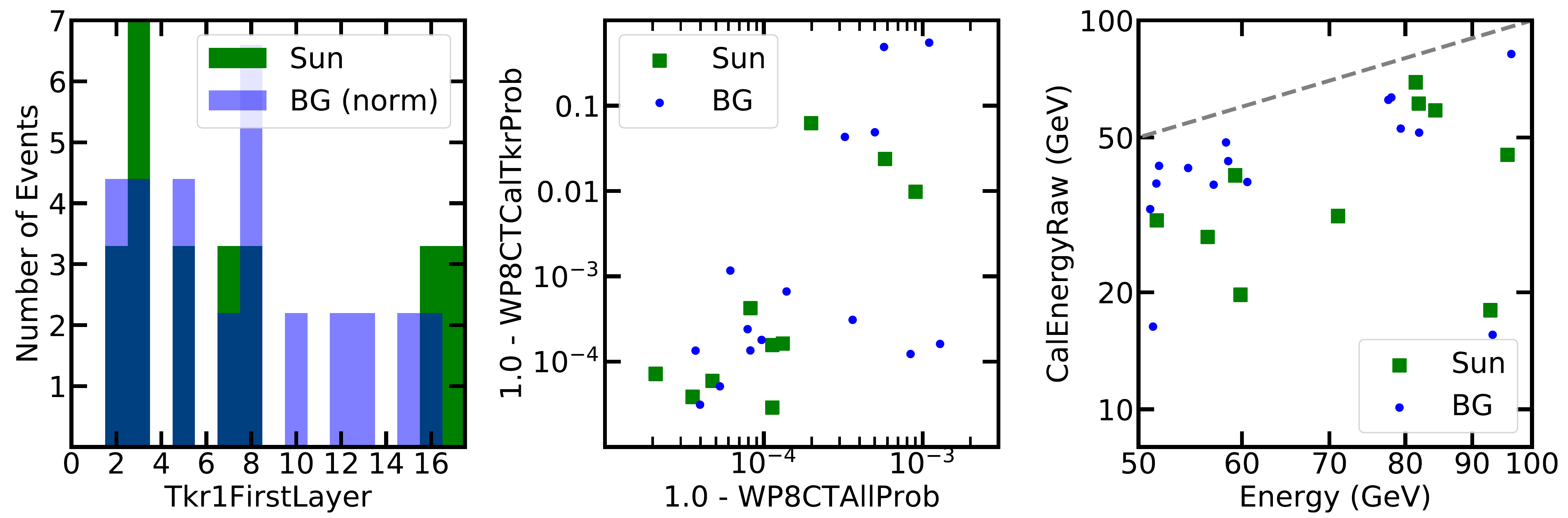}
	\caption{Distributions of Solar (green) and background (blue) events, observed with recorded energies between 50--100~GeV and during the period after solar minimum (top, after 2010-01-21), or before solar minimum (bottom, before 2010-01-21). Events are analyzed in three parameter spaces that might differentiate neutron and gamma-ray emission signals: (Left) The distribution of the first tracker layer (1-17) registering a signal during the event. (Center) The reconstructed probability that the event may be cosmic ray contamination (1-the gamma-ray probability, x-axis), compared to the same probability, but calculated without using information from the anti-Coincidence detector. (Right) The recorded gamma-ray energy (x-axis), compared to the raw energy deposited in the calorimeter. We discuss the potential for each quantity to differentiate neutrons from gamma rays in the main text, but find no evidence here for any significant differences between these distributions.}
	\label{fig:appendix:neutrondip}
\end{figure*}

In this case, $\Delta\chi^{2} = 36.3$ and $40.3$ for the $\theta<0.5^{\circ}$ and $\theta<1.0^{\circ}$ selection, respectively. 
The local significance of the dip can then be obtained by using the $\chi^{2}$ distribution of two degrees of freedom~(width and normalization of the line), which are 5.7$\sigma$ and 6.0$\sigma$ for these two cases.  The global significance can be obtained by considering three degrees of freedom~(including the energy of the dip), which are 5.4$\sigma$ and 5.7$\sigma$, respectively.  For the solar minimum \sag\ flux~(Figure~\ref{fig:vhe}), we apply the same procedure to the full spectrum above 1\,GeV, and obtain $\Delta\chi^{2} = 75.5$, which corresponds to local~(global) significance of 8.4$\sigma$~(8.2$\sigma$). These fits are shown in Figure~\ref{fig:line_significance}. In the main text, we quote the local significance of the dip with this procedure, with the total flux using the $\theta<0.5^{\circ}$ and $E_{\gamma}>10$\,GeV selection.

However, this definition of the significance is not unique.  We can alternatively compare the BPL\,w/\,line $\chi^{2}$ with the best-fitting broken power-law (BPL), which is fit to data that includes the dip points. In this case, the best-fit broken power-law adjusts to fit the dip points, reducing the local~(global) significance of the dip to 3.1$\sigma$~(2.7$\sigma$) for the $\theta<0.5^{\circ}$ 9-year total flux, 2.6$\sigma$~(2.3$\sigma$) for the $\theta<1.0^{\circ}$ 9-year total flux, and 5.9$\sigma$~(5.6$\sigma$) for the solar minimum \sag\ flux.  Intriguingly, the drop in the significance of the 9-year flux is significantly larger than that for the solar minimum flux. This is due to the fact that the hard solar minimum spectrum prevents the broken-power law from falling quickly near the position of the dip in order to account for the dip points. On the other hand, the 9-year averaged spectrum is softer, allowing the best-fit broken power law to more easily adjust to accommodate the dip.

\section{The Width of the Spectral Dip Compared to the Instrumental Energy Dispersion} 
\label{appendix:dispersion}

In the previous section, we have shown that the dip is a statistically significant feature. We now verify that the dip spectrum is not unphysically steep, given the finite energy-resolution of the Fermi-LAT. This is an important diagnostic test, as an extremely steep energy-spectrum would provide a clear indication of instrumental systematic effects. In fact, such a diagnostic was employed in searches of the 130~GeV gamma-ray line to argue for a systematic origin of the line, as the line-feature was more narrow than could be explained by the Fermi-LAT energy dispersion~\citep{Ackermann:2013uma}.  

In Figure~\ref{fig:energydispersion}, we produce three models of a two component solar gamma-ray emission spectrum. The first component is a simple dN/dE~$\propto$~E$^{-2}$ spectrum spanning from 0.1--300~GeV. We then modify this spectrum by placing a ``hole" in the gamma-ray emission centered at an energy of 40~GeV. We test three different combinations of hole widths and depths, including a 10~GeV range with 100\% of the total gamma-ray emission removed, a 20~GeV range with 50\% of the total gamma-ray emission removed, and a 40~GeV range with 25\% of the gamma-ray emission removed. In each case, we smear the injected spectrum by the Fermi-LAT energy dispersion on a photon by photon basis, utilizing distributions in both $\theta$ and instrumental energy-dispersion class that are identical to the photons recorded in our main analysis. This allows us to recover the parameters of the observed spectral dip and compare with our observations. We repeat this process several hundred times to build up a large statistical sample of possible dip configurations, and show the average results.

We note two important results. First, our steepest dip (10~GeV range with 100\% of the injected emission removed), produces an observed spectral feature which is both sharper and deeper than Fermi-LAT observations. This indicates that the dip is not physically impossible, given the finite accuracy of Fermi-LAT energy reconstructions. Second, we note that the dip is best fit not by an emission component which blocks 100\% of the emission in a given energy range, but instead by a component that includes approximately a 50\% inhibition of gamma-ray emission in a certain energy range, and which effects solar photons over an energy range of $\sim$20~GeV. 

\section{Examining Event-Level Photon Quality Near the Spectral Dip}
\label{appendix:neutrondip}

In Section~\ref{sec:neutronhypothesis}, we discussed the possibility that the observed spectral-dip may be due to a contamination of the gamma-ray spectrum by solar neutrons produced in the same hadronic interactions that produce the observed gamma-ray emission. Because neutrons undergo hadronic interactions within the Fermi-LAT detector, the energy-reconstruction of these events may be inaccurate, potentially producing unexpected spectral features. 

\begin{figure}[tbp]
\centering
\includegraphics[width=0.48\textwidth]{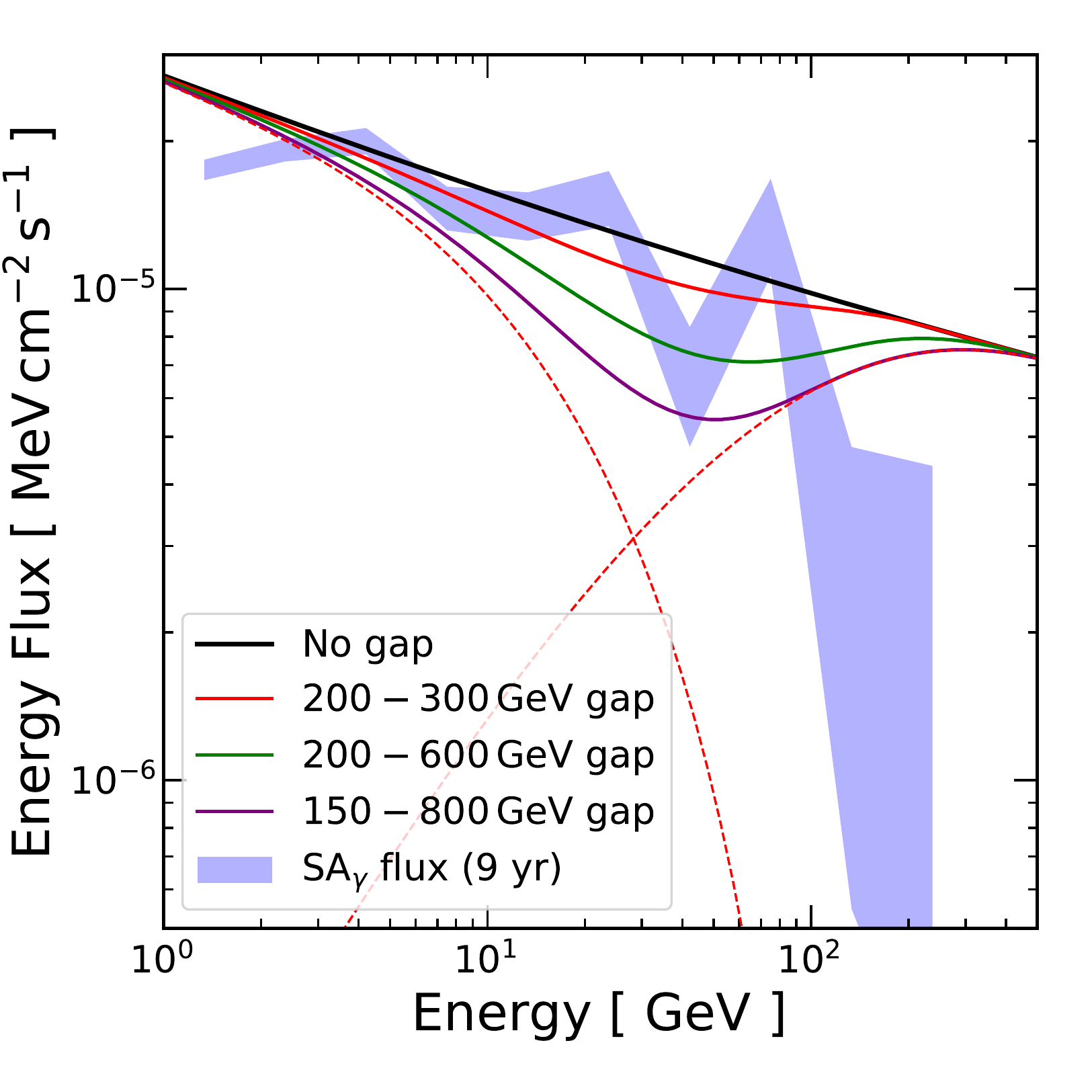}
\caption{
The observed 9-year \sag\ spectrum compared with gamma-ray spectra produced by proton-proton interaction with various proton spectra.  We consider a smooth $E_{\rm p}^{-2.2}$ spectrum~(no gap), and protons spectra that have an infinite gap in $200$--$300$\,GeV, $200$--$600$\,GeV, and $150$--$800$\,GeV.  The dashed curves show the low- and high-energy component separately for the $150$--$800$\,GeV case.  We find that it is difficult to produce the observed sharp dip with two proton spectra. 
}
\label{fig:proton_gap}
\end{figure}

We note that this possibility could be directly tested by comparing the shower reconstruction of events near the position of the 30-50 GeV spectral dip. Unfortunately, the full event-level shower reconstruction information is not publicly available to those outside the Fermi-LAT collaboration. We note that the re-emergence of the solar spectrum above 50~GeV stands as the most puzzling portion of the solar spectrum, as the sharp falloff near 30~GeV could otherwise be interpreted as as a simple exponential suppression, as predicted in the SSG1991 mechanism. In this scenario, the emission above 50~GeV would be produced predominantly by neutrons with inaccurate energy reconstructions. 

In this section, we instead examine several event-level qualities that may be correlated with hadronic interactions within the Fermi-LAT detector. However, we stress here that these qualities are ``black boxes", and we have no pure neutron source to examine in order to determine whether any of these qualities would, in fact, provide reasonable neutron/gamma-ray separation.

In Figure~\ref{fig:appendix:neutrondip}, we examine events with energies between 50--100~GeV, that were recorded after the end of solar minimum (after 2010-01-21). The second cut ensures that no high-energy gamma-rays (observed primarily during solar minimum) were likely to be present in the event analysis. We cut our photon selection between Solar Disk events, recorded within 0.75$^\circ$ of the contemporaneous solar position, and background events, recorded between 0.75$^\circ$ and 5$^\circ$ from the Sun. Because these background events were recorded relatively close to the solar position (but were not produced by the solar disk itself), they will have similar distributions in Fermi-LAT instrumental coordinates, implying that their event reconstructions should be similar. 

For each set of events, we examine three possible parameter spaces that might differentiate neutron and gamma-ray interactions within the Fermi-LAT detector. The first (left) is the parameter Tkr1FirstLayer, which specifies the first silicon layer to record the observed event. Because neutron interaction length is much longer than the gamma-ray radiation length in Tungsten, we would expect the majority of neutron events to be observed predominantly in the final layers of the Fermi-LAT instrument. The second (center) is a comparison of the parameters WP8CTAllProb and WP8CTCalTkProb, which compare the probability of cosmic-ray contamination using, or ignoring, information from the Anti-Coincidence detector, respectively. Because the anti-Coincidence detector is the primary rejection technique for charged cosmic-rays, but is unlikely to be triggered by neutrons, we might expect neutron events to have abnormally large cosmic-ray probabilities when information from the Anti-Coincidence detector is ignored. The third (right) is a comparison of the reconstructed energy of the event, compared to the parameter CalEnergyRaw, which denotes the energy deposited in the calorimeter. We may expect any event (neutron or gamma-ray) that has an overestimated energy, to have abnormally large variations between the recorded and reconstructed gamma-ray energies. 

However, for each test, we find no statistically significant difference between source and background events, indicating that there are no obvious differences between solar disk events and nearby background regions. However, we again caution the reader that these tests do not definitively rule out the neutron hypothesis, as we cannot calculate the sensitivity of any of these tests to a true neutron signal.


\begin{figure*}[tbp]
\begin{center}
\includegraphics[width=0.85\columnwidth]{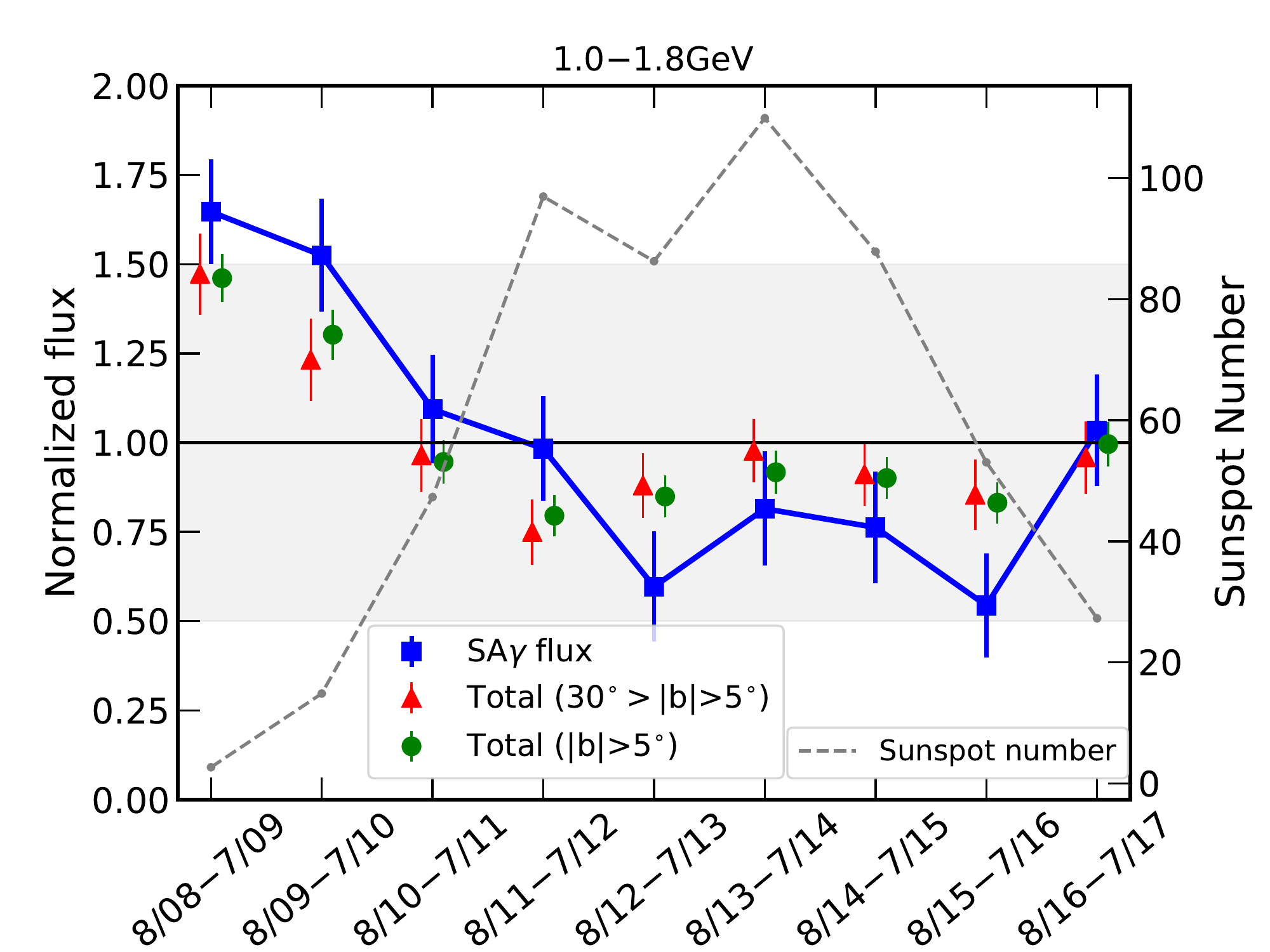}\vspace{-0.05cm}
\includegraphics[width=0.85\columnwidth]{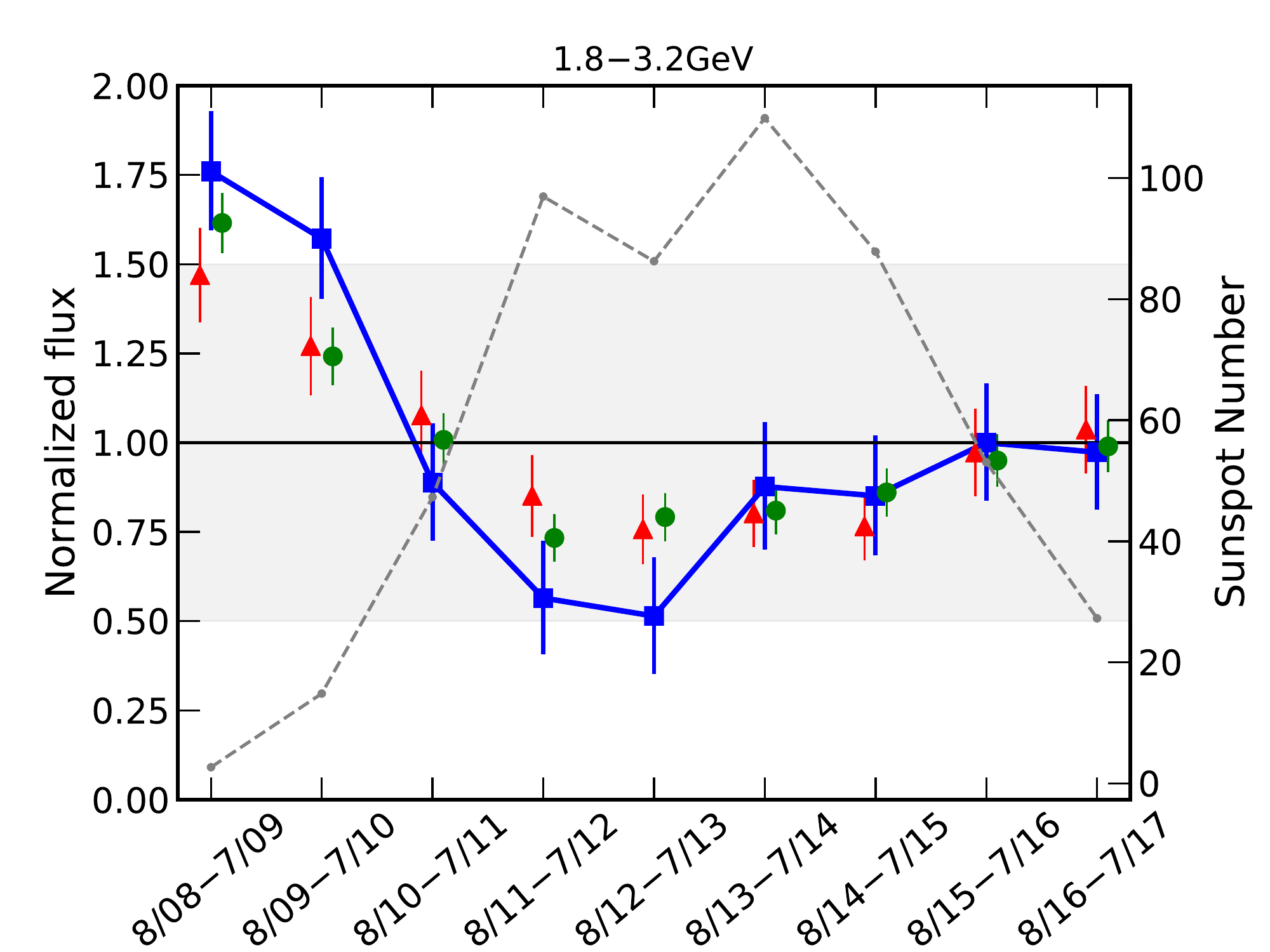}
\includegraphics[width=0.85\columnwidth]{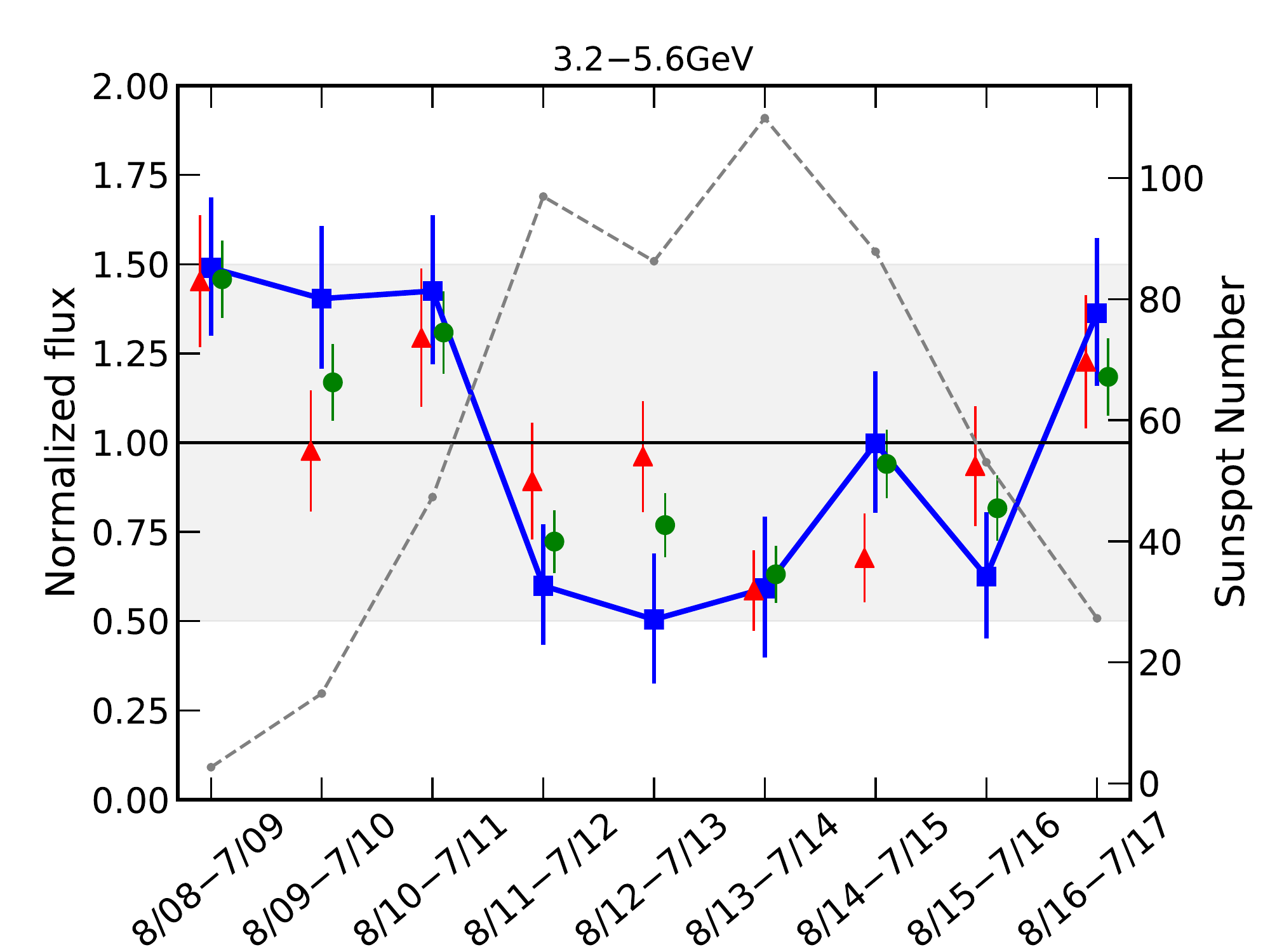}\vspace{-0.05cm}
\includegraphics[width=0.85\columnwidth]{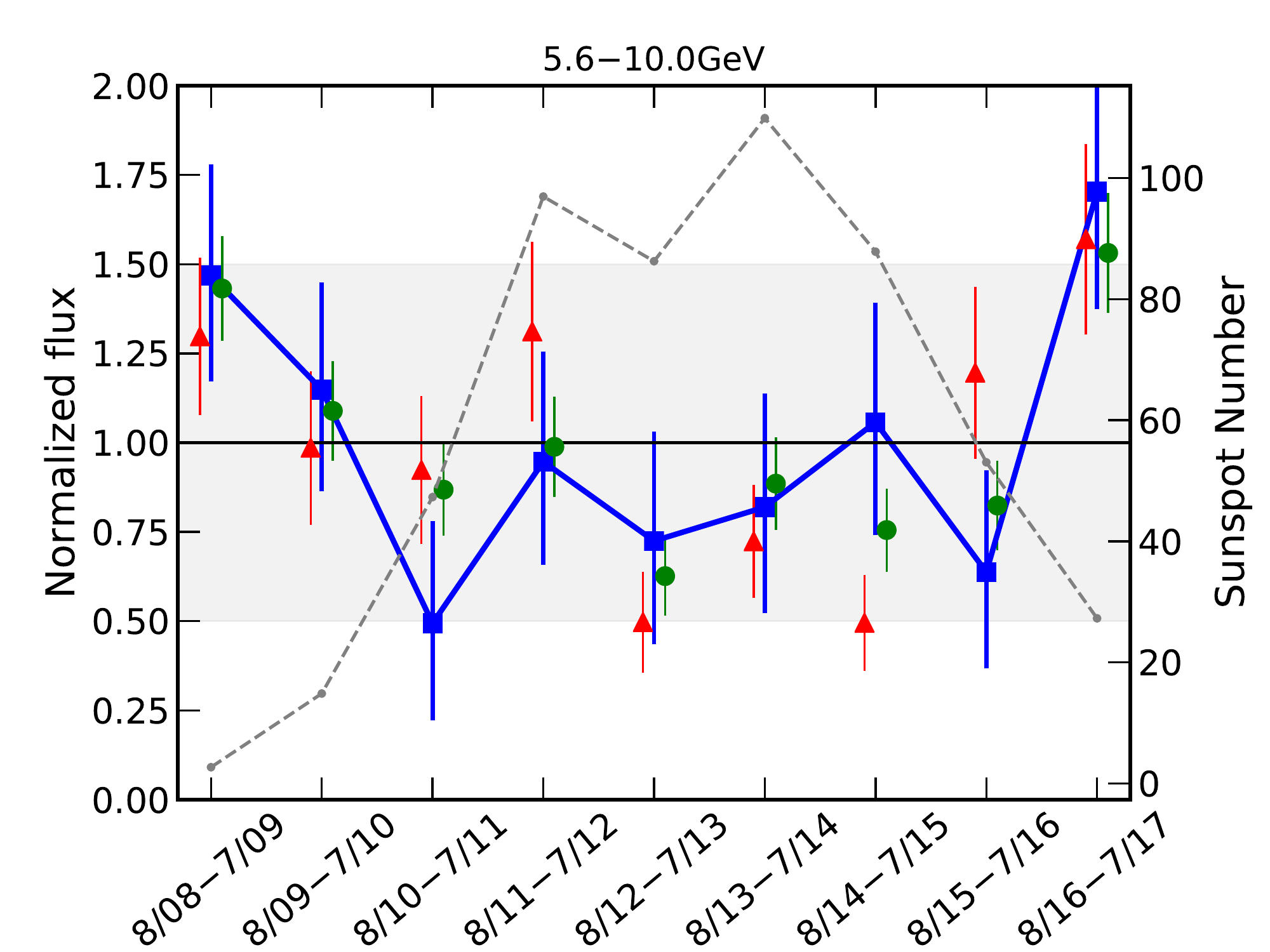}
\includegraphics[width=0.85\columnwidth]{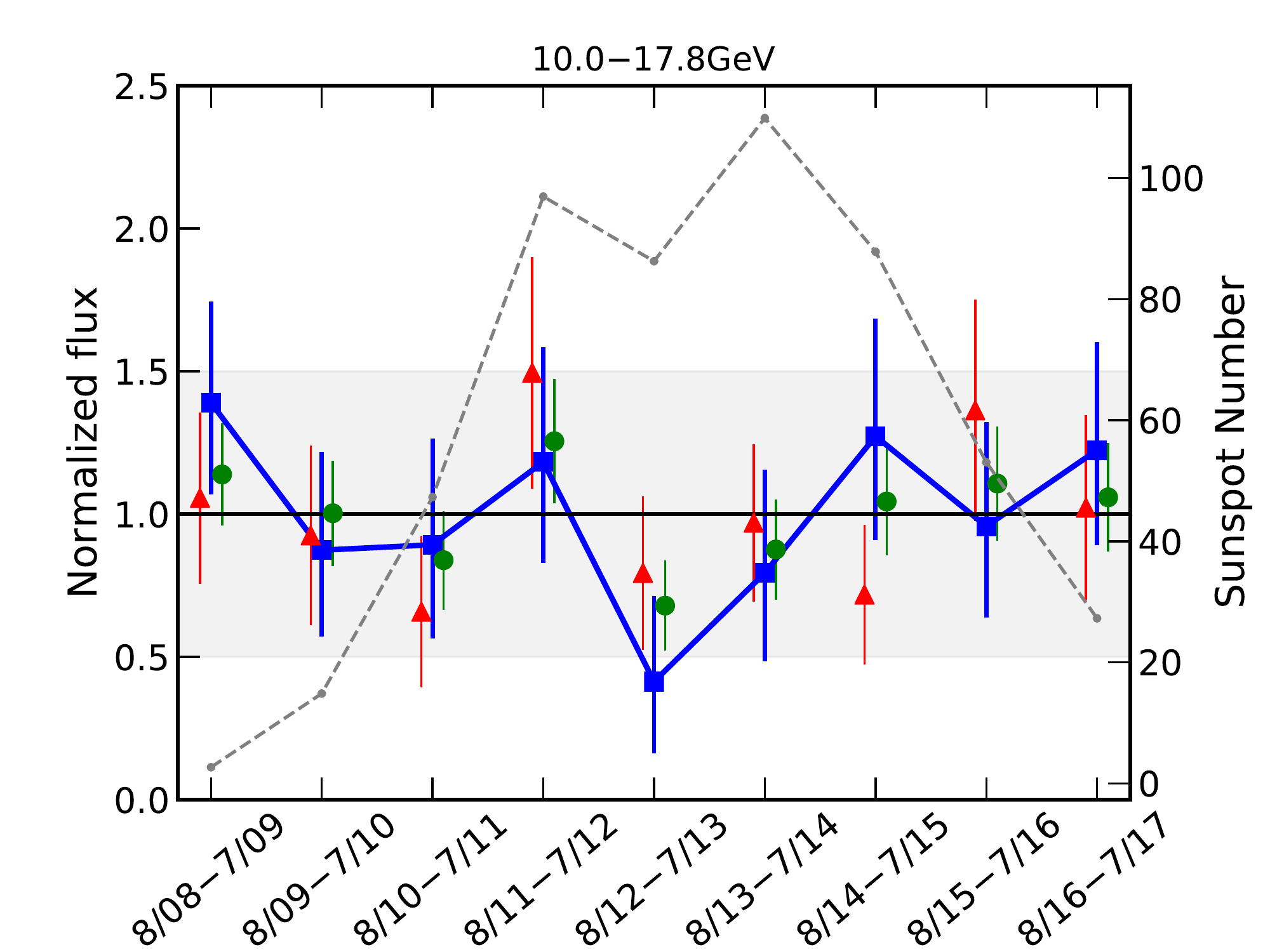}\vspace{-0.05cm}
\includegraphics[width=0.85\columnwidth]{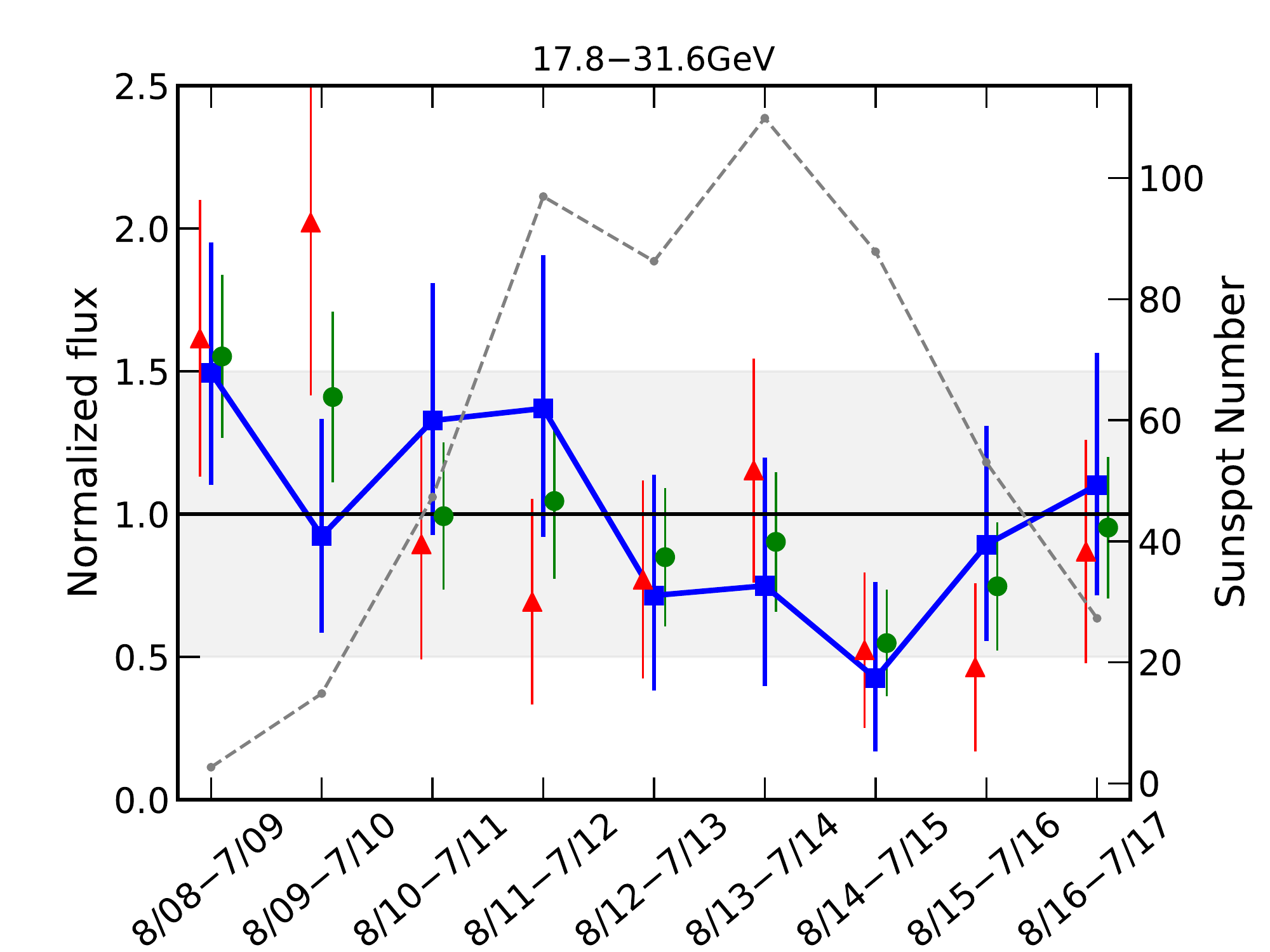}
\includegraphics[width=0.85\columnwidth]{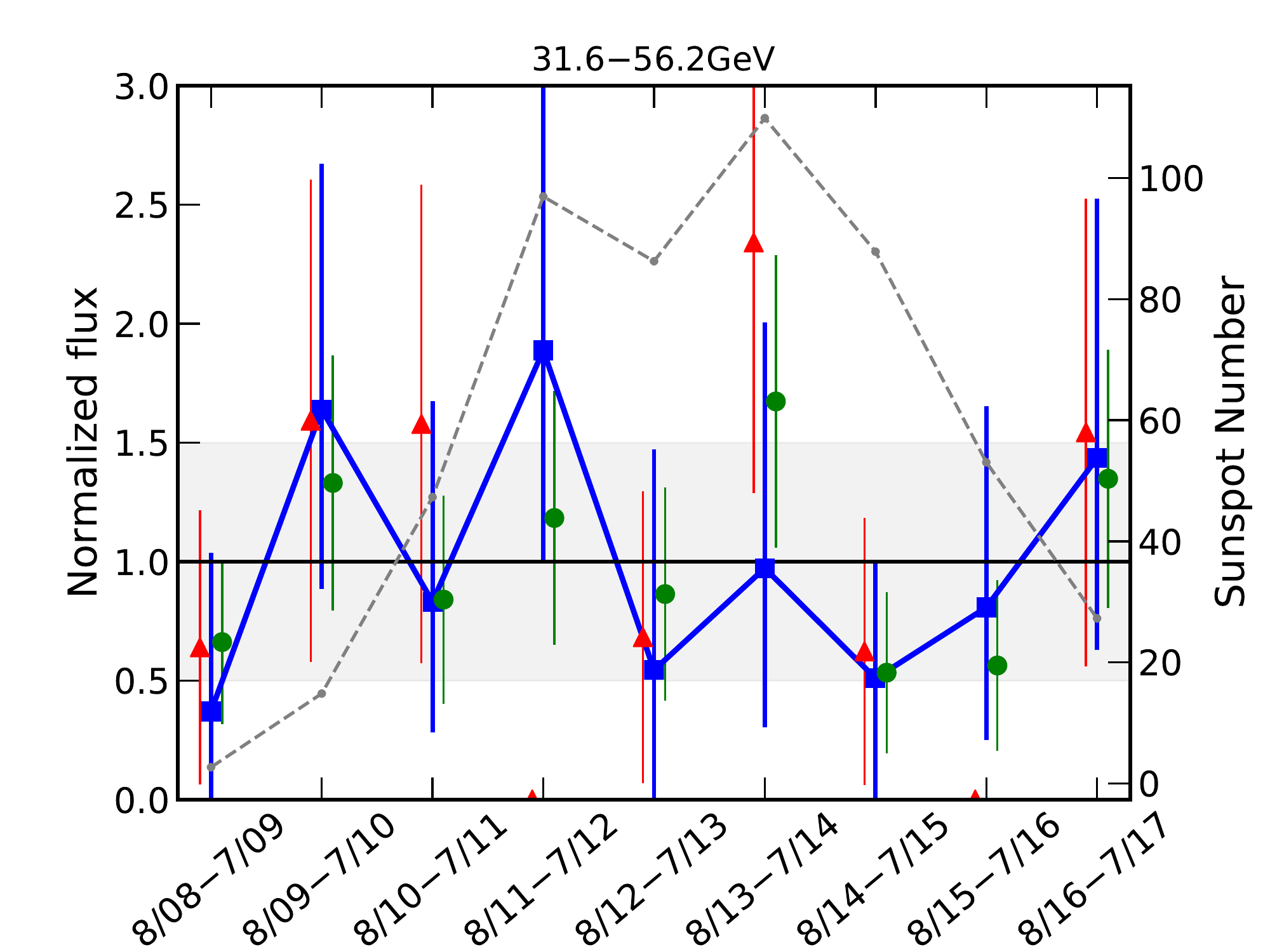}\vspace{-0.05cm}
\includegraphics[width=0.85\columnwidth]{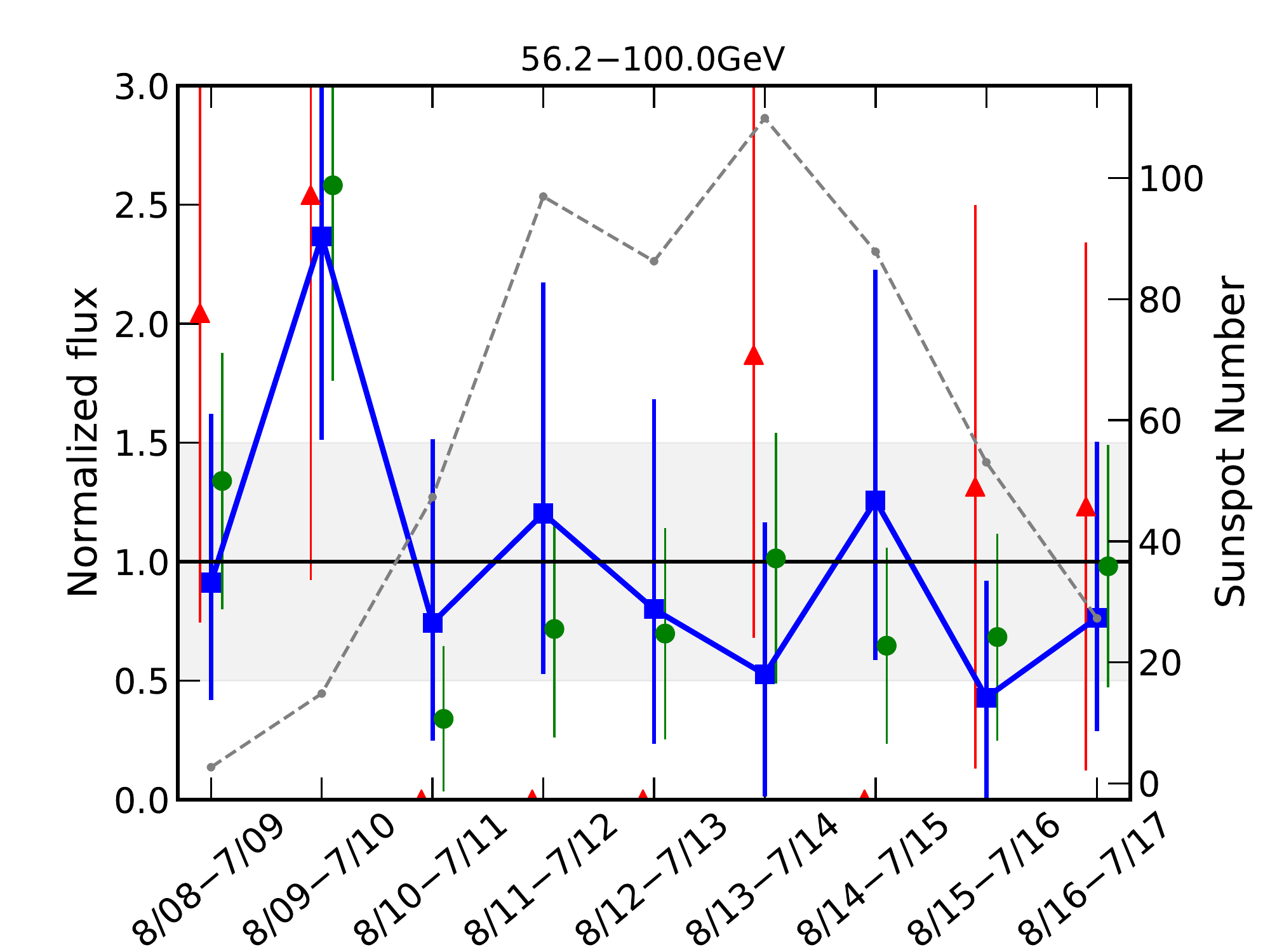}
\caption{ The time variation of the \sag\ flux and the total flux, in 4 logarithmic bins per decade.  The time variation and the anticorrelation with solar activity can be clearly seen between 1--10\,GeV.  For comparison, we also show the time variation by the independent total flux with latitude cut $5^{\circ}<|b|<30^{\circ}$~(red points) and the combined total flux~($|b|>5^{\circ}$).  The amplitude of the variation for the total flux is reduced between 1-3\,GeV due to contamination from time-independent backgrounds.  Otherwise, the total flux is in good agreement with our default \sag\ analysis. }
\label{fig:appendixtimevariation}
\end{center}
\end{figure*}

\section{Testing dip from modified proton spectrum} 
\label{appendix:proton_gap}

In this section, we consider the possibility that the observed gamma-ray dip between 30--50\,GeV is produced by two different proton spectra via hadronic interactions.  We consider various configurations of proton spectrum that may produce the gamma-ray spectrum, with the proton-proton~\cite{Kelner:2006tc} interaction optimistically set in the thin regime~(interaction length $\ll$\,1).

Ignoring the dip, the gamma-ray spectrum below 100\,GeV is roughly $E_\gamma^{-2.2}$, which can be obtained by a input proton spectrum 
$E_{p}^{-2.2}$.  This is shown by the black line in Figure~\ref{fig:proton_gap}~(no gap).  To test the two component hypothesis, we artificially make a infinite-deep gap in the $E_{p}^{-2.2}$ proton spectrum.  The proton gap thus mimics a rapid decay and a rapid rise for the low- and high-energy proton component, respectively.  The resultant spectra, corresponding to proton gaps in $200$--$300$\,GeV, $200$--$600$\,GeV, and $150$--$800$\,GeV, are shown in Figure~\ref{fig:proton_gap}.  We find that in all cases, the dip in the computed spectra are too shallow and wide to explain the observed dip feature.   This is because the $E_{p}$-to-$E_\gamma$ kernel (i.e. gamma-ray spectrum per proton), which is peaked near $E_\gamma \sim 0.1\,E_{p}$, is far too wide (see, e.g., \cite{Kelner:2006tc}), making the high-energy component rises slowly.  

We also test the case that the high-energy proton spectrum has a harder spectral index.  However, we find that the changing the proton shape cannot make gamma-ray dip sharper.  This is because the gamma-ray spectral shape just below the cutoff only depends on the shape of $E_{p}$-to-$E_\gamma$ kernel, not the shape of proton spectrum.  

Lastly, we note that $E_{p}$-to-$E_\gamma$ kernel can be sharper if only contributions from a small forward cone is considered.  This is because the low energy gamma rays mostly emitted at larger emission angle.  However, it is not clear how such a configuration are achieved in a realistic solar atmospheric environment. 

From these studies, we find that it is difficult to produce the depth and the width of the observed gamma-ray dip using two different populations of proton, unless taking extreme kinematic regime.



\section{Time variations} 
\label{appendix:time_variation}

In the main text, we calculated the time variation of the \sag\ flux, using only data with $|b|>$30$^\circ$ and two logarithmic energy bins per decade, in order to present the main result with minimal statistical fluctuations. In Figure~\ref{fig:appendixtimevariation}, we show the same result, instead using four logarithmic bins per decade, and additionally showing the raw (not background subtracted) results for data in both the latitude range $|b|>$30$^\circ$ and $|b|>$5$^\circ$. These latter datasets have fewer systematic assumptions (because no modeling is done to remove either diffuse or IC gamma-rays), but show variations with smaller amplitudes (because these background sources are in steady state).


\newpage

\bibliographystyle{h-physrev}
\bibliography{reference}

\begin{thebibliography}{10}

\bibitem{Zhou:2016ljf}
B.~Zhou, K.~C.~Y. Ng, J.~F. Beacom, and A.~H.~G. Peter,
\newblock Phys. Rev. {\bf D96}, 023015 (2017), 1612.02420.

\bibitem{Seckel:1991ffa}
D.~Seckel, T.~Stanev, and T.~K. Gaisser,
\newblock Astrophys. J. {\bf 382}, 652 (1991).

\bibitem{Potgieter:2013pdj}
M.~Potgieter,
\newblock Living Rev. Solar Phys. {\bf 10}, 3 (2013), 1306.4421.

\bibitem{2014A&ARv..22...78W}
T.~{Wiegelmann}, J.~K. {Thalmann}, and S.~K. {Solanki},
\newblock A\&A Rev. {\bf 22}, 78 (2014), 1410.4214.

\bibitem{2017PDU....17...13B}
S.~{Bertolucci}, K.~{Zioutas}, S.~{Hofmann}, and M.~{Maroudas},
\newblock Physics of the Dark Universe {\bf 17}, 13 (2017), 1602.03666.

\bibitem{Leane:2017vag}
R.~K. Leane, K.~C.~Y. Ng, and J.~F. Beacom,
\newblock Phys. Rev. {\bf D95}, 123016 (2017), 1703.04629.

\bibitem{Arina:2017sng}
C.~Arina, M.~Backovi\'c, J.~Heisig, and M.~Lucente,
\newblock Phys. Rev. {\bf D96}, 063010 (2017), 1703.08087.

\bibitem{Orlando:2006zs}
E.~Orlando and A.~Strong,
\newblock Astrophys. Space Sci. {\bf 309}, 359 (2007), astro-ph/0607563.

\bibitem{Moskalenko:2006ta}
I.~V. Moskalenko, T.~A. Porter, and S.~W. Digel,
\newblock Astrophys. J. {\bf 652}, L65 (2006), astro-ph/0607521,
\newblock [Erratum: Astrophys. J. 664, L143 (2007)].

\bibitem{Orlando:2013pza}
E.~Orlando and A.~Strong,
\newblock (2013), 1307.6798.

\bibitem{Kafexhiu:2018wmh}
E.~Kafexhiu, C.~Romoli, A.~M. Taylor, and F.~Aharonian,
\newblock (2018), 1803.02635.

\bibitem{Fermi-LAT:2013cla}
Fermi-LAT Collaboration, M.~Ajello {\em et~al.},
\newblock Astrophys. J. {\bf 789}, 20 (2014), 1304.5559.

\bibitem{Pesce-Rollins:2015hpa}
M.~Pesce-Rollins {\em et~al.},
\newblock Astrophys. J. {\bf 805}, L15 (2015), 1505.03480.

\bibitem{Ackermann:2017uer}
Fermi-LAT, M.~Ackermann {\em et~al.},
\newblock Astrophys. J. {\bf 835}, 219 (2017), 1702.00577.

\bibitem{Orlando:2008uk}
E.~Orlando and A.~W. Strong,
\newblock Astron. Astrophys. {\bf 480}, 847 (2008), 0801.2178.

\bibitem{ROG:ROG56}
J.~F. Dolan and G.~G. Fazio,
\newblock Reviews of Geophysics {\bf 3}, 319 (1965).

\bibitem{JGRA:JGRA13592}
D.~J. Thompson, D.~L. Bertsch, D.~J. Morris, and R.~Mukherjee,
\newblock Journal of Geophysical Research: Space Physics {\bf 102}, 14735
  (1997).

\bibitem{Abdo:2011xn}
Fermi-LAT Collaboration, A.~A. Abdo {\em et~al.},
\newblock Astrophys. J. {\bf 734}, 116 (2011), 1104.2093.

\bibitem{Ng:2015gya}
K.~C.~Y. Ng, J.~F. Beacom, A.~H.~G. Peter, and C.~Rott,
\newblock Phys. Rev. {\bf D94}, 023004 (2016), 1508.06276.

\bibitem{Aielli:2006cj}
Argo-YBJ, G.~Aielli {\em et~al.},
\newblock Nucl. Instrum. Meth. {\bf A562}, 92 (2006).

\bibitem{Hibino:1988er}
TIBET ASgamma Collaboration, K.~Hibino {\em et~al.},
\newblock Nucl. Phys. Proc. Suppl. {\bf 10B}, 219 (1989).

\bibitem{Abeysekara:2013tza}
A.~U. Abeysekara {\em et~al.},
\newblock Astropart. Phys. {\bf 50-52}, 26 (2013), 1306.5800.

\bibitem{Abeysekara:2017mjj}
A.~U. Abeysekara {\em et~al.},
\newblock Astrophys. J. {\bf 843}, 39 (2017), 1701.01778.

\bibitem{Zhen:2014zpa}
Z.~Cao,
\newblock Frascati Phys. Ser. {\bf 58}, 331 (2014).

\bibitem{He:2016del}
LHAASO Collaboration, H.~He,
\newblock PoS {\bf ICRC2015}, 1010 (2016).

\bibitem{Moskalenko:1991hm}
I.~V. Moskalenko, S.~Karakula, and W.~Tkaczyk,
\newblock Astron. Astrophys. {\bf 248}, L5 (1991).

\bibitem{Moskalenko:1993ke}
I.~V. Moskalenko and S.~Karakula,
\newblock J. Phys. {\bf G19}, 1399 (1993).

\bibitem{Ingelman:1996mj}
G.~Ingelman and M.~Thunman,
\newblock Phys. Rev. {\bf D54}, 4385 (1996), hep-ph/9604288.

\bibitem{Hettlage:1999zr}
C.~Hettlage, K.~Mannheim, and J.~G. Learned,
\newblock Astropart. Phys. {\bf 13}, 45 (2000), astro-ph/9910208.

\bibitem{Fogli:2006jk}
G.~L. Fogli, E.~Lisi, A.~Mirizzi, D.~Montanino, and P.~D. Serpico,
\newblock Phys. Rev. {\bf D74}, 093004 (2006), hep-ph/0608321.

\bibitem{Arguelles:2017eao}
C.~A. Argüelles, G.~de~Wasseige, A.~Fedynitch, and B.~J.~P. Jones,
\newblock JCAP {\bf 1707}, 024 (2017), 1703.07798.

\bibitem{Ng:2017aur}
K.~C.~Y. Ng, J.~F. Beacom, A.~H.~G. Peter, and C.~Rott,
\newblock Phys. Rev. {\bf D96}, 103006 (2017), 1703.10280.

\bibitem{Edsjo:2017kjk}
J.~Edsjo, J.~Elevant, R.~Enberg, and C.~Niblaeus,
\newblock JCAP {\bf 1706}, 033 (2017), 1704.02892.

\bibitem{Masip:2017gvw}
M.~Masip,
\newblock Astropart. Phys. {\bf 97}, 63 (2018), 1706.01290.

\bibitem{icecubesolar}
IceCube, S.~In and C.~Rott,
\newblock {Solar atmospheric neutrino search with IceCube},
\newblock in {\em {Proceedings, 35th International Cosmic Ray Conference (ICRC
  2017): Bexco, Busan, Korea, July 12-20, 2017}}, 2017, 1710.01194.

\bibitem{Adrian-Martinez:2016fdl}
KM3Net, S.~Adrian-Martinez {\em et~al.},
\newblock J. Phys. {\bf G43}, 084001 (2016), 1601.07459.

\bibitem{Linden:2018}
T.~Linden {\em et~al.},
\newblock Phys. Rev. Lett. {\bf 121}, 131103 (2018), 1803.05436.

\bibitem{flare_list}
http://hesperia.gsfc.nasa.gov/fermi/lat/qlook/lat$\_$events\\.txt .

\bibitem{Rolke:2004mj}
W.~A. Rolke, A.~M. Lopez, and J.~Conrad,
\newblock Nucl. Instrum. Meth. A {\bf 551}, 493 (2005), arXiv:physics/0403059.

\bibitem{Cowan:2010js}
G.~Cowan, K.~Cranmer, E.~Gross, and O.~Vitells,
\newblock Eur. Phys. J. C {\bf 71}, 1554 (2011), arXiv:1007.1727.

\bibitem{Ackermann:2014usa}
Fermi-LAT, M.~Ackermann {\em et~al.},
\newblock Astrophys. J. {\bf 799}, 86 (2015), 1410.3696.

\bibitem{Orlando:2013nga}
E.~Orlando and A.~W. Strong,
\newblock Nucl. Phys. Proc. Suppl. {\bf 239-240}, 266 (2013), 1303.5491.

\bibitem{Note1}
http://www.sws.bom.gov.au/Solar/1/6.

\bibitem{2012ApJS..203....4A}
M.~{Ackermann} {\em et~al.},
\newblock ApJS {\bf 203}, 4 (2012), 1206.1896.

\bibitem{fermifl8y}
https://fermi.gsfc.nasa.gov/ssc/data/access/lat/fl8y/\\FL8Y\_description\_v7.pdf
  .

\bibitem{fermicaveat}
https://fermi.gsfc.nasa.gov/ssc/data/analysis/LAT\_\\caveats.html .

\bibitem{Nisa:2017xpf}
HAWC, M.~U. Nisa,
\newblock {Probing Cosmic-ray Propagation with TeV Gamma Rays from the Sun
  Using the HAWC Observatory},
\newblock in {\em {Proceedings, 35th International Cosmic Ray Conference (ICRC
  2017): Bexco, Busan, Korea, July 12-20, 2017}}, 2017, 1708.03732.

\bibitem{argo:2016}
Argo-YBJ, Z.~Li, S.~Z. Chen, and H.~H. He,
\newblock 7th Workshop on Air Shower Detection at High Altitude .

\bibitem{Amenomori:2013own}
Tibet ASgamma, M.~Amenomori {\em et~al.},
\newblock Phys. Rev. Lett. {\bf 111}, 011101 (2013), 1306.3009.

\bibitem{Patrignani:2016xqp}
Particle Data Group, C.~Patrignani {\em et~al.},
\newblock Chin. Phys. {\bf C40}, 100001 (2016).

\bibitem{2016SoPh..291.1241M}
Y.~{Muraki} {\em et~al.},
\newblock Solar Physics {\bf 291}, 1241 (2016), 1508.04923.

\bibitem{Muraki:2017mhm}
Y.~Muraki {\em et~al.},
\newblock (2017), 1706.09082.

\bibitem{Kistler:2006hp}
M.~D. Kistler and J.~F. Beacom,
\newblock Phys. Rev. {\bf D74}, 063007 (2006), astro-ph/0607082.

\bibitem{Aartsen:2016zhm}
IceCube, M.~G. Aartsen {\em et~al.},
\newblock Eur. Phys. J. {\bf C77}, 146 (2017), 1612.05949.

\bibitem{Bagnaschi:2017tru}
E.~Bagnaschi {\em et~al.},
\newblock (2017), 1710.11091.

\bibitem{Costa:2017gup}
J.~C. Costa {\em et~al.},
\newblock Eur. Phys. J. {\bf C78}, 158 (2018), 1711.00458.

\bibitem{Bell:2011sn}
N.~F. Bell and K.~Petraki,
\newblock JCAP {\bf 1104}, 003 (2011), 1102.2958.

\bibitem{Feng:2016ijc}
J.~L. Feng, J.~Smolinsky, and P.~Tanedo,
\newblock Phys. Rev. {\bf D93}, 115036 (2016), 1602.01465.

\bibitem{fermi_performance}
https://www.slac.stanford.edu/exp/glast/groups/canda\\/lat\_Performance.htm .

\bibitem{Ackermann:2013uma}
Fermi-LAT, M.~Ackermann {\em et~al.},
\newblock Phys. Rev. {\bf D88}, 082002 (2013), 1305.5597.

\bibitem{Kelner:2006tc}
S.~R. Kelner, F.~A. Aharonian, and V.~V. Bugayov,
\newblock Phys. Rev. {\bf D74}, 034018 (2006), astro-ph/0606058,
\newblock [Erratum: Phys. Rev. D79, 039901 (2009)].

\end{thebibliography}

\end{document}